\documentclass[11pt, letterpaper]{article}

\usepackage[utf8]{inputenc}
\usepackage[T1]{fontenc}
\usepackage{microtype}
\usepackage{amsmath}
\usepackage{mleftright}
\usepackage{amsthm}
\usepackage{natbib}
\usepackage{hyperref}
\usepackage{enumitem}
\usepackage[margin=1in]{geometry}
\usepackage{xcolor}
\usepackage{amssymb}
\hypersetup{
	colorlinks,
	linkcolor={red!50!black},
	citecolor={blue!50!black},
	urlcolor={blue!80!black}
}
\usepackage{graphicx}
\usepackage{hyperref}
\usepackage{cleveref}
\usepackage[utf8]{inputenc}
\usepackage{natbib}
\usepackage{mathtools}
\usepackage{tabularx}  
\usepackage{xspace}
\usepackage{enumitem}
\usepackage[noend]{algpseudocode}
\usepackage{algorithm}
\usepackage{tikz}
\usepackage{booktabs}
\usepackage{comment}
\usepackage{soul}
\usepackage{csquotes}
\usepackage{color-edits}
\usepackage{dsfont}
\usepackage{stmaryrd}
\usepackage[normalem]{ulem}

\newtheorem{theorem}{Theorem}[section]

\newtheorem{lemma}{Lemma}[section]
\newtheorem{remark}{Remark}[section]
\newtheorem{corollary}{Corollary}[section]
\newtheorem{proposition}{Proposition}[section]
\newtheorem{definition}{Definition}[section]

\newcommand{\newreptheorem}[2]
{\newtheorem*{rep@#1}{\rep@title}
\newenvironment{rep#1}[1]
{\def\rep@title{#2 \ref*{##1}}
\begin{rep@#1}}{\end{rep@#1}}}

\newreptheorem{theorem}{Theorem}
\newreptheorem{lemma}{Lemma}
\newreptheorem{claim}{Claim}
\newreptheorem{observation}{Observation}
\newreptheorem{proposition}{Proposition}
\newreptheorem{corollary}{Corollary}

\newcommand{\E}{\mathbb{E}}
\newcommand{\R}{\mathbb{R}}
\def\N{\mathbb{N}}
\def\eps{\varepsilon}

\def\cS{\mathcal{S}}
\def\cA{\mathcal{A}}

\def\cP{\mathcal{P}}

\def\Var{\operatorname{Var}}

\def\from{\leftarrow}
\def\dif#1{\mathop{d #1}}
\def\der#1#2{\frac{\dif{#1}}{\dif{#2}}}

\def\argmax{\operatornamewithlimits{arg\,max}}

\newcommand{\Ber}[1]{\operatorname{Ber}\prn{#1}}
\newcommand{\Bin}[1]{\operatorname{Bin}\prn{#1}}
\newcommand{\Poi}[1]{\operatorname{Poi}\prn{#1}}
\newcommand{\PB}[1]{\operatorname{PB}\prn{#1}}

\def\AR{\operatorname{AR}}

\def\bigO#1{O\prn{#1}}

\def\bigOm#1{\operatorname{\Omega}\prn{#1}}

\def\floor#1{\left\lfloor #1 \right\rfloor}

\def\set#1{\left\{ #1 \right\}}
\def\abs#1{\left| #1 \right|}

\newcommand{\prn}[2][\!]{\mleft({#2}\mright)}%
\def\brk#1{\left[#1\right]}
\def\1{\mathbf{1}}		
\def\midd{\:\middle|\:}

\addauthor[shaddin]{sd}{green}
\usepackage[textsize=small,textwidth=2cm]{todonotes}

\definecolor{TODOcolor}{cmyk}{0.05,0,0,0}
\definecolor{TODOtxtcolor}{cmyk}{0,1,1,0}

\newsavebox{\TODOBox}

\usepackage{silence}
\allowdisplaybreaks

\title{Optimal Contention Resolution from Samples}

\author{
Shaddin Dughmi
\thanks{Department of Computer Science, University of Southern California. {\tt shaddin@usc.edu}. Supported by the Air Force Office of Scientific Research under award number FA9550-24-1-0261. Part of this work was done while the author was on sabbatical as the Carter and Tania Neild visiting professor at Northwestern University, as well as a visiting professor in the Data Science Institute at the University of Chicago.}
\and
Vasilis Livanos
\thanks{Department of Computer Science, University of Southern California. {\tt vas.livanos@gmail.com}. Supported by the Air Force Office of Scientific
Research under award number FA9550-24-1-0261.}
\and
Pranav Nuti
\thanks{Booth School of Business, University of Chicago. {\tt pranav.nuti@chicagobooth.edu}.}
}

\date{}

\begin{document}
\maketitle

\begin{abstract} 
Contention resolution schemes (CRSs) convert random sets which are ex-ante feasible for some constraint into sets which are feasible ex post, and their online variants (ROCRSs and OCRSs) have broad applications in Bayesian and stochastic optimization, including prophet inequalities and stochastic probing. In many such applications, however, full distributional knowledge is an unattainable requirement. This motivates the study of contention resolution schemes which employ a limited number of samples from the underlying distributions. 

In this paper, we study sample-based contention resolution for $k$-uniform matroids,  for both uniformly random and adversarial arrival orders, and compare their balance guarantees to the best possible with full distributional knowledge. For a single-item constraint, we show that one sample per element suffices for an optimal $(1-1/e)$-balanced ROCRS. For adversarial arrivals, however, we prove that no single-sample OCRS can be $1/2$-balanced, even for known arrival orders, separating contention resolution from prophet inequalities in the single-sample setting. Perhaps surprisingly, we show that two samples are necessary and sufficient for the optimal $1/2$ guarantee. We also complement our results with a single-sample $0.48$-balanced scheme.

For general $k$, we give a simple single-sample ROCRS that achieves an optimal balance ratio of $1-e^{-k}k^k/k!$. In adversarial order, we obtain a single-sample $\left(1 - \operatorname{O}(1/\sqrt{k})\right)$-balanced scheme, matching the optimal asymptotic dependence on $k$.

Finally, we complete the picture by characterizing the minimum number of samples needed to match the optimal balance ratio achievable with full distributional knowledge. For random arrivals, one sample suffices for $k \in \set{1,2,3,4}$, and zero samples suffice for $k \geq 5$. For an adversarial arrival order, even known in advance, and any $k \geq 2$, we show that no constant number of samples suffices for achieving optimal balance.
\end{abstract}

\medskip
\noindent
{\small \textbf{Keywords:} contention resolution; decision making under uncertainty; online algorithms; rounding algorithms.
}
\medskip
\medskip
\medskip
\noindent

\clearpage

\section{Introduction}\label{sec:introduction}

Contention resolution has emerged as a fundamental algorithmic primitive, abstracting the familiar task of turning a random set which satisfies a system of constraints ex-ante,  i.e., on average, into one which satisfies them ex-post, i.e., always. Originally formulated as a tool for enforcing feasibility in randomized rounding of linear programs~\cite{crs}, it has since found many more applications to online and stochastic decision making problems such as prophet inequalities, stochastic probing, secretary problems, and more~(e.g.~\cite{ocrs,rocrs,lee-singla,dughmi2025contention,delegated_probing,sparsification}). The broad reach of contention resolution arguably stems from the fact that it provides per-element guarantees: a \emph{$c$-balanced contention resolution scheme (CRS)} selects an element $i$ with probability at least $c$ when $i$ is \emph{active}, meaning it is available for selection. This lends itself to objective-agnostic guarantees that hold across problems: all linear objectives, and many more, are preserved up to the same $c$ factor. 

Despite the aforementioned generality, the literature on contention resolution relies on a strong assumption: the distribution of active elements---which is described fully by the marginal activation probabilities $p_i$ for product distributions---is known and given up front. Indeed, activation probabilities feature prominently in the design of optimal contention resolution schemes, even for the simplest of constraint systems. This precludes applying contention resolution to settings where the distribution may be difficult to describe exactly, or must be learned from (perhaps limited) historical data. Somewhat surprisingly, much remains open with regards to optimal contention resolution from samples, even for product distributions and the simplest of constraints, despite the fact that the analogous question has been studied extensively for the closely-related setting of \emph{prophet inequalities}~\cite{azar2014prophet,rubinstein_samples,fu_samplebased,cristi2024prophet,ezra2026prophet}.

In this paper, we study contention resolution from samples for the simplest of settings: product distributions with unknown activation probabilities $p_i$ for each element $i$, for $k$-uniform matroid constraints (also known as cardinality constraints), paying special attention to the single-item ($k = 1$) setting. We consider online arrivals in adversarial or random order and ask the following question:
\vspace{1em}

\emph{How many samples are needed, per element, to obtain an optimal online contention resolution scheme?}

\subsection{Results and Techniques}

We consider online contention resolution in both the adversarial and random order settings, for single-item and $k$-uniform matroid constraints. For the precise problem statement and relevant definitions, see Section~\ref{sec:preliminaries}.

\paragraph{Single-item.}
For the simplest case of a single-item constraint with known activation probabilities, there is a $(1 - 1/e)$-balanced Random Order Contention Resolution Scheme (ROCRS) \cite{rocrs} and a $1/2$-balanced Online Contention Resolution Scheme (OCRS) when the elements arrive in adversarial order \cite{ocrs}. Both guarantees are the best possible in their respective settings, and in fact the $1 - 1/e$ guarantee is optimal even for offline contention resolution.

We show a  dichotomy between the random and adversarial order settings: we design a $(1 - 1/e)$-balanced single-item ROCRS that uses a single sample per element, whereas we show that every single-sample single-item OCRS has its balance ratio bounded away from the optimal ratio of $1/2$. Our ROCRS is quite simple: active elements are accepted greedily, except that each element whose sample is also active is discarded with probability $1/e$. Our impossibility result for adversarial arrivals is through a case analysis on a small example. 

Though a single sample does not suffice for an optimal OCRS, we show, perhaps surprisingly, that two samples do suffice. Our two-sample $1/2$-balanced OCRS is based on the following simple yet powerful observation: a scheme which chooses a start time $T$ such that $\Pr[T = 0] = 1/2$ and $\Pr[T = i] = p_i / 2$ for all $i$, and accepts the first active element arriving strictly after $T$, is $1/2$-balanced. We show how to sample such a $T$ using only the two samples per element. This is closely related to the \emph{Bernoulli race} problem studied in \cite{dughmi2021bernoulli}, and more broadly to the literature on \emph{combinatorial Bernoulli factories} \cite{niazadeh2021combinatorial,niazadeh2023combinatorial,niazadeh2024bernoulli}, though we are not aware of any  constant sample factories for such problems.\footnote{The connection between our problem and that of the two-sample Bernoulli race runs quite deep, with reductions in both directions. We defer the interesting details to the full version of this paper, or perhaps a future paper! } Our approach to sampling $T$ is different and interesting in its own right, however: we reduce sampling $T$ to none other than a different contention resolution problem, this time offline and with a single sample! In particular, by using one set of samples to design a $1/2$-balanced CRS for the other, and enforcing the balance guarantee with equality, we obtain the desired $p_i/2$ probabilities. We lean on a CRS of Feige and Vondr\'{a}k~\cite{feige-vondrak-fair-crs}, modifying it appropriately to use a single sample and enforce the balance of $1/2$ with equality, to accomplish just that.

Finally, we  show that a single sample per element can be used to approximate this distribution of $T$ and obtain a $0.48$-balanced OCRS. While this result is not tight, our aforementioned impossibility provides an upper bound strictly smaller than $1/2$.

\paragraph{General $k$.}

We next turn our attention to general $k$-uniform matroid constraints. When activation probabilities are known, these admit an optimal $(1 - e^{-k} k^k/k!)$-balanced ROCRS \cite{willma-random} which is optimal even for the offline setting of contention resolution \cite{crs,yan-correlation-gap}. When the elements arrive in adversarial order, no general closed-form expression in $k$ is known for the optimal balance ratio, but it is implicitly characterized by a factor-revealing linear program~\cite{jiang-ma-zhang-multi-unit}.

For random arrivals, we show that a single sample per element again suffices to recover the optimal $(1 - e^{-k} k^k/k!)$ guarantee for every $k$. Our ROCRS accepts the first $k$ active elements, except that the $k$-th is discarded with probability $k/(2k+1)$ if its sample is also active; if discarded, it is replaced by the next active element. In other words, our ROCRS greedily selects the first $k-1$ active elements without looking at their samples, and only uses the sample of the $k$-th active element to decide between that or the $(k+1)$-th active element. We also give an alternative proof in Appendix~\ref{app:alternative-proof}, using a different scheme that directly generalizes our single-item ROCRS.

For adversarial arrivals, we design a single-sample $\prn{1 - \bigO{1/\sqrt{k}}}$-balanced OCRS, matching the asymptotically optimal dependence on $k$~\cite{alaei,jiang-ma-zhang-multi-unit}. Our approach is a single-sample analogue of the OCRS of Dinev and Weinberg~\cite{dinev-weinberg-ocrs}: we use the samples in place of the known activation probabilities to control how quickly we accept elements. More specifically, we admit an active element as a candidate if the resulting number of candidates does not exceed the number of active samples among the elements that have arrived, plus a random allowance of order $\sqrt{k}$. We then select among the candidates online, giving each the same acceptance probability while ensuring that we accept at most $k$ elements.

\paragraph{Sample complexity.}

Finally, we investigate the exact sample complexity of optimal schemes, in particular as we vary $k$. We find a surprising contrast between random and adversarial arrivals:  increasing $k$ reduces the required number of samples per element in the case of random arrivals, but does the reverse for adversarial arrivals. 

For random order arrivals, we determine the minimum number of samples per element needed to achieve the optimal balance ratio: one for $k \in \set{1,2,3,4}$, and zero for $k \geq 5$. The latter follows directly from a result of Arnosti and Ma~\cite[Lemma~6]{willma-random}, as greedily accepting the first $k$ active elements already yields the optimal guarantee when $k \geq 5$. For $k \leq 4$, we show that no scheme using zero samples can be better than $k/(k+1)$-balanced, strictly below the optimum. This impossibility also holds for offline contention resolution, and therefore so does the characterization. 

For adversarial arrivals, we show a striking separation between the single-item case and larger cardinality constraints, where the latter require more samples (reversing the trend observed for random arrivals). Whereas two samples per element suffice to recover the optimal guarantee for known activation probabilities when $k = 1$, no constant number of samples suffices for any $k \geq 2$! This impossibility holds even when the arrival order is known in advance, and even if the number of samples is allowed to depend on $k$.  Our proof builds on the characterization of Jiang, Ma, and Zhang~\cite{jiang-ma-zhang-multi-unit}: we construct two families of instances for which optimality requires the probability of accepting at least two elements in an initial segment to tend to zero in one family and stay bounded away from zero in the other. We then show that a fixed number of samples cannot support this difference.

\subsection{Additional Discussion of Related Work}

There is a large literature on contention resolution and closely-related stochastic decision problems, most notably prophet inequalities, with known prior distributions. We cannot hope to do justice to this vast literature, and instead refer the reader to related work discussions in~\cite{dughmi2025contention} and~\cite{stationary-ocrs}. Below, we overview work on contention resolution and related problems from samples, as that is most relevant to our focus in this paper.

\paragraph{Contention Resolution.} To our knowledge, there are few papers discussing contention resolution from samples. The work of Fu, Lu, Tang, Turkieltaub, Wu, Wu and Zhang~\cite{oblivious-ocrs} considers \emph{oblivious} schemes that have neither distributional knowledge nor access to any samples. They design a $1/e$-balanced oblivious OCRS for the single-item case with adversarial arrivals, and show that this is the best possible without distributional information, implying a separation from the setting with known activation probabilities. The same work also considers schemes with sample access, and shows that certain matroid constraints prohibit constant balance guarantees with constantly-many samples per element. Subsequent work by Fu, Lu, Tang, Wu, Wu and Zhang~\cite{fu_samplebased} designed a $(1/4-\eps)$-balanced OCRS for matroid constraints with $\operatorname{O}_{\eps}(\prn{\log{n}}^4)$ samples per element, for any $\eps > 0$. Recently, Feldman, Svensson and Zenklusen~\cite{matroid-ocrs-samples} improved upon this by achieving the same balance guarantee using only $\bigO{\log{r} (\log{\log{r}})^2}$ many samples, almost matching the $\bigOm{\log{r}}$ lower bound due to \cite{fu_samplebased}, where $r$ is the rank of the matroid.

\paragraph{Prophet Inequalities and Related Problems.} Closely related to online contention resolution are prophet inequality problems. Here, elements arrive online in adversarial order, and instead of per-element activations we get per-element rewards $v_i$. The goal is to achieve a high total reward, competing with an omniscient prophet. In contrast to contention resolution, there has been a large body of work studying prophet inequalities from samples. A single sample suffices for an optimal factor of $1/2$ in the single-choice setting \cite{rubinstein_samples,threshold-dynamics}, as well as for an asymptotically-optimal guarantee in the $k$-choice setting as $k$ grows large~\cite{azar}. The same question has also been studied for more elaborate constraints, as well as random order arrivals (known as prophet secretary problems). Most recently, \cite{abdi-linear-msp} exhibit a single-sample matroid prophet inequality achieving an optimal competitive ratio of $1/2$ when the order is adversarial yet fixed, and \cite{openAI} achieve a constant competitive ratio with respect to the more powerful almighty adversary which adaptively chooses the arrival order.

One matter worth addressing is the relationship between prophet inequality problems and contention resolution. Indeed, for known prior distributions the problems are almost equivalent, with reductions in both directions \cite{lee-singla}.%
\footnote{To be precise, the equivalence is with prophet inequalities measured against the ex-ante relaxation, which is a stronger benchmark than the expected offline optimum.}
One may therefore wonder whether our results can be obtained by reducing to the corresponding prophet (or prophet secretary) problem from samples. The answer is emphatically no. The reductions of \cite{lee-singla} are sensitive to knowing the distribution, and our impossibility results for both single-sample, single-item and finite-sample, $k$-uniform OCRSs serve to formally separate the two problems; in particular, a single sample permits a $1/2$ prophet inequality, but not a $1/2$-balanced OCRS.
\section{Preliminaries}
\label{sec:preliminaries}

Let $\set{1,\dots,n}$ denote a ground set of elements, $k \geq 1$ be an integer capacity, and $p = (p_1, \dots, p_n) \in [0,1]^n$ be an activation probability vector satisfying $\sum_{i = 1}^n p_i \leq k$. We sometimes use $[n]$ to denote $\set{1, \dots, n}$. A subset of $[n]$ is \emph{feasible} if it contains at most $k$ elements; this is a $k$-uniform matroid constraint---also known as a cardinality constraint---with the \emph{single-item} case corresponding to $k = 1$. We denote the $k$-uniform matroid polytope by $\cP_k \coloneqq \set{x \in [0,1]^n \midd \sum_{i = 1}^n x_i \leq k}$. Let $\text{support}(p) \coloneqq \set{i \in [n] \midd p_i > 0}$.

\begin{definition}[Contention Resolution Scheme]
An (offline) {\em Contention Resolution Scheme (CRS)} $\pi$ for the polytope $\cP_k$ is a procedure that, given a vector $p \in \cP_k$ and $W \subseteq [n]$, returns a random set $\pi_p(W) \subseteq W \cap \text{support}(p)$ such that $|\pi_p(W)| \leq k$ with probability $1$, for all $W \subseteq [n], p \in \cP_k$. The scheme is said to be \emph{monotone} if $\Pr[i \in \pi_p(A_1)] \geq \Pr[i \in \pi_p(A_2)]$ for every $i \in A_1 \subseteq A_2$.
\end{definition}

Contention resolution schemes were introduced in \cite{crs} as a tool for rounding fractional solutions for submodular maximization, though have found a variety of other uses since. Often $p$ denotes a fractional linear programming solution and we draw a random subset $W(p)$ of $[n]$ where each $i \in [n]$ appears independently with probability $p_i$. The issue is that $W(p)$ might not be feasible for the given polytope. A CRS is then applied to $W(p)$ to return a set $\pi_p(W(p)) \subseteq W(p)$ that is feasible with probability $1$; in our case, this means it satisfies $|\pi_p(W(p))| \leq k$. 

In the description above, $\pi$ is given all elements of $W(p)$ and can then choose $\pi_p(W(p))$. Feldman, Svensson and Zenklusen~\cite{ocrs} introduce an \emph{Online Contention Resolution Scheme (OCRS)} where the random set $W(p)$ is obtained in the same way, but the indicator $i \in W(p)$ is revealed sequentially to $\pi$ in an adversarial order fixed independently of the events $i \in W(p)$, the samples and any internal randomness of $\pi$. After each arrival, the OCRS has to immediately and irrevocably decide whether or not to accept $i \in W(p)$ into $\pi_p(W(p))$, subject to feasibility. Later, Adamczyk and W{\l}odarczyk~\cite{rocrs} introduced \emph{Random Order Contention Resolution Schemes (ROCRSs)} which are OCRSs with the arrival order chosen uniformly at random. We use $X_i \sim \Ber{p_i}$, where $\Ber{p_i}$ denotes the Bernoulli distribution with success probability $p_i$, to denote the random variable indicating whether or not $i \in W(p)$. We call $i$ \emph{online-active} if $X_i = 1$ and \emph{online-inactive} otherwise.

In this work, we study ROCRSs and OCRSs that are given $n$ and $k$, but not the activation probability vector $p$. Instead, before observing the online activations, they receive $m$ samples per element for some $m \geq 0$. We write $Y_i^{(s)} \sim \Ber{p_i}$ for the $s$-th sample of element $i$. For a sample-based scheme on an arbitrary input set $W$, we require the output to be a feasible subset of $W$; we do not additionally require it to lie in $\text{support}(p)$. On the random input $W(p)$, the latter property holds automatically. A sample-based scheme is called monotone if $\Pr[i \in \pi_p(A_1)] \geq \Pr[i \in \pi_p(A_2)]$ holds for all $p \in \cP_k$ and $i \in A_1 \subseteq A_2$, when the samples $Y^{(\ell)}_j$ are independently drawn from $\Ber{p_j}$ for all $j \in [n]$. In the single-sample setting, we write $Y_i$ and call $i$ \emph{sample-active} if $Y_i = 1$ and \emph{sample-inactive} otherwise. All samples and online activations are mutually independent. For adversarial arrivals, we index elements in arrival order, without loss of generality, so the online observations are $X_1,\dots, X_n$. For random arrivals, indices identify fixed elements, rather than their positions in the arrival order.

For a scheme $\pi$, let $A_{i, \pi}$ be the event that $\pi$ accepts element $i$. For $p_i > 0$, let
\[
    C_{i,\pi}(p) \coloneqq \Pr\brk{A_{i,\pi} \midd X_i = 1}.
\]
For $c \in [0,1]$, we say that $\pi$ is \emph{$c$-balanced}, if $C_{i,\pi}(p) \geq c$ for every activation vector $p$ and every element $i$ with $p_i > 0$; for an ROCRS, this probability is also taken over the randomness in the arrival order. Equivalently, $\Pr[A_{i,\pi}] \geq c \: p_i$ for every $i \in [n]$. When $\pi$ is clear from context, we write $A_i$ and $C_i(p)$; when $p$ is also clear, we write $C_i$.

For an online scheme $\pi$, let $R_{i,\pi}$ be the event that $\pi$ \emph{reaches} element $i$, meaning that it has accepted fewer than $k$ elements before $i$ arrives. We write $R_i$ when $\pi$ is clear from context. Observe that, in the single-item case with adversarial arrivals,
\[
    R_{i,\pi} = \prn{\bigcup_{j<i} A_{j,\pi}}^c,
\]
where $E^c$ denotes the complement of an event $E$.

For a set $S$ and an element $i$, write $S+i \coloneqq S \cup \set{i}$ and $S-i \coloneqq S \setminus \set{i}$. For an event $E$, $\1_E$ is its indicator, and for $x \in \R$, $(x)^+ \coloneqq \max\set{0,x}$. We write $\Bin{m,q}$ for the binomial distribution with $m$ trials and success probability $q$, and $\Poi{\lambda}$ for the Poisson distribution with mean $\lambda$.

\begin{remark}\label{rem:monotone}
In the subsequent sections, whenever it simplifies our arguments, we assume that $\sum_{i = 1}^n p_i = k$. As we show in Appendix~\ref{app:monotonicity}, all contention resolution schemes we design are monotone. Furthermore, we explain how we can add ``dummy'' elements to account for any missing mass for the random-order schemes, and why the assumption is justified by direct calculations for the adversarial-order schemes. Therefore, this assumption can be made without loss of generality.
\end{remark}
\section{The Single-Item Case}
\label{sec:single-item}

We start with the single-item case. Here, we show that a single sample suffices to design an optimal $(1 - 1/e)$-balanced ROCRS (and thus also an offline CRS). We also show that this is not the case for adversarial arrivals, giving an impossibility of $0.499$. We then proceed to design an optimal $1/2$-balanced OCRS using two samples per element, as well as a $0.48$-balanced OCRS with one sample.

\subsection{Random Order}
\label{sec:single-item:random-order}

We start with the random-order single-item case. Though we obtain a more general result for arbitrary $k$ in Section~\ref{sec:cardinality:random-order}, the algorithm we present here, and its analysis, are appealing due to their simplicity. We prove the guarantee when $\sum_i p_i = 1$; its extension to the case $\sum_i p_i \leq 1$ is justified in Appendix~\ref{app:monotonicity} (see also~Remark \ref{rem:monotone}). As in Section~\ref{sec:preliminaries}, $X_i$ indicates whether element $i$ is online-active, and $Y_i$ independently indicates whether it is sample-active.

Our single-item ROCRS is shown in Algorithm~\ref{alg:random-order:single-item}. For every online-active element $i$, this ROCRS tosses an independent coin and accepts $i$ with some probability depending on the sample $Y_i$. We prove the following.

\begin{algorithm}[ht]
  \caption{The Conservative Scaling algorithm for $k = 1$.}
  \label{alg:random-order:single-item}
  \begin{algorithmic}[1]
    \Require $\set{Y_1, \dots, Y_n}$
    \For{elements $i = 1, \ldots, n$ in order of arrival}
        \If{$X_i = 1$ and $Y_i = 0$}
            \State $q_i \from 1$
        \ElsIf{$X_i = 1$ and $Y_i = 1$}
            \State $q_i \from 1 - 1/e$
        \Else
            \State $q_i \from 0$
        \EndIf
        \State $Q_i \from \Ber{q_i}$
        \If{$Q_i = 1$}
            \State Accept $i$ and stop
        \Else
            \State Reject $i$
        \EndIf
    \EndFor
  \end{algorithmic}
\end{algorithm}

\begin{theorem}\label{thm:random-order-k-equals-1}
For every $n \in \N$, (unknown) probability vector $p \in [0,1]^n$ with $\sum_{i = 1}^n p_i = 1$, and every element $i \in [n]$ with $p_i > 0$, Algorithm~\ref{alg:random-order:single-item} always accepts at most one element and satisfies
\[
    C_i \geq 1 - \frac{1}{e}.
\]
\end{theorem}

Feasibility is obvious by the algorithm's description.
Next, notice that the probability that element $i$ is online-active and its coin succeeds is $p_i(1 - p_i/e)$. This implies an alternative description of the algorithm: it observes a sequence of independent Bernoulli random variables $Z_1, \dots, Z_n$ where $Z_i \sim \Ber{p_i(1 - p_i/e)}$ and accepts the first $i$ such that $Z_i = 1$. Since the arrival order is uniformly random, the probability that we accept element $i$, conditional on $X_i = 1$, can be written as
\[
    C_i = \prn{1 - \frac{p_i}{e}}\E\brk{\frac{1}{1 + \sum_{j \neq i} Z_j}},
\]
since this is the probability that $Z_i = 1$ and $i$ arrives first from a sequence of $1 + \sum_{j \neq i} Z_j$ ones.

Next, notice that if all $p_j$ for $j \neq i$ were close to $0$, then $\sum_{j \neq i} Z_j \to \Poi{1 - p_i}$, and the probability that $i$ is accepted conditional on $X_i = 1$ would be
\[
    \prn{1 - \frac{p_i}{e}} \frac{1 - e^{-(1 - p_i)}}{1 - p_i} \geq 1 - \frac{1}{e},
\]
where the inequality follows from elementary manipulations, interpreting, at $p_i = 1$, the fraction by its continuous value $1$. Thus, it suffices to show that this is the worst case. In other words, it suffices to show that, for $Z'_j \sim \Poi{p_j}$,
\begin{equation}\label{eq:k1-peeling}
    \E\brk{\frac{1}{1 + \sum_{j \neq i} Z_j}} \geq \E\brk{\frac{1}{1 + \sum_{j \neq i} Z'_j}}.
\end{equation}

By linearity of expectation, we have
\begin{align*}
    \E\brk{\frac{1}{1 + \sum_{j \neq i} Z_j}} &= \int_0^1 \E\brk{t^{\sum_{j \neq i} Z_j}} \dif t.
\end{align*}
Taking an arbitrary $j_1 \neq i$ and defining $F(t) = \E\brk{t^{\sum_{j \neq i, j_1} Z_j}}$, we get
\begin{align*}
     \E\brk{\frac{1}{1 + \sum_{j \neq i} Z_j}} 
     &= \int_0^1 \E\brk{t^{Z_{j_1}}} F(t) \dif t = \int_0^1 \prn{1 \cdot \prn{1 - p_{j_1} + \frac{p^2_{j_1}}{e}} + t \cdot \prn{p_{j_1} - \frac{p^2_{j_1}}{e}}} F(t) \dif t \\
    &= \int_0^1 \prn{1 - (1 - t) p_{j_1} + (1 - t)\frac{p^2_{j_1}}{e}} F(t) \dif t = \int_0^1 \prn{1 - tp_{j_1} +t \frac{p^2_{j_1}}{e}} F(1-t) \dif t. \\
\end{align*}
Note that
\[
    \int_0^1 \prn{\frac{t}{3} - \frac{t^2}{2} }F(1-t) \dif t = \int_{0}^1 \prn{\frac{t^2}{6} -\frac{t^3}{6} }F'(1-t) \dif t \geq 0,
\]
where the equality follows by integration-by-parts, and the inequality from noting that $F$ is increasing. Thus,
\begin{equation}\label{eq:byparts}
    \int_0^1 \frac{t}{3} F(1-t) \dif t \geq \int_0^1 \frac{t^2}{2}F(1-t) \dif t.
\end{equation}
We conclude that
\begin{align*}
    \E\brk{\frac{1}{1 + \sum_{j \neq i} Z_j}} &= \int_0^1 \prn{1 - t p_{j_1} +t \frac{p^2_{j_1}}{e}} F(1-t) \dif t \geq \int_0^1 \prn{1 - tp_{j_1} +t\frac{p^2_{j_1}}{3}} F(1-t) \dif t \\
    &\geq \int_0^1 \prn{1 - tp_{j_1} +\frac{t^2p^2_{j_1}}{2}} F(1-t) \dif t 
\end{align*}
where the last inequality follows from \eqref{eq:byparts}. Employing the Taylor series for the exponential, we get
\[
    \E\brk{\frac{1}{1 + \sum_{j \neq i} Z_j}} \geq \int_0^1 e^{- p_{j_1} t} F(1 - t) \dif t.
\]
Finally, observe that
\[
    e^{-p_{j_1} t} = e^{-p_{j_1}} e^{p_{j_1} (1 - t)} = e^{- p_{j_1}} \sum_{\ell = 0}^\infty \frac{(p_{j_1}(1 - t))^\ell}{\ell!} = e^{- p_{j_1}} \sum_{\ell = 0}^\infty (1 - t)^\ell \frac{p^\ell_{j_1}}{\ell!} = \E\brk{(1 - t)^{Z'_{j_1}}}.
\]
Therefore,
\[
    \E\brk{\frac{1}{1 + \sum_{j \neq i} Z_j}} \geq \int_0^1 \E\brk{(1 - t)^{Z'_{j_1}}} F(1 - t) \dif t = \E\brk{\frac{1}{1 + Z'_{j_1} + \sum_{j \neq i, j_1} Z_j}},
\]
where the last equality follows by linearity of expectation and the fact that, for every $r \in \N$,
\[
    \int_0^1 (1 - t)^r \dif t = \frac{1}{1 + r}.
\]
Repeating the same process for the remaining $j \neq i$ yields \eqref{eq:k1-peeling}.

\subsection{Adversarial Order: Impossibility}
\label{sec:single-item:adversarial:impossibility}

Next, we construct a hard instance showing that, for adversarial arrivals and $k = 1$, there does not exist a single-sample $1/2$-balanced OCRS. In the proof, we use the fact that for an OCRS $\pi$ to accept element $i$, it has to first reach $i$, and thus for every probability vector $p$
\begin{equation}\label{basic-balancedness}
    C_i(p) \leq \Pr\brk{R_{i,\pi} \midd X_i = 1} = 1 - \sum_{j < i} p_j C_j(p).
\end{equation}
Here, we present a proof sketch -- the full proof can be found in Appendix~\ref{app:missing:impossibility}.

\begin{theorem}\label{thm:adv-order:half-impossibility}
There does not exist a single-sample, $1/2$-balanced OCRS in the single-item setting, when the elements arrive in adversarial order.
\end{theorem}
\begin{proof}[Proof Sketch of Theorem~\ref{thm:adv-order:half-impossibility}]
Suppose, by way of contradiction, that there exists a single-sample OCRS $\pi$ that is $1/2$-balanced. We consider the following instances $p^{(i)}$ on $6$ elements, and study how well the algorithm performs on them, not knowing which instance it is actually facing.
\[
\begin{array}{c|cccccc}
\text{} & 1 & 2 & 3 & 4 & 5 & 6 \\
\hline
p^{(1)} & 1/2 & 1/2 & 0 & 0 & 0 & 0 \\
p^{(2)} & 0 & 0 & 1 & 0 & 0 & 0 \\
p^{(3)} & 0 & 0 & 0 & 1 & 0 & 0 \\
p^{(4)} & 0 & 0 & 0 & 0 & 1 & 0 \\
p^{(5)} & 0 & 0 & 1/3 & 1/3 & 1/3 & 0
\end{array}
\]
Each instance above is a limiting instance. In each row, the zero values are actually $\eps > 0$ and the non-zero values are appropriately adjusted so that the total remains one. For the fixed algorithm, every conditional acceptance probability extends continuously to each $p^{(i)}$, and  $C_j(p^{(i)})$ denotes this limiting value.

For each $S \subseteq \set{3,4,5}$, let $a_S$ be the probability of $A_3$, conditional on elements $1$ and $2$ being online-inactive, element $3$ being online-active, the sample-active elements among $\set{3,4,5}$ being exactly $S$, and all other elements being sample-inactive.

In the limit, $C_j(p^{(i)})\geq1/2$ for all $i,j$, and \eqref{basic-balancedness} applied for $i = 6$ yields
\[
    1 \leq \frac{1}{2} \sum_{j = 1}^5 p_j + \frac{1}{2} \leq \sum_{j = 1}^5 p_j C_j(p) + C_6(p) \leq 1,
\]
and equality in this chain forces $C_j(p^{(i)})=1/2$ whenever the corresponding $p_j^{(i)}$ is positive. In particular, under $p^{(1)}$ we have $C_1(p^{(1)})=C_2(p^{(1)})=1/2$, so $\Pr[R_3]=1/2$. Since $1/2\leq C_3(p^{(1)})\leq\Pr[R_3]$, this also gives $C_3(p^{(1)})=1/2$. Thus, whenever $R_3$ occurs and element $3$ is online-active, the algorithm must accept element $3$. Since under $p^{(1)}$ the elements $\set{3,4,5}$ are all sample-inactive, this implies $a_{\emptyset} = 1$.

Next consider the fact that the algorithm might be facing $p^{(2)},p^{(3)}$ or $p^{(4)}$. In these cases, elements $1$ and $2$ are online-inactive, and the sample-active set on $\set{3,4,5}$ is
respectively $\set{3}$, $\set{4}$, and $\set{5}$. The $1/2$-balancedness constraint for element $3$ in these three instances therefore gives:
\[
    a_{\set{3}} \geq \frac12,
    \quad
    a_{\set{4}} \geq \frac12,
    \quad
    a_{\set{5}} \geq \frac12.
\]

Finally consider the fact that the algorithm might be facing $p^{(5)}$. Let $S$ be the random set of sample-active elements among
$\set{3,4,5}$. Since elements $1$ and $2$ are again online-inactive,
\[
    C_3(p^{(5)}) = \E\brk{a_S}.
\]
Note that $a_S \geq 1 - |S|/2$, with the inequality being strict if $|S| = 3$. We conclude that:
\[
    C_3(p^{(5)})
=\E\brk{a_S} > \E\brk{1 - |S|/2} =
    1/2,
\]
which contradicts $C_3(p^{(5)}) = 1/2$.
\end{proof}

\subsection{Adversarial Order: Two-Sample \texorpdfstring{$1/2$}{1/2}-balanced OCRS}
\label{sec:single-item:adversarial:two-samples}

In this section, we use two independent samples $Y_i^{(1)}, Y_i^{(2)}\sim\Ber{p_i}$ per element to design a $1/2$-balanced OCRS for the single-item setting. We start with a useful reduction to a simple threshold construction.

\begin{lemma}\label{lem:adv-order:ideal-threshold}
Fix $n \in \N$ and a probability vector $p \in [0,1]^n$ with $\sum_{i = 1}^n p_i = 1$. Let $T$ be a random variable independent of the online activations $X_1, \dots, X_n$ that satisfies
\[
    \Pr[T = 0] = \frac{1}{2}, \qquad \text{and} \qquad 
    \Pr[T = i] = \frac{p_i}{2} \qquad\text{for every }i \in [n].
\]
The OCRS $\pi$ that accepts the first online-active element strictly after $T$, if there exists one, is $1/2$-balanced.
\end{lemma}
\begin{proof}
It suffices to show that $C_i \geq 1/2$ for every element $i$ with $p_i > 0$. Conditional on $X_i = 1$, the event $A_i$ occurs exactly when $T < i$ and $X_{T+1} = \dots = X_{i-1} = 0$. Thus,
\begin{align*}
    C_i &= \Pr[T = 0] \Pr[X_1 = \dots = X_{i-1} = 0] +  \sum_{j =1}^{i-1} \Pr[T = j] \Pr[X_{j+1} = \dots = X_{i-1} = 0] \\
    &= \Pr[T = 0] \prod_{\ell =1}^{i-1} (1 - p_\ell) +  \sum_{j =1}^{i-1} \Pr[T = j] \prod_{j < \ell < i} (1 - p_\ell) \\
    &= \frac{1}{2} \prod_{\ell=1}^{i-1} (1 - p_\ell) + \frac{1}{2} \sum_{j =1}^{i-1} p_j \prod_{j < \ell < i} (1 - p_\ell) = \frac{1}{2},
\end{align*}
where the last equality follows from the fact that  $\prod_{\ell=1}^{i-1} (1 - p_\ell)$ is the probability that none of the first $i-1$ elements are active, and  $p_j \prod_{j < \ell < i} (1 - p_\ell)$ is the probability that the $j$th element is the last active element among the first $i-1$, which are mutually exclusive and exhaustive events.
\end{proof}

By Lemma~\ref{lem:adv-order:ideal-threshold} if, using only the samples, we can construct a threshold $T \in \set{0, 1, \dots, n}$ such that
\[
    \Pr[T = 0] = \frac{1}{2}, \qquad \text{and} \qquad \Pr[T = i] = \frac{p_i}{2} \qquad \text{for every } i \in [n],
\]
the result follows. Next, we show how to construct such a threshold by reducing it in turn to the design of a single-sample offline CRS with certain guarantees, with the first set of samples playing the role of online arrivals and the second set of samples serving as samples. In more detail, we choose $T$ by applying a modified version of the $(1 - 1/e)$-balanced (offline) CRS of Feige and Vondr\'ak~\cite[Section~1.2]{feige-vondrak-fair-crs} to $\set{i \midd Y_i^{(1)} = 1}$, which we also implement using only a single sample per element -- in this case the $Y^{(2)}_i$'s. Our final OCRS simply accepts the first online-active element strictly after $T$.

\begin{algorithm}[ht]
  \caption{Threshold $T$ from a single-sample (offline) CRS on $\cS$.}
  \label{alg:adv-order:two-samples}
  \begin{algorithmic}[1]
    \Require $\set{Y_1^{(1)}, \dots, Y_n^{(1)}}, \set{Y_1^{(2)}, \dots, Y_n^{(2)}}$
    \Ensure $T$
    \State $\cS \from \set{i \in [n] \midd Y_i^{(1)} = 1}$
    \If{$\cS = \emptyset$}
        \State Set $T \from 0$
    \ElsIf{$\cS = \set{j}$ for some $j$}
        \If{$Y^{(2)}_j = 1$}
            \State With probability $\frac{1}{2}$, set $T \from 0$, otherwise set $T \from j$
        \Else
            \State Set $T \from j$
        \EndIf
    \Else
        \State $Q \from \Ber{\frac{1}{2}}$
        \If{$Q = 1$}
            \State Set $T \from 0$
        \Else
            \State Let $J$ be a uniformly random element of $\cS$
            \If{$Y^{(2)}_J = 1$}
                \State Set $T \from J'$ where $J'$ is chosen uniformly at random from $\cS - J$
            \Else
                \State $T \from J$
            \EndIf
        \EndIf
    \EndIf
  \end{algorithmic}
  
\end{algorithm}

\begin{theorem}\label{thm:adv-order:two-samples}
For every $n \in \N$ and (unknown) probability vector $p \in [0,1]^n$ with $\sum_{i = 1}^n p_i = 1$, the threshold $T$ in Algorithm~\ref{alg:adv-order:two-samples} satisfies
\[
    \Pr[T = 0] = \frac{1}{2}, \qquad \text{and} \qquad \Pr[T = i] = \frac{p_i}{2} \qquad \text{for every } i \in [n].
\]
\end{theorem}

Below, we analyze Algorithm~\ref{alg:adv-order:two-samples} using the Feige and Vondr\'ak (offline) CRS~\cite[Section~1.2]{feige-vondrak-fair-crs} as a black-box and describe the main ideas behind the proof of Theorem~\ref{thm:adv-order:two-samples}. The formal proof of the theorem is self-contained and can be found in Appendix~\ref{app:missing:self-contained}.

For $i \in [n]$ and $S \subseteq [n]$, let
\[
    r_i(S) =
    \begin{cases}
        0, & \forall i \notin S, \\
        1, & \text{if } S = \set{i}, \\
        \displaystyle
        \frac{1}{|S|} \sum_{j \notin S} p_j + \frac{1}{|S| - 1} \sum_{j \in S - i} p_j, & \forall i \in S,\: \text{if } |S| \geq 2.
    \end{cases}
\]
The Feige--Vondr\'ak CRS~\cite[Section~1.2]{feige-vondrak-fair-crs}, applied to the $Y^{(1)}_j$'s, accepts an element $j \in [n]$ with probability $r_j(S)$ when $\set{\ell \midd Y^{(1)}_\ell = 1} = S$. We show that, for sets $S$ such that $i \in S$ and $|S| \geq 2$, we have
\[
    \Pr\brk{J = i \wedge Y^{(2)}_i = 0 \midd \cS = S} + \Pr\brk{J \neq i \wedge Y^{(2)}_J = 1 \wedge J' = i \midd \cS = S} = r_i(S),
\]
and use this to write
\[
    \Pr[T = i] = \prn{1 - \frac{p_i}{2}} \Pr[\cS = \set{i}] + \frac{1}{2} \sum_{\substack{S \subseteq [n] \\ i \in S, |S| \geq 2}} \Pr[\cS = S] r_i(S),
\]
which can be re-arranged as
\[
    \Pr[T = i] = \frac{1}{2} \E[r_i(\cS)] + \frac{1 - p_i}{2} \Pr[\cS = \set{i}] = \frac{p_i}{2} \Pr[\cS \neq \emptyset] + \frac{p_i}{2} \Pr[\cS = \emptyset] = \frac{p_i}{2},
\]
where the second equality follows from $\E\brk{r_i(\cS)} = p_i \Pr[\cS \neq \emptyset]$~\cite[Lemma 1.5]{feige-vondrak-fair-crs} and the fact that
\[
    (1 - p_i) \Pr[\cS = \set{i}] = (1 - p_i) p_i \prod_{j \neq i} (1 - p_j) = p_i \prod_{j = 1}^n (1 - p_j) = p_i \Pr[\cS = \emptyset].
\]
Finally, since $T \in \set{0, 1, \dots, n}$ and $\sum_{i = 1}^n p_i = 1$, we have $\Pr[T = 0] = 1 - \sum_{i = 1}^n \Pr[T = i] = 1 - (1/2) \sum_{i = 1}^n p_i = 1/2$.

\subsection{Adversarial Order: Single-Sample \texorpdfstring{$0.48$}{0.48}-balanced OCRS}
\label{sec:single-item:adversarial:single-sample}

Next, we design a single-sample, $0.48$-balanced OCRS for the single-item setting and adversarial arrivals. The idea behind our OCRS is to use the samples to construct a threshold $T$ that is close to the ideal one that satisfies $\Pr[T = 0] = 1/2$ and $\Pr[T = i] = p_i/2$.

\begin{algorithm}[ht]
  \caption{A single-sample $0.48$-balanced OCRS.}
  \label{alg:adv-order:single-sample}
  \begin{algorithmic}[1]
    \Require $\set{Y_1, \dots, Y_n}$
    \State $\cS \from \set{i \in [n] \midd Y_i = 1}$
    \If{$\cS = \emptyset$}
        \State Set $T \from 0$
    \ElsIf{$\cS = \set{j}$ for some $j$}
        \State With probability $\frac{1}{2}$, set $T \from 0$, otherwise set $T \from j$
    \Else
        \State Set $T \from I$, where $I$ is chosen uniformly at random from $\cS$
    \EndIf
    \For{elements $i = 1, \dots, n$ in order of arrival}
        \If{$i \leq T$ or $X_i = 0$}
            \State Reject $i$
        \Else
            \State Accept $i$ and stop
        \EndIf
    \EndFor
  \end{algorithmic}
\end{algorithm}

\begin{theorem}\label{thm:adv-order:single-sample}
For every $n \in \N$, (unknown) probability vector $p \in [0,1]^n$ with $\sum_{i = 1}^n p_i = 1$ and every element $i \in [n]$ with $p_i > 0$, Algorithm~\ref{alg:adv-order:single-sample} always accepts at most one element and satisfies $C_i > 0.48$.
\end{theorem}

Feasibility is obvious by the algorithm's description.
If, for some $i$ we have $p_i = 1$, Algorithm~\ref{alg:adv-order:single-sample} accepts element $i$ with probability $1/2$, so assume that $p_i < 1$ for all $i \in [n]$.

Let $T^*$ be the ideal threshold in Lemma~\ref{lem:adv-order:ideal-threshold}, where $\Pr[T^* = 0] = 1/2$ and $\Pr[T^* = i] = p_i/2$. For $i \in [n]$, denote the error between the ideal distribution for $T$ and our actual distribution by 
\[
    \eps_i \coloneqq \frac{p_i}{2} - \Pr[T = i] \geq 0,
\]
where the inequality follows from the fact that, conditional on $Y_i = 1$, Algorithm~\ref{alg:adv-order:single-sample} chooses $T = i$ with probability at most $1/2$. Thus, the distribution of $T$ is obtained from that of $T^*$ by moving $\eps_i$ mass from each $i$ to $0$.

For a deterministic threshold $t$, let $g_\ell(t)$ be the probability of $A_\ell$ conditional on $X_\ell = 1$, when the algorithm uses threshold $t$. Under this conditioning, if $t < \ell$, the event $A_\ell$ occurs exactly when every element strictly between $t$ and $\ell$ is online-inactive. If $t \geq \ell$, the algorithm rejects $\ell$. Therefore,
\[
    g_\ell(t)=
    \begin{cases}
    \displaystyle
    \prod_{t < j < \ell}(1 - p_j), & 0 \leq t < \ell, \\
    0, & \ell \leq t \leq n.
    \end{cases}
\]
It follows that
\[
    C_\ell = \E[g_\ell(T)], \qquad \text{and} \qquad
    \E[g_\ell(T^*)] = \frac{1}{2},
\]
where the second equality follows from Lemma~\ref{lem:adv-order:ideal-threshold}.

The proof of Theorem~\ref{thm:adv-order:single-sample} uses the following two lemmas, which we prove in Appendices~\ref{app:missing:threshold-deficit-lemma} and~\ref{app:missing:g-bound-lemma}, respectively. First, we show that the distribution of $T$ is not too far from that of $T^*$.

\begin{lemma}\label{lem:adv-order:threshold-deficit}
For every element $i$,
\[
    \eps_i \leq \frac{1-1/e}{12} \: p_i(1 - p_i).
\]
\end{lemma}

Next, we provide a way to move from the error estimate on the threshold to error estimates on $C_\ell$, for a fixed element $\ell$, via a bound on the difference between different values of $g_\ell$.

\begin{lemma}\label{lem:adv-order:g-bound}
For every $p \in [0,1]^n$ and $\ell \in [n]$ with $\sum_{i = 1}^\ell p_i \leq 1$, we have
\[
    \sum_{i = 1}^{\ell-1} p_i(1 - p_i)(g_\ell(i) - g_\ell(0)) \leq \frac{3}{8}.
\]
\end{lemma}

We are ready to prove the theorem.
\begin{proof}[Proof of Theorem~\ref{thm:adv-order:single-sample}]
Fix an element $\ell$ with $p_\ell > 0$. Observe that, for $1 \leq i < \ell$, we have
\[
    g_\ell(0) = \prn{\prod_{j = 1}^i (1 - p_j)} g_\ell(i) \leq g_\ell(i).
\]
Thus, changing threshold $i < \ell$ to $0$ decreases the conditional acceptance probability from $g_\ell(i)$ to $g_\ell(0)$, since the additional earlier elements can prevent $\ell$ from being accepted. If $i \geq \ell$, the same change makes $\ell$ eligible and increases its conditional acceptance probability from $0$ to $g_\ell(0)$.

Therefore, moving probability $\eps_i$ mass from threshold $i$ to $0$ changes the expectation of $g_\ell$ by $\eps_i (g_\ell(0) - g_\ell(i))$. Summing over all $i$ yields
\begin{align*}
    \frac{1}{2} - C_\ell
    &= \E[g_\ell(T^*)] - \E[g_\ell(T)] = \sum_{i = 1}^n \eps_i \prn{g_\ell(i) - g_\ell(0)} \\
    &= \sum_{i = 1}^{\ell - 1} \eps_i \prn{g_\ell(i) - g_\ell(0)} - g_\ell(0) \sum_{i = \ell}^n \eps_i \\
    &\leq \sum_{i = 1}^{\ell-1} \eps_i \prn{g_\ell(i) - g_\ell(0)} \leq \frac{1-1/e}{12} \sum_{i = 1}^{\ell-1} p_i(1 - p_i) \prn{g_\ell(i) - g_\ell(0)} \\
    &\leq \frac{1-1/e}{32},
\end{align*}
where the second inequality follows from Lemma~\ref{lem:adv-order:threshold-deficit} and the third from Lemma~\ref{lem:adv-order:g-bound}. Thus,
\[
    C_\ell \geq \frac{1}{2} - \frac{1 - 1/e}{32} > 0.48. \qedhere
\]
\end{proof}
\section{General \texorpdfstring{$k$}{k}}
\label{sec:cardinality}

We continue with the cardinality case for general $k \in \N$. Here we show that, perhaps surprisingly, for uniformly random arrivals a single sample suffices to obtain an optimal $(1 - e^{-k} k^k/k!)$-balanced ROCRS (and thus also offline CRS) for every $k$, generalizing our result from Section~\ref{sec:single-item:random-order}. We then move on to adversarial arrivals. Even though a single sample does not suffice for an optimal OCRS for $k = 1$, we show that, for general $k$, a single sample does suffice to obtain an asymptotically optimal $1 - \bigO{1/\sqrt{k}}$-balanced OCRS.

\subsection{Random Order}
\label{sec:cardinality:random-order}

We start with the random order case. The ROCRS we design, shown in Algorithm~\ref{alg:cardinality:random-order}, is surprisingly simple: greedily accept the first $k-1$ online-active elements and when observing the $k$-th online-active element $i$, if $Y_i = 0$, accept it. Else, if $Y_i = 1$, accept $i$ with probability $(k+1)/(2k+1)$, else reject it and greedily accept the next online-active element. We call this the \emph{Almost-Greedy} ROCRS. Notice that it satisfies an interesting property: it only ever looks at a single sample $Y_i$, but it cannot a priori know which $Y_i$ is the useful one so it needs a sample per element. We prove the following.

\begin{algorithm}[ht]
  \caption{The Almost-Greedy algorithm.}
  \label{alg:cardinality:random-order}
  \begin{algorithmic}[1]
    \Require $\set{Y_1, \dots, Y_n}$
    \Ensure $\cA$
    \State $\cA \from \emptyset$, flag $\from$ \textsc{False}
    \For{elements $i = 1, \dots, n$ in order of arrival}
        \If{$X_i = 1$ and $|\cA| < k-1$}
            \State Accept $i$
            \State $\cA \from \cA + i$
        \ElsIf{$X_i = 1, \: |\cA| = k-1$ and flag $=$ \textsc{False}}
            \If{$Y_i = 0$}
                \State Accept $i$
                \State $\cA \from \cA + i$
            \Else
                \State flag $\from$ \textsc{True}
                \State $Q \from \Ber{\dfrac{k+1}{2k+1}}$
                \If{$Q = 1$}
                    \State Accept $i$
                    \State $\cA \from \cA + i$
                \Else
                    \State Reject $i$
                \EndIf
            \EndIf
        \ElsIf{$X_i = 1, \: |\cA| = k-1$ and flag $=$ \textsc{True}}
            \State Accept $i$
            \State $\cA \from \cA + i$
        \Else
            \State Reject $i$
        \EndIf
    \EndFor
  \end{algorithmic}
\end{algorithm}

\begin{theorem}\label{thm:cardinality:random-order}
For every $n, k \in \N$, (unknown) probability vector $p \in [0,1]^n$ with $\sum_{i = 1}^n p_i = k$ and every element $i \in [n]$ with $p_i > 0$, Algorithm~\ref{alg:cardinality:random-order} always accepts at most $k$ elements and satisfies
\[
    C_i \geq 1 - e^{-k} \frac{k^k}{k!}.
\]
\end{theorem}
\begin{proof}
Feasibility follows easily from the algorithm's description.
Fix an element $i$ with $p_i > 0$, and let $S_{-i} = \sum_{j \neq i} X_j$ be the number of other online-active elements. First we analyze the greedy rule which accepts the first $k$ online-active elements, before comparing our algorithm to this rule in order to lower-bound $C_i$.

Condition on $X_i = 1$ and on the other online activations $\set{X_j}_{j \neq i}$. The arrival order of the $S_{-i}+1$ online-active elements is uniformly random, so each possible rank of $i$ has probability $1/(S_{-i}+1)$. The greedy rule accepts $i$ exactly when its rank is at most $k$. Thus, its conditional probability of accepting $i$ is $\AR_k(S_{-i})$ where, for $s \geq 0$
\[
    \AR_k(s) \coloneqq \min\set{1, \frac{k}{s+1}}.
\]
Notice that the Almost-Greedy ROCRS differs from this greedy rule only at the $k$-th and $(k+1)$-st online-active elements. If $i$ has rank $k$, we reject it with probability $kp_i/(2k+1)$, since $Y_i$ is independent of the online activations and the arrival order and element $i$ is sample-active with probability $p_i$. Therefore, when $S_{-i} < k-1$, the conditional probability of accepting $i$ does not change but, when $S_{-i} \geq k-1$, it decreases by $kp_i/((2k+1)(S_{-i}+1))$.

On the other hand, if $i$ has rank $k+1$, we accept it exactly when the $k$-th online-active element is rejected. Conditional on $i$ having rank $k+1$, each of the other $S_{-i}$ online-active elements is equally likely to have rank $k$. If the element of rank $k$ is $j$, it is rejected with probability $kp_j/(2k+1)$ and thus, when $S_{-i} \geq k$, the conditional probability of accepting $i$ increases by
\[
    \frac{1}{S_{-i}+1}\sum_{j \neq i}\frac{X_j}{S_{-i}}\cdot \frac{kp_j}{2k+1}
    = \frac{k}{2k+1}\sum_{j \neq i}\frac{p_jX_j}{S_{-i}(S_{-i}+1)},
\]
whereas when $S_{-i} < k$ it does not change since $i$ cannot have rank $k+1$. Since the other online activations are independent of $X_i$, averaging over their original distributions gives
\begin{align}\label{eq:almost-greedy:rank}
    C_i &= \E[\AR_k(S_{-i})] - \frac{kp_i}{2k+1}\E\brk{\frac{\1_{\set{S_{-i} \geq k-1}}}{S_{-i}+1}} \nonumber \\
    &\qquad + \frac{k}{2k+1}\sum_{j \neq i} p_j \E\brk{\frac{X_j \cdot \1_{\set{S_{-i} \geq k}}}{S_{-i}(S_{-i}+1)}},
\end{align}
where by convention, the last fraction is taken to be zero when $S_{-i} = 0$.

Next, we rewrite the two correction terms in \eqref{eq:almost-greedy:rank} in terms of $S_{-i}$. For the last term, observe that
\[
    \AR_k(S_{-i}-X_j)-\AR_k(S_{-i})
    = \frac{k \: X_j \cdot \1_{\set{S_{-i} \geq k}}}{S_{-i}(S_{-i}+1)}.
\]
Indeed, both sides vanish if $X_j=0$ or $S_{-i}<k$, while for $X_j=1$ and $S_{-i} \geq k$ the left side is
\[
    \frac{k}{S_{-i}} - \frac{k}{S_{-i}+1} = \frac{k}{S_{-i}(S_{-i}+1)}.
\]
Furthermore, $X_j$ is independent of $S_{-i}-X_j$, and hence
\begin{align*}
    \sum_{j \neq i}p_j\E[\AR_k(S_{-i}-X_j)]
    &= \sum_{j \neq i}\E[X_j\AR_k(S_{-i}-X_j)]
     = \E\brk{\min\set{k,S_{-i}}},
\end{align*}
where the last equality follows because, whenever $S_{-i} > 0$, each of the $S_{-i}$ non-zero summands for which $X_j = 1$ equals $\min\set{1,k/S_{-i}}$. Combining these identities and using $\sum_{j \neq i}p_j=k-p_i$, we obtain
\begin{align*}
    k\sum_{j \neq i}p_j\E\brk{\frac{X_j \cdot \1_{\set{S_{-i} \geq k}}}{S_{-i}(S_{-i}+1)}}
    &= \sum_{j \neq i}p_j\E\brk{\AR_k(S_{-i}-X_j)-\AR_k(S_{-i})} \\
    &= \E\brk{\min\set{k,S_{-i}}}-(k-p_i)\E[\AR_k(S_{-i})].
\end{align*}
For the second term in \eqref{eq:almost-greedy:rank}, notice that
\begin{align*}
    \E[\AR_k(S_{-i})] &= \E[1 \cdot \1_{\set{S_{-i} \leq k-2}}] + k \; \E\brk{\frac{\1_{\set{S_{-i} \geq k-1}}}{S_{-i}+1}} &\iff \\
    k \; \E\brk{\frac{\1_{\set{S_{-i} \geq k-1}}}{S_{-i}+1}}
    &= \E[\AR_k(S_{-i})]-\Pr[S_{-i} \leq k-2].
\end{align*}
Substituting the two identities into \eqref{eq:almost-greedy:rank}, we obtain
\begin{align*}
    C_i
    &= \E[\AR_k(S_{-i})]
       -\frac{p_i}{2k+1}\prn{\E[\AR_k(S_{-i})]-\Pr[S_{-i} \leq k-2]} \\
    &\qquad +\frac{1}{2k+1}\prn{\E\brk{\min\set{k,S_{-i}}}-(k-p_i)\E[\AR_k(S_{-i})]} \\
    &= \frac{k+1}{2k+1}\E[\AR_k(S_{-i})]
       +\frac{1}{2k+1}\E\brk{\min\set{k,S_{-i}}}
       +\frac{p_i}{2k+1}\Pr[S_{-i} \leq k-2].
\end{align*}
Next, let
\[
    h(s) \coloneqq \frac{(k+1)\AR_k(s)+\min\set{k,s}}{2k+1}.
\]
We can thus write the conditional acceptance probability as
\begin{equation}\label{eq:almost-greedy:balance}
    C_i = \E[h(S_{-i})] + \frac{p_i}{2k+1}\Pr[S_{-i} \leq k-2].
\end{equation}

To see why this rewriting is useful, consider $Z \sim \Poi{k}$. We will show that $\E[h(Z)] = 1 - e^{-k} k^k/k!$ and that $C_i \geq \E[h(Z)]$. First, using the fact that
\[
    \Pr[Z = s+1] = e^{-k} \frac{k^{s+1}}{(s+1)!} = \frac{k}{s+1} e^{-k} \frac{k^s}{s!} = \frac{k}{s+1} \Pr[Z = s],
\]
we have
\begin{align*}
    \E[\AR_k(Z)]
    &= \Pr[Z \leq k-1]
       +\sum_{s=k}^{\infty}\frac{k}{s+1}\Pr[Z=s] = \Pr[Z \leq k-1]
       +\sum_{s=k}^{\infty}\Pr[Z=s+1] \\
    &= 1-\Pr[Z=k],
\end{align*}
and
\begin{align}
    \frac{1}{k}\E\brk{\min\set{k,Z}}
    &= \sum_{s=1}^{k}\frac{s}{k}\Pr[Z=s]+\Pr[Z \geq k+1] = \Pr[Z \leq k-1]+\Pr[Z \geq k+1] \nonumber \\
    \label{eq:min-k-s-poisson} &= 1-\Pr[Z=k].
\end{align}
Next, notice that
\[
    h(s) = \frac{(k+1)\AR_k(s)+\min\set{k,s}}{2k+1} = \frac{k+1}{2k+1} \cdot \AR_k(s) + \frac{k}{2k+1} \cdot \frac{\min\set{k,s}}{k},
\]
and thus $h(s)$ is a convex combination of $\AR_k(s)$ and $\min\set{k,s}/k$. Since both these values are equal on expectation, it follows that
\[
    \E[h(Z)] = 1-\Pr[Z=k] = 1-e^{-k}\frac{k^k}{k!}.
\]
Thus, it suffices to show that the expression in \eqref{eq:almost-greedy:balance} is at least $\E[h(Z)]$.

To prove this, the idea is to use the additional term in \eqref{eq:almost-greedy:balance} to include $i$ in the count. Let $S = S_{-i}+X_i = \sum_{j=1}^n X_j$ be the unconditioned number of online-active elements, so that $\E[S]=k$. We will show that $C_i \geq \E[h(S)]$, reducing the problem to comparing $\E[h(S)]$ and $\E[h(Z)]$, i.e. a sum of independent Bernoulli random variables with a Poisson random variable of the same mean.

Observe that, for every integer $s \geq 0$,
\begin{equation}\label{eq:almost-greedy:differences}
    h(s+1)-h(s) =
    \begin{cases}
        \dfrac{1}{2k+1}, & 0 \leq s \leq k-2, \\
        0, & s=k-1, \\
        -\dfrac{k(k+1)}{(2k+1)(s+1)(s+2)}, & s \geq k,
    \end{cases}
\end{equation}
and thus
\[
    \E\brk{h(S_{-i}+1)-h(S_{-i})} \leq \frac{1}{2k+1} \Pr[S_{-i} \leq k-2].
\]
Also, since $X_i \sim \Ber{p_i}$ is independent of $S_{-i}$, this implies
\begin{align*}
    \E[h(S)]
    &= \E[h(S_{-i})] + p_i\E\brk{h(S_{-i}+1)-h(S_{-i})} \\
    &\leq \E[h(S_{-i})] + \frac{p_i}{2k+1}\Pr[S_{-i} \leq k-2]
     = C_i,
\end{align*}
where the last equality follows from \eqref{eq:almost-greedy:balance}.

It remains to show that $\E[h(S)] \geq \E[h(Z)]$. Notice from \eqref{eq:almost-greedy:differences} that $h$ increases up to $k-1$, has the same value at $k-1$ and $k$, and decreases thereafter. Therefore, for every integer $N \geq n$ and every $t \in [0,1]$, the integers $s \in \set{0,\dots,N}$ satisfying $h(s) \geq t$ form either an empty set or an interval containing $k$.

By Hoeffding's theorem~\cite[Theorem~5]{hoeffding-poisson-binomial}, among sums of $N$ independent Bernoulli random variables with mean $k$, equal probabilities minimize the probability of each integer interval containing $k$. Append $N-n$ zero-probability summands to $S$, which changes neither its distribution nor its mean, and let $B_N \sim \Bin{N,k/N}$. Applying Hoeffding's theorem to the intervals above gives, for every $t \in [0,1]$,
\[
    \Pr\brk{h(S) \geq t} \geq \Pr\brk{h(B_N) \geq t}.
\]
Since $0 \leq h(s) \leq 1$, we have $h(s)=\int_0^1\1_{\set{h(s) \geq t}}\dif t$. Integrating the last inequality over $t$ and using linearity of expectation, we obtain
\[
    \E[h(S)] = \int_0^1 \Pr\brk{h(S) \geq t}\dif t
    \geq \int_0^1 \Pr\brk{h(B_N) \geq t}\dif t
    = \E[h(B_N)].
\]
Finally, $B_N$ converges in distribution to $Z$ as $N \to \infty$. Since $0 \leq h \leq 1$, taking the limit yields $\E[h(S)] \geq \E[h(Z)]$. Therefore,
\[
    C_i \geq \E[h(S)] \geq \E[h(Z)]
    = 1-e^{-k}\frac{k^k}{k!}. \qedhere
\]
\end{proof}

\subsection{Adversarial Order}
\label{sec:cardinality:adv-order}

We continue with the adversarial order case. Without loss of generality, let $1, \dots, n$ be the arrival order and $p$ be a probability vector such that $\sum_{i = 1}^n p_i \leq k$. The idea behind our OCRS, shown in Algorithm~\ref{alg:cardinality:adv-order}, is the following: for each online-active element, we call it a candidate if it passes a certain \textit{count test}, maintaining a counter $C$ of candidates seen and a counter $A$ of number of accepted elements. Afterwards, we accept every candidate with a certain probability that changes as the algorithm progresses, depending on $C$ and $A$. We call this the \emph{Count Test} OCRS.

\begin{algorithm}[ht]
  \caption{The Count Test algorithm.}
  \label{alg:cardinality:adv-order}
  \begin{algorithmic}[1]
    \Require $r$, $\set{Y_1, \dots, Y_n}$
    \Ensure $\cA$
    \State $\cA \from \emptyset, C \from 0, A \from 0$
    \State Draw $R$ uniformly from $\set{0,\dots,r}$ and set $N \from R + \sum_{i = 1}^n Y_i$
    \For{elements $i = 1, \dots, n$ in order of arrival}
        \If{$X_i = 1$ and $C + 1 \leq R + \sum_{j \leq i} Y_j$}
            \State $Q \from \Ber{\dfrac{\min\set{k, N} - A}{N - C}}$
            \State $C \from C + 1$
            \If{$Q = 1$}
                \State Accept $i$
                \State $\cA \from \cA + i$
                \State $A \from A + 1$
            \Else
                \State Reject $i$
            \EndIf
        \Else
            \State Reject $i$
        \EndIf
    \EndFor
  \end{algorithmic}
\end{algorithm}

The OCRS we design is closely related to that of Dinev and Weinberg~\cite{dinev-weinberg-ocrs} for known activation probabilities. However, it differs in a few ways. First, our algorithm uses a random $R$ instead of fixing it at $\sqrt{k}$ as in \cite{dinev-weinberg-ocrs}. Second, the acceptance probability of our algorithm changes as $C$ and $A$ change, instead of being fixed at $1 - 1/\sqrt{k}$ as in \cite{dinev-weinberg-ocrs}. Third, we frame the algorithm in terms of a running count of candidates and have a slightly different acceptance rule than that of \cite{dinev-weinberg-ocrs}.

All of these three changes are actually not necessary: we can simply use $R = \floor{\sqrt{k}}$ deterministically, we can accept at a flat rate of $\min\set{1, (k - \sqrt{k})/N}$ until we run out of space and we can also remove the candidates and set $q = 1$ if $N = R$ and $q = \min \set{1, (k - \sqrt{k})/{(N-R)}}$ otherwise, accepting an online-active element $i$ with probability $q$ only if $A+1 \leq R + q \sum_{j \leq i} Y_j$; this is exactly equivalent to the OCRS of \cite{dinev-weinberg-ocrs}, with the (unknown) $p_i$'s replaced by the $Y_i$'s. All these changes, however, significantly simplify the analysis and thus were made for the sake of clarity of exposition.

\begin{theorem}\label{thm:cardinality:adv-order}
For every $n, k \in \N$, integer $r \geq 1$, (unknown) probability vector $p \in [0,1]^n$ with $\sum_{i = 1}^n p_i \leq k$ and every element $i \in [n]$ with $p_i > 0$, Algorithm~\ref{alg:cardinality:adv-order} always accepts at most $k$ elements and satisfies
\[
    C_i \geq 1-\frac{1}{r+1}-\frac{r}{2k}-\frac{1}{2\sqrt{k}}.
\]
\end{theorem}

Before we prove the theorem, observe that, for $r = \floor{\sqrt{2k}}$, the guarantee becomes
\[
    C_i \geq 1 - \frac{\sqrt{2}+1/2}{\sqrt{k}} = 1 - \bigO{\frac{1}{\sqrt{k}}}.
\]

\begin{proof}[Proof of Theorem~\ref{thm:cardinality:adv-order}]
We first give an equivalent description of the acceptance rule in Algorithm~\ref{alg:cardinality:adv-order}. Condition on $R$, the samples, and the online activations. The list of candidates is then fixed, since the count test does not depend on the acceptance decisions. Moreover, the count test ensures that there are at most $N = R + \sum_{j = 1}^n Y_j$ candidates.

Consider choosing $\min\set{k,N}$ of the $N$ possible candidate positions uniformly at random, and accepting each candidate exactly when its position is chosen. Any chosen positions beyond the end of the list are unused. After processing $C$ candidates, of which $A$ were accepted, there are $N-C$ positions left, of which $\min\set{k,N}-A$ are chosen. Conditional on the choices for the first $C$ positions, the chosen positions among the remaining ones are still uniformly distributed. Thus, the conditional probability of accepting the next candidate is
\[
    \frac{\min\set{k,N}-A}{N-C},
\]
which is exactly the acceptance probability in Algorithm~\ref{alg:cardinality:adv-order}. This description shows that the acceptance probabilities are well-defined and that the algorithm accepts at most $k$ elements. If $N = 0$, the count test admits no candidates and no acceptance decision is needed.

Fix an element $i$ with $p_i > 0$. Conditional on $X_i = 1$, the algorithm rejects $i$ either because it fails the count test or because it is a candidate whose position is not chosen. We first use the description above to bound the probability of the latter event.

For fixed $R$, samples, and online activations, the position of each candidate is fixed, while each of the $N$ positions is equally likely to be chosen. Since exactly $\min\set{k,N}$ positions are chosen, each candidate is accepted with probability $\min\set{k,N}/N$. Therefore, when $N \leq k$, every candidate is accepted, while for $N > k$ the rejection probability is
\[
    1-\frac{k}{N} = \frac{N-k}{N} \leq \frac{N-k}{k}.
\]
Overall, for every fixed realization of $R$, the samples, and the online activations, this rejection probability is at most $(N-k)^+/k$. Averaging over $R$, the samples, and the other online activations gives
\[
    \Pr\brk{\set{i \text{ is a candidate}} \cap A_i^c \midd X_i = 1} \leq \frac{1}{k}\E\brk{(N-k)^+ \midd X_i = 1} = \frac{1}{k}\E\brk{(N-k)^+}.
\]
Taking also into account the probability that $i$ fails the count test, we obtain
\begin{equation}\label{eq:cardinality:adv-order-rejection}
    1 - C_i \leq \Pr\brk{i \text{ is not a candidate} \midd X_i = 1} + \frac{1}{k}\E\brk{(N-k)^+}.
\end{equation}
Thus, it suffices to bound the probability of failing the count test and the expected value of $(N-k)^+$.

For the first term, let $B_t$ be the difference between $R + \sum_{j = 1}^t Y_j$ and the number of candidates among the first $t$ elements. The count test ensures that $B_t \geq 0$. Upon observing element $t$, we first add $Y_t$ to $B_{t-1}$. If $X_t = 1$ and $B_{t-1}+Y_t \geq 1$, the element becomes a candidate, and we subtract one. Otherwise, the number of candidates does not change. Therefore,
\[
    B_0 = R, \qquad \text{and} \qquad
    B_t = (B_{t-1} + Y_t - X_t)^+ \qquad \text{for } t > 0.
\]
Conditional on $X_i = 1$, element $i$ fails the count test exactly when $B_{i-1}+Y_i < 1$, or equivalently, when $B_{i-1} = 0$ and $Y_i = 0$. Furthermore, $B_{i-1}$ depends only on $R$ and on the samples and online activations of the elements up to $i-1$. It is therefore independent of both $Y_i$ and $X_i$, and thus
\[
    \Pr\brk{i \text{ is not a candidate} \midd X_i = 1}
    = (1-p_i) \Pr[B_{i-1} = 0] < \Pr[B_{i-1} = 0].
\]
Next, we show that $\Pr[B_{i-1} = 0] \leq 1/(r+1)$. In fact, we bound all point probabilities of $B_t$ together and show something stronger: for every $t \in \set{0, 1, \dots, n}$ and integer $m \geq 0$,
\[
    \Pr[B_t = m] \leq \frac{1}{r+1}.
\]
We prove this by induction on $t$. For $t = 0$, it follows because $R$ is uniform on $\set{0,\dots,r}$. For the induction step, since $X_t$ and $Y_t$ are independent of $B_{t-1}$ and of each other, $Y_t - X_t$ equals either $1$ or $-1$ with probability $p_t(1 - p_t)$ each, and $0$ with probability $1 - 2 p_t(1 - p_t)$. If $B_{t-1} = 0$, then $B_t = (Y_t - X_t)^+$. If $B_{t-1} = 1$, then $B_t = 0$ exactly when $Y_t - X_t = -1$. Thus,
\[
    \Pr[B_t = 0] = \prn{1-p_t(1-p_t)} \Pr[B_{t-1} = 0]
    + p_t(1-p_t)\Pr[B_{t-1} = 1].
\]
Furthermore, for $B_t = m$ where $m \geq 1$, the previous value must be $m-1$, $m$, or $m+1$, and the corresponding increment must be $1$, $0$, or $-1$, respectively. Therefore,
\begin{align*}
    \Pr[B_t = m] &= p_t(1-p_t)\Pr[B_{t-1} = m-1] \\
    &\qquad + \prn{1-2p_t(1-p_t)}\Pr[B_{t-1} = m]
    + p_t(1-p_t)\Pr[B_{t-1} = m+1].
\end{align*}
Each expression is a convex combination of probabilities at time $t-1$, all of which are at most $1/(r+1)$ by the induction hypothesis. This completes the induction, and hence
\[
    \Pr\brk{i \text{ is not a candidate} \midd X_i = 1}
    \leq \frac{1}{r+1}.
\]

It remains to bound the second term in \eqref{eq:cardinality:adv-order-rejection}. Let $S = \sum_{j = 1}^n Y_j$ be the number of sample-active elements, so that $N = R+S$. Since $\sum_{j = 1}^n p_j \leq k$, we have
\[
    \E[S] = \sum_{j = 1}^n p_j \leq k,
    \qquad \text{and} \qquad
    \Var(S) = \sum_{j = 1}^n p_j(1-p_j) \leq k.
\]
Thus,
\[
    (N-k)^+ = (R + S - k)^+ \leq R + (S - \E[S])^+.
\]
To bound the second term, use the identity $(x)^+ = (x+\abs{x})/2$. Since $S - \E[S]$ has mean zero, this gives
\begin{align*}
    \E\brk{(S - \E[S])^+} &= \frac{\E\brk{S - \E[S]}}{2} + \frac{\E\brk{\abs{S - \E[S]}}}{2} = \frac{\E\brk{\abs{S-\E[S]}}}{2} \\
    &\leq \frac{\sqrt{\E\brk{\prn{S - \E[S]}^2}}}{2} = \frac{\sqrt{\Var(S)}}{2},
\end{align*}
where the inequality follows from the Cauchy--Schwarz inequality. Combining the above and using the fact that $\E[R] = r/2$, we obtain
\[
    \frac{\E\brk{(N - k)^+}}{k} \leq \frac{\E[R]}{k} + \frac{\E\brk{(S - \E[S])^+}}{k} \leq \frac{r}{2k} + \frac{\sqrt{\Var(S)}}{2k}
    \leq \frac{r}{2k} + \frac{1}{2\sqrt{k}}.
\]
Substituting the two bounds into \eqref{eq:cardinality:adv-order-rejection}, we obtain
\[
    C_i \geq 1-\frac{1}{r+1}-\frac{r}{2k}-\frac{1}{2\sqrt{k}}. \qedhere
\]
\end{proof}
\section{Sample Complexity}
\label{sec:sample-complexity}

Next, we investigate whether the samples used by our schemes are necessary to obtain optimal guarantees. For random arrivals, the answer depends on $k$: one sample per element is necessary for $k \in \set{1,2,3,4}$, whereas no samples are needed for $k \geq 5$. The latter fact follows directly from a lemma by Arnosti and Ma~\cite{willma-random}. Perhaps surprisingly, for adversarial arrivals and every fixed $k \geq 2$, we show that no fixed finite number of samples per element suffices to attain the optimal guarantee for known activation probabilities.

\subsection{Random Order}
\label{sec:sample-complexity:random-order}

We start with the random order case. Consider the greedy rule that accepts the first $k$ online-active elements. Clearly, this rule uses no samples and does not need to know $p$. The following guarantee is a direct consequence of~\cite[Lemma 6]{willma-random}. 

\begin{corollary}\label{cor:sample-complexity:greedy}
For every $n,k \geq 1$, probability vector $p \in [0,1]^n$ with $\sum_{i = 1}^n p_i \leq k$, and element $i$ with $p_i>0$, the ROCRS that greedily accepts the first $k$ active elements satisfies
\[
    C_i \geq \min\set{\frac{k}{k+1},\; 1-e^{-k}\frac{k^k}{k!}}.
\]
In particular, it is an optimal $(1-e^{-k}k^k/k!)$-balanced ROCRS for $k \geq 5$.
\end{corollary}
\begin{proof}
Firstly, the set returned by the ROCRS is trivially feasible. Assume that $\sum_{i = 1}^n p_i = k$, and let $S = \sum_{j=1}^n X_j$ and $S_{-i}=\sum_{j\neq i}X_j$. By~\cite[Lemma 6]{willma-random}, for every sum $S$ of independent Bernoulli random variables with mean $k$, we have
\[
    \E[\AR_k(S)] \geq \min\set{\frac{k}{k+1},\;1-e^{-k}\frac{k^k}{k!}}.
\]
Conditional on $X_i = 1$ and the other online activations, the rank of $i$ among the $S_{-i}+1$ online-active elements is uniform. Thus, using the fact that $\AR_k$ is non-increasing,
\[
    C_i=\E[\AR_k(S_{-i})] \geq \E[\AR_k(S)] \geq \min\set{\frac{k}{k+1},\;1-e^{-k}\frac{k^k}{k!}}.
\]
Finally, if $\sum_i p_i < k$, we can simply add independent Bernoulli random variables to $S$ such that $\E[S] = k$. Since $\AR_k$ is non-increasing, this can only decrease $\E[\AR_k(S)]$, and thus the same bound follows.
\end{proof}

For $k \in \set{1,2,3,4}$, we have $k/(k+1) < 1-e^{-k}k^k/k!$, whereas the inequality is reversed for $k \geq 5$. Next, we show that the guarantee in Corollary~\ref{cor:sample-complexity:greedy} is tight for schemes with no samples, even in the offline setting. Together with Theorem~\ref{thm:cardinality:random-order}, this characterizes exactly the minimum number of samples per element needed for an optimal ROCRS: one for $k \in \set{1,2,3,4}$ and zero for $k \geq 5$.

\begin{proposition}\label{prp:sample-complexity:zero-samples}
Fix an integer $k\geq1$. If an ROCRS (respectively, an offline CRS)
that uses no samples is $c$-balanced uniformly over all ground-set
sizes and all $p \in \cP_k$, then
\[
    c \leq \min\set{\frac{k}{k+1}, \; 1-e^{-k}\frac{k^k}{k!}}.
\]
\end{proposition}
\begin{proof}
First we show that $c \leq k/(k+1)$. Let $n = k+1$. Whenever all elements are online-active, feasibility implies that some element is accepted with probability at most $k/(k+1)$; let $i$ be this element. This probability is taken over the uniformly random arrival order and any internal randomness of $\pi$. Since $\pi$ receives no samples and does not know $p$, it is the same for every probability vector $p \in (0,1]^n$.

For $\eps \in (0,1)$, consider
\[
    p_i=\eps,\qquad p_j=1-\frac{\eps}{k}\quad\text{for every }j\neq i.
\]
Observe that $\sum_j p_j = k$. Conditional on $X_i = 1$, all other elements are online-active with probability $(1 - \eps/k)^k$. Therefore,
\[
    C_i \leq \frac{k}{k+1}\prn{1-\frac{\eps}{k}}^k + 1 - \prn{1-\frac{\eps}{k}}^k = 1-\frac{1}{k+1}\prn{1-\frac{\eps}{k}}^k \leq \frac{k+\eps}{k+1},
\]
where the last inequality uses $(1-\eps/k)^k \geq 1-\eps$. Since $c\leq C_i$ on every such instance, letting $\eps\to0$ gives $c\leq k/(k+1)$. The same argument applies to an offline CRS since, even after observing the entire online-active set, it must reject some element with probability at least $1/(k+1)$.

For the other upper bound, consider $n \geq k$ elements with $p_i = k/n$ for every $i \in [n]$, and let $S = \sum_{i=1}^n X_i\sim\Bin{n,k/n}$. Suppose that $\pi$ is $c$-balanced for some $c \in [0,1]$. Then,
\[
    c k \leq \sum_{i = 1}^n \Pr[A_i] \leq \E\brk{\min\set{k,S}},
\]
where the second inequality follows from the fact that $\pi$ has to output a feasible set. As $n \to \infty$, $S$ converges in distribution to $Z\sim\Poi{k}$. Taking the limit and using \eqref{eq:min-k-s-poisson} yields
\[
    c \leq \frac{1}{k}\E\brk{\min\set{k,Z}} = 1 - e^{-k}\frac{k^k}{k!}. \qedhere
\]
\end{proof}

\subsection{Adversarial Order}
\label{sec:sample-complexity:adversarial-order}

We continue with adversarial arrivals. As shown in Section~\ref{sec:single-item:adversarial:two-samples}, two samples per element suffice to recover the optimal $1/2$-balanced OCRS for $k = 1$. We show that for $k \geq 2$ even an arbitrarily large fixed number of samples per element does not suffice to attain the optimal guarantee for known activation probabilities.

Fix an integer $k \geq 2$. We use $\gamma$ (and drop the dependency on $k$) to denote the optimal guarantee of an OCRS that knows $p$, uniformly over all $n$ and all probability vectors with $\sum_i p_i = k$; this is characterized for every $k$ by Jiang, Ma, and Zhang~\cite{jiang-ma-zhang-multi-unit}.

\begin{theorem}\label{thm:sample-complexity:finite-impossibility}
For every fixed integer $k \geq 2$, finite integer $m \geq 0$, and every OCRS $\pi$ using at most $m$ independent samples per element, there exist a ground-set size $n$ and a probability vector $p \in [0,1]^n$ with $\sum_i p_i = k$ such that
\[
    \min_{\set{i \midd p_i > 0}} C_{i, \pi}(p) < \gamma.
\]
This holds even if the fixed arrival order is given to $\pi$ in advance.
\end{theorem}

The idea behind the proof is to compare two instances $p^{(1)}, p^{(2)}$ on which optimality requires different behavior before the same element arrives. In one instance, the probability of having accepted two elements must tend to zero, whereas in the other it must stay bounded away from zero. We first derive this restriction from the characterization of the optimal OCRS when $p$ is known. We then show that a fixed number of samples cannot support these two different behaviors.

Let $\tilde y_\ell:[0,k]\to[0,1]$ denote the functions $\tilde y_{\ell,\gamma}$ of \cite[Definition~3 and Appendix~C.8]{jiang-ma-zhang-multi-unit}, with the fixed parameter $\gamma$ denoted by $\gamma_k^*$ in \cite{jiang-ma-zhang-multi-unit}. The equations of \cite{jiang-ma-zhang-multi-unit} can be written as
\begin{equation}\label{eq:sample-complexity:tail-benchmark}
\begin{aligned}
    \tilde y_\ell'(t)
       &=\prn{\gamma-1+\tilde y_{\ell-1}(t)}^+
         -\prn{\gamma-1+\tilde y_\ell(t)}^+
         &&(1\leq\ell<k),\\
    \tilde y_k'(t)
       &=\prn{\gamma-1+\tilde y_{k-1}(t)}^+,
\end{aligned}
\end{equation}
where $\tilde y_0(t)=1$ and $\tilde y_\ell(0)=0$ for $1\leq\ell\leq k$. Let
\[
    h_\ell(t)\coloneqq\sum_{r=\ell}^k\tilde y_r(t),
    \qquad \text{and} \qquad
    h'_\ell(t)\coloneqq\sum_{r=\ell}^k\tilde y'_r(t),
    \qquad (1\leq\ell\leq k).
\]

The functions $\tilde y_\ell$ are non-decreasing and, at the optimal parameter $\gamma$, satisfy
\[
    1=\tilde y_0(t)\geq\tilde y_1(t)\geq\dots\geq\tilde y_k(t)\geq0.
\]
Since $h_k=\tilde y_k$, the characterization of \cite{jiang-ma-zhang-multi-unit} gives the following identity.

\begin{proposition}[Jiang, Ma, and Zhang~{\cite[Theorem~2]{jiang-ma-zhang-multi-unit}}]\label{prop:sample-complexity:benchmark}
The optimal balance satisfies $0<\gamma<1$ and
\[
    h_k(k)=\tilde y_k(k)=1-\gamma.
\]
\end{proposition}

Next, observe that the right-hand sides of \eqref{eq:sample-complexity:tail-benchmark} telescope when summed over $r=\ell,\dots,k$. For $\ell=1$, this gives $h_1'(t)=\gamma$.  Let $t_2 \coloneqq (1-\gamma)/\gamma$. For $\ell \geq 2$, using $\tilde y_{\ell-1}=h_{\ell-1}-h_\ell$, we obtain
\begin{equation}\label{eq:sample-complexity:benchmark}
    h_1(t)=\gamma t,\qquad
    h_\ell(0)=0,\qquad
    h_\ell'(t)=\prn{\gamma-1+h_{\ell-1}(t)-h_\ell(t)}^+
    \quad(2\leq\ell\leq k).
\end{equation}
Solving \eqref{eq:sample-complexity:benchmark} for $\ell=2$ gives
\begin{equation}\label{eq:sample-complexity:h2}
    h_2(t)=
    \begin{cases}
        0,&0\leq t\leq t_2,\\
        \gamma\prn{t-t_2-1+e^{-(t-t_2)}},&t>t_2.
    \end{cases}
\end{equation}
Thus $h_2(t)>0$ for $t>t_2$. Also, $h_2(k)\geq h_k(k)=1-\gamma>0$, so $0<t_2<k$. The time $t_2$ is the first time that $\tilde y_1$ reaches $1-\gamma$.

Next, we compare these functions with the behavior of an arbitrary $\gamma$-balanced OCRS.

For an OCRS $\pi$ and an instance with $p_i>0$ for all $i$ and $\sum_i p_i=k$, let
\[
    N_j\coloneqq\sum_{i=1}^j\1_{A_i},\qquad
    s_j\coloneqq\sum_{i=1}^j p_i,\qquad
    H_{\ell,j}\coloneqq\E\brk{(N_j-\ell+1)^+}
    \quad(1\leq\ell\leq k),
\]
with $N_0=s_0=H_{\ell,0}=0$. Thus, $N_j$ counts the acceptances among the first $j$ elements, and $H_{\ell,j}$ is their expected number beyond the first $\ell-1$. Also, let $\delta\coloneqq\max_i p_i$, and $H_\ell^{(\delta)}:[0,k]\to\R$ be the linear interpolation of the values $H_{\ell,j}$ at the points $s_j$. The proof of the following lemma can be found in Appendix~\ref{app:missing:rigidity}.

\begin{lemma}\label{lem:sample-complexity:rigidity}
Consider any sequence of $\gamma$-balanced OCRSs and instances with positive activation probabilities summing to $k$, such that $\delta\to0$. Then, $H_2^{(\delta)}$ converges uniformly to $h_2$ on $[0,k]$. In particular, for any choice of prefix lengths $j_r$ along the sequence of instances $p^{(r)}$ such that
\[
    s_{j_r}^{(r)} = \sum_{i=1}^{j_r} p_i^{(r)} \longrightarrow t,
\]
we have
\[
    \E\brk{(N_{j_r}^{(r)}-1)^+}\longrightarrow h_2(t).
\]
\end{lemma}

We are ready to prove the impossibility theorem.
\begin{proof}[Proof of Theorem~\ref{thm:sample-complexity:finite-impossibility}]
Suppose, towards contradiction, that there exists a $\gamma$-balanced OCRS $\pi$ using at most $m$ samples per element, for a fixed finite $m$.

Since $0<t_2<k$, we can choose $a,b$ such that $0<a<t_2<b<k$. For sufficiently large $n$, consider the following two instances on $2n$ elements, both arriving in order $1,\dots,2n$:
\[
\begin{array}{c|cc}
    &1,\dots,n&n+1,\dots,2n\\
    \hline
    p^{(1)}&a/n&(k-a)/n\\
    p^{(2)}&b/n&(k-b)/n
\end{array}
\]
Both vectors have total activation mass $k$ and their maximum coordinates tend to zero as $n \to \infty$. Abusing notation, we write $\Pr_1$ and $\E_1$ (resp. $\Pr_2$ and $\E_2$) for probabilities and expectation under instance $p^{(1)}$ (resp. $p^{(2)}$).

Consider the event $E_n=\set{N_n\geq2}$ that $\pi$ accepts at least two of the first $n$ elements. First, since the activation mass of the first $n$ elements is $a < t_2$ in the first instance, by Lemma~\ref{lem:sample-complexity:rigidity}, we have
\[
    \E_1\brk{(N_n - 1)^+} \to h_2(a) = 0,
\]
where the equality follows from \eqref{eq:sample-complexity:h2}.
Furthermore, observe that $\1_{E_n} \leq (N_n-1)^+$, which implies $\Pr_1[E_n] \to 0$, as $n \to \infty$. Next, since the activation mass of the first $n$ elements is $b > t_2$ in the second instance, by Lemma~\ref{lem:sample-complexity:rigidity}, we have
\[
    \E_2\brk{(N_n - 1)^+} \to h_2(b) > 0,
\]
where the inequality follows from \eqref{eq:sample-complexity:h2}.
Furthermore, $(N_n-1)^+ \leq (k-1) \1_{E_n}$, which implies $\liminf_{n \to \infty} \Pr_2[E_n] \geq h_2(b)/(k-1) > 0$. Overall, we have
\begin{equation}\label{eq:sample-complexity:two-instances}
    \Pr_1[E_n] \longrightarrow 0,
    \qquad \text{and} \qquad
    \liminf_{n \to \infty} \Pr_2[E_n] \geq \frac{h_2(b)}{k-1} > 0.
\end{equation}
Next, we show that the first limit in \eqref{eq:sample-complexity:two-instances} forces $\Pr_2[E_n]\to0$ as well. Let $P_1, P_2$ be the distributions of the full array of $m$ sample rows and one online row, together with an independent random seed for the algorithm, for instances $p^{(1)}$ and $p^{(2)}$, respectively. The seed has the same law under both instances. Let $L_n = \dif P_2/ \dif P_1$ be the likelihood ratio under $p^{(2)}$ relative to $p^{(1)}$ of all samples, all online activations, and the internal random seed of $\pi$. In other words, $L_n$ measures how much more likely an observation record is under instance $p^{(2)}$ compared to $p^{(1)}$. The seed has the same distribution in both instances and is independent of the samples and online activations.

For one Bernoulli variable whose success probability is $p$ in the first instance and $q$ in the second, the second moment of its likelihood ratio under the first instance is
\[
    p\prn{\frac{q}{p}}^2
       +(1-p)\prn{\frac{1-q}{1-p}}^2
       =1+\frac{(q-p)^2}{p(1-p)}.
\]
In the first block of $n$ elements, we have $p=a/n$ and $q=b/n$. Substituting gives
\[
    1 + \frac{(b-a)^2}{an(1-a/n)} = 1 + \frac{(b-a)^2}{an} + \bigO{n^{-2}},
\]
because $1/(1-a/n)=1 + \bigO{n^{-1}}$. In the second block of $n$ elements, $p=(k-a)/n$ and $q=(k-b)/n$, so the same calculation gives
\[
    1+\frac{(b-a)^2}{(k-a)n\bigl(1-(k-a)/n\bigr)} = 1+\frac{(b-a)^2}{(k-a)n}+\bigO{n^{-2}}.
\]
Each block has $n$ elements, and each element supplies its $m$ samples and its online activation for a total of $m+1$ independent Bernoulli observations. Since the random variables of each block are independent, the full likelihood ratio is the product of the individual likelihood ratios for each block. Thus, for
\[
    c_1 = \frac{(b-a)^2}{a}, \qquad \text{and} \qquad c_2 = \frac{(b-a)^2}{k-a},
\]
we obtain
\begin{align}
    \E_1[L_n^2] &= \prod_{i=1}^{2n} \prn{1+\frac{(p_i^{(2)}-p_i^{(1)})^2} {p_i^{(1)}(1-p_i^{(1)})}}^{m+1} \nonumber \\
    \label{eq:sample-complexity:likelihood} &= \prn{1 + \frac{c_1}{n} + \bigO{n^{-2}}}^{n(m+1)} \prn{1 + \frac{c_2}{n} + \bigO{n^{-2}}}^{n(m+1)} \longrightarrow \exp\prn{(m+1)(c_1 + c_2)} < \infty,
\end{align}
where we use that $\prn{1+c/n+\bigO{n^{-2}}}^n \to e^c$ and the fact that the random seed contributes a likelihood-ratio factor of one because it is independent of the observations and has the same distribution in both instances.

Next, let $B_n$ denote all the samples, online activations, and the algorithm's random seed. Since $\pi$ does not know $p$, the event $E_n$ is determined by the same function of $B_n$. Thus, there exists a function $f_n$, common to both instances, such that $\1_{E_n} = f_n(B_n)$. Therefore, by the change-of-measure identity, we have
\[
    \Pr_2[E_n] = \E_2[\1_{E_n}] = \E_1\brk{\frac{\dif P_2}{\dif P_1} \cdot \1_{E_n}} = \E_1[L_n \1_{E_n}],
\]
and, by the Cauchy--Schwarz inequality, we obtain
\[
    \Pr_2[E_n] = \E_1[L_n \1_{E_n}] \leq \sqrt{\E_1[L_n^2]\Pr_1[E_n]} \longrightarrow 0,
\]
where the limit follows from \eqref{eq:sample-complexity:likelihood} and $\Pr_1[E_n]\to0$. This contradicts the positive lower bound in \eqref{eq:sample-complexity:two-instances}.
\end{proof}

The same argument gives a positive gap from $\gamma$ for each fixed sample bound. For $k = 1$, we have $\gamma=1/2$ and $t_2=1=k$, so the choice $a<t_2<b<k$ is impossible, which is why there exists an optimal two-sample OCRS like the one we present in Section~\ref{sec:single-item:adversarial:two-samples}.

\begin{corollary}\label{cor:sample-complexity:positive-gap}
For every fixed $k\geq2$ and finite $m\geq0$, there is $\eps_{k,m}>0$ such that every OCRS using at most $m$ samples per element has worst-case balancedness at most $\gamma-\eps_{k,m}$, even on instances with $\sum_i p_i=k$.
\end{corollary}
\begin{proof}
Suppose, towards contradiction, that there is a sequence of OCRSs with balancedness guarantees $c_r\to\gamma$, all using at most $m$ samples per element. Apply the $r$-th OCRS to the two instances above with block size $n_r\to\infty$. The inequalities for a fixed instance in the proof of Lemma~\ref{lem:sample-complexity:rigidity} hold with $c_r$ in place of $\gamma$. Since $c_r\to\gamma$, taking the limit gives the same inequalities for $H_\ell$ and the same bound $H_k(k)\leq1-\gamma$. That proof therefore still gives \eqref{eq:sample-complexity:two-instances}. The second moments in \eqref{eq:sample-complexity:likelihood} remain bounded because $m$ is fixed, and the same Cauchy--Schwarz argument that yields a contradiction still holds.
\end{proof}

\section*{AI Disclosure Statement}

The problem formulation and the proof of Theorem~\ref{thm:random-order-k-equals-1} were developed by the authors. The authors also developed the original ROCRS in Appendix~\ref{app:alternative-proof}. Subsequently, a simpler ROCRS was obtained with the use of ChatGPT (GPT 6 Astra); this is the scheme presented in Section~\ref{sec:cardinality:random-order}. The remaining proofs were developed by ChatGPT (GPT 5.6 Sol and GPT 6 Astra) after lengthy discussions with the authors who provided guidance and feedback for the AI's approaches. The paper was written by the authors, who take full responsibility for its correctness and logical integrity.

\bibliographystyle{plain}
\bibliography{refs}

\appendix
\section{Missing Proofs}
\label{app:missing-proofs}

\subsection{Proof of Theorem~\ref{thm:adv-order:half-impossibility}}
\label{app:missing:impossibility}

\begin{reptheorem}{thm:adv-order:half-impossibility}
There does not exist a single-sample, $1/2$-balanced OCRS in the single-item setting, when the elements arrive in adversarial order.
\end{reptheorem}
\begin{proof}
Suppose, towards contradiction, that there exists a single-sample $1/2$-balanced OCRS $\pi$. Fix $\eps = 1/1000$. We consider the following five instances $p^{(1)}, \dots, p^{(5)}$ on $6$ elements and study how well the algorithm performs on them, not knowing which instance it is actually facing.
\[
\begin{array}{c|cccccc}
\text{} & 1 & 2 & 3 & 4 & 5 & 6 \\
\hline
p^{(1)} & 1/2-\eps & 1/2 & \eps & 0 & 0 & 0 \\
p^{(2)} & 0 & 0 & 1 & 0 & 0 & 0 \\
p^{(3)} & 0 & 0 & \eps & 1-\eps & 0 & 0 \\
p^{(4)} & 0 & 0 & \eps & 0 & 1-\eps & 0 \\
p^{(5)} & 0 & 0 & 1/3 & 1/3 & 1/3-\eps & \eps
\end{array}
\]
For each $S \subseteq \set{3,4,5}$, let $a_S$ be the probability that $\pi$ accepts element $3$,
conditional on elements $1$ and $2$ being
online-inactive, element $3$ being online-active, the sample-active elements
among $\set{3,4,5}$ being exactly $S$, and all other elements being
sample-inactive. In other words,
\[
    a_S \coloneqq \Pr\brk{A_3 \midd X_1 = X_2 = 0, X_3 = 1, (Y_3, Y_4, Y_5) = \1_S, (Y_1, Y_2, Y_6) = \mathbf{0}}.
\]
Since the algorithm does not know $p$ and $a_S$ depends only on the fixed observations up to element $3$, the values $a_S$ are the same across all five instances.

First, consider the fact that the algorithm might be facing $p^{(1)}$. Since $\pi$ is $1/2$-balanced, $C_1(p^{(1)}), C_2(p^{(1)}) \geq 1/2$ and we have
\begin{align*}
    \Pr\brk{R_3 \midd X_3 = 1} &= 1 - \Pr\brk{A_1 \midd X_3 = 1} - \Pr\brk{A_2 \midd X_3 = 1} = 1 - \Pr[A_1] - \Pr[A_2] \\
    &= 1 - \prn{\frac{1}{2} - \eps} C_1(p^{(1)}) - \frac{1}{2} C_2(p^{(1)}) \leq 1 - \prn{\frac{1}{2} - \eps}\frac{1}{2} - \frac{1}{2} \cdot \frac{1}{2} \\
    &= \frac{1}{2} + \frac{\eps}{2}.
\end{align*}
Let $D_3$ be the event that element $3$ is online-active, $\pi$ reaches it, and $\pi$ rejects it, i.e. $D_3 \coloneqq \set{X_3 = 1} \cap R_3 \cap A_3^c$, and observe that $A_3\subseteq R_3\cap\set{X_3=1}$. Since $\pi$ is $1/2$-balanced, $C_3(p^{(1)}) \geq 1/2$, and we have
\begin{align*}
    \Pr\brk{D_3 \midd X_3 = 1} &= \Pr\brk{R_3 \midd X_3 = 1} - \Pr\brk{A_3 \midd X_3 = 1} \\
    &= \Pr\brk{R_3 \midd X_3 = 1} - C_3(p^{(1)}) \leq \frac{\eps}{2}.
\end{align*}
Next, consider the event $E = \set{X_1 = X_2 = 0, \; Y_1 = Y_2 = \dots = Y_6 = 0}$. Whenever $E$ happens, we cannot accept elements $1$ and $2$, thus $E \subseteq R_3$. Furthermore,
\[
	\Pr\brk{A_3 \midd E, X_3 = 1} = a_{\emptyset},
\]
and also $\Pr[E] = (1/4) (1/2 + \eps)^2 (1 - \eps) \geq 1/16$. Whenever $E$ occurs, element $3$ is online-active, and the algorithm rejects it, then the event $D_3$ occurs. Therefore,
\begin{align*}
    \Pr\brk{D_3 \midd X_3 = 1} &\geq \Pr\brk{E \cap A_3^c \midd X_3 = 1} = \Pr\brk{E \midd X_3 = 1} \Pr\brk{A_3^c \midd E, X_3 = 1} = \Pr[E] \: (1 - a_{\emptyset}) \\
    &\geq \frac{1}{16} \: (1 - a_{\emptyset}),
\end{align*}
which implies that
\[
	\frac{1}{16}(1 - a_{\emptyset}) \leq \frac{\eps}{2} \iff a_{\emptyset} \geq 1 - 8 \eps.
\]

Next consider the fact that the algorithm might be facing $p^{(2)}, p^{(3)}$ or $p^{(4)}$. Under $p^{(2)}$, elements $1$ and $2$ are online-inactive and the sample-active set is exactly $\set{3}$, thus
\[
    a_{\set{3}} = C_3(p^{(2)}) \geq \frac{1}{2}.
\]
For each $j \in \set{4,5}$, under $p^{(j-1)}$ elements $1$ and $2$ are again online-inactive, and the sample-active set is exactly $\set{j}$ with probability $(1 - \eps)^2$, independently of $X_3$. Since the probability of accepting element $3$ on any other sample outcome is at most $1$,
\[
    \frac{1}{2} \leq C_3(p^{(j-1)}) \leq (1-\eps)^2 a_{\set{j}} + 1 - (1 - \eps)^2 \leq a_{\set{j}} + 2 \eps.
\]
Consequently, $a_{\set{4}},a_{\set{5}} \geq 1/2 - 2 \eps$.

Finally, consider $p^{(5)}$, and let $S$ be the random set of sample-active elements among $\set{3,4,5}$. Elements $1$ and $2$ are both online-inactive and sample-inactive, while $Y_6$ is independent of $S$ and $X_3$. Therefore,
\[
    C_3(p^{(5)}) \geq (1 - \eps) \Pr\brk{A_3 \midd Y_6 = 0, X_3 = 1} = (1 - \eps) \E[a_S].
\]
Also, when facing $p^{(5)}$,
\begin{align*}
    \Pr[S = \emptyset] &= \prn{\frac{2}{3}}^2 \prn{\frac{2}{3} + \eps} \geq \frac{8}{27}, \qquad \text{and} \\
    \Pr[|S| = 1] &= 2 \cdot \frac{1}{3} \cdot \frac{2}{3} \prn{\frac{2}{3} + \eps} + \prn{\frac{1}{3} - \eps} \prn{\frac{2}{3}}^2 = \frac{4}{9}.
\end{align*}
Therefore,
\begin{equation}\label{eq:adv-order:single-sample-contradiction}
    C_3(p^{(5)}) \geq (1 - \eps) \prn{\frac{8}{27}(1 - 8 \eps) + \frac{4}{9} \prn{\frac{1}{2} - 2 \eps}} \geq \frac{14}{27} - 4 \eps > \frac{1}{2} + \frac{3 \eps}{2},
\end{equation}
where the last inequality uses $\eps = 1/1000$.

Furthermore, the $1/2$-balancedness for elements $4$, $5$, and $6$ together with \eqref{basic-balancedness} for $i = 6$, imply that
\begin{align*}
    \frac{1}{3} C_3(p^{(5)}) &\leq 1 - C_6(p^{(5)}) - \frac{1}{3} C_4(p^{(5)}) - \prn{\frac{1}{3} - \eps} C_5(p^{(5)}) \\
    &\leq 1 - \frac{1}{2} - \frac{1}{6} - \prn{\frac{1}{6} - \frac{\eps}{2}} = \frac{1}{6} + \frac{\eps}{2}.
\end{align*}
Thus, $C_3(p^{(5)}) \leq 1/2 + 3\eps/2$, contradicting \eqref{eq:adv-order:single-sample-contradiction}. Therefore, there does not exist a single-sample, $1/2$-balanced OCRS. \qedhere
\end{proof}

In fact, this construction shows an upper bound on the balancedness of single-sample OCRSs of $0.499$. To see this, set $c = 0.499$ and use the same five instances with $\eps = 1/10000$. If the scheme were $c$-balanced, the preceding reach and rejection calculation would give
\[
    a_{\emptyset}\geq1-16\prn{1-(2-\eps)c},
    \qquad
    a_{\{3\}} \geq c,
    \qquad
    a_{\{4\}},a_{\{5\}}\geq c-2\eps.
\]
Thus,
\[
    C_3(p^{(5)}) \geq (1-\eps) \prn{\frac8{27}\prn{1-16(1-(2-\eps)c)} + \frac49(c-2\eps)} > 0.5082.
\]
However, the constraint for element $6$ gives
\[
    C_3(p^{(5)}) \leq 3 - (5-3\eps) c < 0.5052,
\]
a contradiction.

\subsection{Proof of Lemma~\ref{lem:adv-order:threshold-deficit}}
\label{app:missing:threshold-deficit-lemma}

\begin{replemma}{lem:adv-order:threshold-deficit}
For every element $i$,
\[
    \eps_i \leq \frac{1-1/e}{12} \: p_i(1 - p_i).
\]
\end{replemma}
\begin{proof}
If $p_i = 0$, then $\eps_i = 0$; thus assume $p_i > 0$. Condition on $Y_i = 1$ and let
\[
    |\cS - i| = \sum_{j \neq i} Y_j, \qquad \text{and} \qquad q_i \coloneqq \E[|\cS - i|] = \sum_{j \neq i} p_j = 1 - p_i.
\]
If $|\cS - i| = 0, 1$, Algorithm~\ref{alg:adv-order:single-sample} sets $T = i$ with probability $1/2$. If, on the other hand, $|\cS - i| = z \geq 2$, $T$ is chosen uniformly at random among the $z+1$ sample-active elements; thus, it sets $T = i$ with probability $1/(z+1)$. Define $\varphi : \N \to \R$ as
\[
    \varphi(z) \coloneqq \begin{cases}
        0 & \text{for } z \leq 1, \\
        \displaystyle
        \frac{1}{2} - \frac{1}{z+1} & \text{for } z \geq 2,
    \end{cases}
\]
and notice that
\[
    \eps_i = p_i \E\brk{\varphi \prn{|\cS - i|} }.
\]
Observe that, for every integer $z \geq 2$,
\[
    \frac{1}{2} - \frac{1}{z + 1} \leq \frac{z}{12},
\]
since this inequality is equivalent to $(z-2)(z-3) \geq 0$. Therefore,
\begin{align}
    \frac{\eps_i}{p_i} &= \sum_{z \geq 2} \prn{\frac{1}{2} - \frac{1}{z+1}} \Pr\brk{|\cS - i| = z}
    \leq \frac{1}{12} \sum_{z \geq 2} z \Pr\brk{|\cS - i| = z} \nonumber \\
    \label{eq:adv-threshold-singleton-deficit} &= \frac{1}{12} \prn{\E\brk{|\cS - i|} - \Pr\brk{|\cS - i|=1}} = \frac{q_i-\Pr\brk{|\cS - i|=1}}{12}.
\end{align}

If $q_i = 0$, the desired inequality follows trivially. Otherwise, $0 < q_i < 1$ since $p_i > 0$. By Hoeffding's theorem~\cite[Corollary~2.1]{hoeffding-poisson-binomial}, the minimum singleton probability at fixed mean $q_i$ is attained by $r$ equal non-zero probabilities $q_i/r$. Thus,
\[
    \Pr[|\cS - i| = 1] \geq q_i(1 - q_i/r)^{r-1} \geq q_i e^{-q_i},
\]
where the last inequality follows from $(1 - x/r)^{r-1} \geq e^{-x}$ for $0 \leq x \leq1$. Substituting this into \eqref{eq:adv-threshold-singleton-deficit}, we get that
\[
    \eps_i \leq \frac{1 - e^{- q_i}}{12} p_i q_i = \frac{1 - e^{-q_i}}{12} p_i (1 - p_i) \leq \frac{1 - 1/e}{12} p_i (1 - p_i),
\]
where the last inequality follows from $q_i \leq 1$.\footnote{Note that we could also apply Hoeffding's theorem directly to the function $\varphi$ and obtain a better constant, but we avoid this for the sake of expositional simplicity.}
\end{proof}

\subsection{Proof of Lemma~\ref{lem:adv-order:g-bound}}
\label{app:missing:g-bound-lemma}

\begin{replemma}{lem:adv-order:g-bound}
For every $p \in [0,1]^n$ and $\ell \in [n]$ with $\sum_{i = 1}^\ell p_i \leq 1$, we have
\[
    \sum_{i = 1}^{\ell-1} p_i(1 - p_i)(g_\ell(i) - g_\ell(0)) \leq \frac{3}{8}.
\]
\end{replemma}
\begin{proof}
For $\ell = 1$, the sum is empty and the statement follows trivially, so assume $\ell \geq 2$. Let $\cS_{< \ell} \coloneqq \cS \cap [\ell-1]$ be the set of sample-active elements before $\ell$, and let $J$ be its last element, with $J = 0$ if $\cS_{< \ell} = \emptyset$. Observe that, for every $i < \ell$,
\begin{align*}
    \Pr\brk{J = i} &= p_i \prod_{i < j < \ell}(1 - p_j) = p_i g_\ell(i), \qquad \text{and} \\
    (1 - p_i)\Pr\brk{J = i \text{ and } |\cS_{<\ell}| = 1} &= (1 - p_i)p_i \prod_{\substack{j < \ell \\ j \neq i}}(1 - p_j) = p_i g_\ell(0),
\end{align*}
since $\set{J = i, \: |\cS_{<\ell}| = 1}$ is the event $\cS_{< \ell} = \set{i}$. Therefore,
\begin{align*}
    & \sum_{i = 1}^{\ell - 1} p_i (1 - p_i) \prn{g_\ell(i) - g_\ell(0)} = \sum_{i < \ell} (1 - p_i) \prn{ p_i g_\ell(i) - p_i g_\ell(0) } \\
    &\qquad = \sum_{i = 1}^{\ell-1} (1 - p_i) \prn{ \Pr\brk{J = i} - (1 - p_i) \Pr\brk{J = i, |\cS_{<\ell}| = 1} } \\
    &\qquad = \sum_{i = 1}^{\ell-1} (1 - p_i) \Pr\brk{J = i, |\cS_{<\ell}| \geq 2} + \sum_{i = 1}^{\ell-1} \prn{(1 - p_i) - (1 - p_i)^2} \Pr\brk{J = i, |\cS_{<\ell}| = 1} \\
    &\qquad = \sum_{i = 1}^{\ell-1} (1 - p_i) \Pr\brk{J = i, |\cS_{<\ell}| \geq 2} + \sum_{i = 1}^{\ell-1} p_i (1 - p_i) \Pr\brk{J = i, |\cS_{<\ell}| = 1} \\
    &\qquad \leq \sum_{i = 1}^{\ell-1} \Pr\brk{J = i, |\cS_{<\ell}| \geq 2} + \frac{1}{4} \sum_{i = 1}^{\ell-1}\Pr\brk{J = i, |\cS_{<\ell}| = 1} \\
    &\qquad \leq \Pr\brk{|\cS_{<\ell}| \geq 2} + \frac{1}{4} \Pr\brk{|\cS_{<\ell}| = 1} \\
    &\qquad = 1 - \Pr\brk{|\cS_{<\ell}| = 0} - \frac{3}{4}\Pr\brk{|\cS_{< \ell}| = 1},
\end{align*}
where the third equality follows from the fact that $\Pr[J = i] = \Pr[J = i, |\cS_{<\ell}| = 1] + \Pr[J = i, |\cS_{<\ell}| \geq 2]$, the fourth equality follows from the fact that $(1 - p_i) - (1 - p_i)^2 = p_i (1 - p_i)$ and the inequality follows from the facts that $1 - p_i \leq 1$ and $p_i (1 - p_i) \leq 1/4$, together with the law of total probability. Thus, it suffices to show that
\[
    \Pr\brk{|\cS_{< \ell}| = 0} + \frac{3}{4}\Pr\brk{|\cS_{< \ell}| = 1} \geq \frac{5}{8}.
\]
Define $\psi : \N \to \R$ as
\[
    \psi(z) \coloneqq \begin{cases}
        1 & \text{for } z = 0, \\
        \dfrac{3}{4} & \text{for } z = 1, \\
        0 & \text{for } z \geq 2,
    \end{cases}
\]
and observe that $\Pr\brk{|\cS_{< \ell}| = 0} + (3/4) \Pr\brk{|\cS_{< \ell}| = 1} = \E\brk{\psi\prn{ |\cS_{< \ell}| }}$. Furthermore, $\psi$ is non-increasing and thus increasing any $p_i$ for $i < \ell$ cannot increase $\E\brk{\psi\prn{ |\cS_{< \ell}| }}$. Therefore, $\E\brk{\psi\prn{ |\cS_{< \ell}| }}$ is minimized for $\sum_{i = 1}^{\ell-1} p_i = 1$.

By Hoeffding's theorem~\cite[Corollary~2.1]{hoeffding-poisson-binomial}, we can choose such a minimizing vector with probabilities $p_i \in \set{0, a ,1}$ for all $i \in [\ell-1]$ and for some $a \in (0, 1)$. If there exists $p_i = 1$ then $p_j = 0$ for all $j \neq i$ and $|\cS_{<\ell}| = 1$ with probability one, in which case $\E\brk{\psi\prn{ |\cS_{< \ell}| }} = 3/4 > 5/8$. Assume that there are $r \geq 2$ non-zero probabilities, all equal to $a = 1/r$, and thus $|\cS_{<\ell}| \sim \Bin{r, 1/r}$. In this case,
\[
    \E\brk{\psi\prn{ |\cS_{< \ell}| }} = \prn{1 - \frac{1}{r}}^r + \frac{3}{4}\prn{1 - \frac{1}{r}}^{r-1} = \prn{1 - \frac{1}{r}}^{r-1}\prn{\frac{7}{4} - \frac{1}{r}}.
\]
Let
\[
    f(r) \coloneqq \log\prn{\prn{1 - \frac{1}{r}}^{r-1}\prn{\frac{7}{4} - \frac{1}{r}}} = (r-1) \log\prn{1 - \frac{1}{r}} + \log\prn{\frac{7}{4} - \frac{1}{r}}.
\]
For $r \geq 2$, we have
\[
    f'(r) = \log\prn{1 - \frac{1}{r}} + \frac{1}{r} + \frac{4}{r(7r - 4)}, \qquad \text{and} \qquad f''(r) = \frac{16 - 7r}{r(r - 1)(7r - 4)^2}.
\]
Thus $f'$ increases on $[2, 16/7]$ and decreases to $0$ afterwards. Since $f'(2) = 7/10 - \log 2 > 0$, it follows that $e^f$ is increasing on $[2,\infty)$, and hence $e^{f(r)} \geq e^{f(2)} = \frac{5}{8}$. Therefore,
\[
    \sum_{i = 1}^{\ell - 1}p_i(1 - p_i)\prn{g_\ell(i) - g_\ell(0)} \leq 1 - \frac{5}{8} = \frac{3}{8}. \qedhere
\]
\end{proof}

\subsection{A Self-Contained Proof of Theorem~\ref{thm:adv-order:two-samples}}
\label{app:missing:self-contained}

\begin{reptheorem}{thm:adv-order:two-samples}
For every $p \in [0,1]^n$ with $\sum_{i = 1}^n p_i = 1$, the threshold $T$ in Algorithm~\ref{alg:adv-order:two-samples} satisfies
\[
    \Pr[T = 0] = \frac{1}{2}, \qquad \text{and} \qquad \Pr[T = i] = \frac{p_i}{2} \qquad \text{for every } i \in [n].
\]
\end{reptheorem}
\begin{proof}
Recall that $\cS = \set{j \in [n] \midd Y_j^{(1)} = 1}$ is the set of sample-active elements in the first samples, where the $Y_j^{(1)} \sim \Ber{p_j}$ are independent. Algorithm~\ref{alg:adv-order:two-samples} can choose $T = i$ only when $Y_i^{(1)} = 1$, so $\Pr[T = i] = 0$ whenever $p_i = 0$. If $p_i = 1$, then $\cS = \set{i}$ almost surely and the singleton branch chooses $i$ with probability $1/2$, as required. Fix an element $i$ with $0 < p_i < 1$. It suffices to show that
\[
    \Pr\brk{T = i \midd Y_i^{(1)} = 1} = \frac{1}{2}.
\]
Condition on $Y_i^{(1)} = 1$ and on the other first samples. The probabilities below are over the second samples and the algorithm's internal randomness. If $\cS -i = \emptyset$, the algorithm chooses $T = i$ with probability $1 - p_i/2$. Otherwise, conditioning on $\cS - i \neq \emptyset$, we have
\begin{align}
    \Pr\brk{T = i \midd \cS - i \neq \emptyset} &= \frac{1}{2} \Pr\brk{J = i \wedge Y^{(2)}_i = 0 \midd \cS - i \neq \emptyset} \nonumber \\
    \label{eq:adv-order:two-samples-0} &\quad + \frac{1}{2} \Pr\brk{J \neq i \wedge Y^{(2)}_J = 1 \wedge J' = i \midd \cS - i \neq \emptyset}
\end{align}
Fix $S$ with $i \in S$ and $|S| \geq 2$, and condition on $\cS = S$.
The choice of $J$ is uniformly random from $S$ and, conditional on $J = j$, the choice $J'$ is uniform on $S - j$. These choices use random coins that are independent of the $Y^{(2)}_j$'s and of each other. Thus, we have
\[
    \Pr\brk{J = i \wedge Y^{(2)}_i = 0 \midd \cS = S} + \Pr\brk{J \neq i \wedge Y^{(2)}_J = 1 \wedge J' = i \midd \cS = S} = \frac{1 - p_i}{|S|} + \sum_{j \in S - i} \frac{1}{|S|} \cdot \frac{p_j}{|S| - 1}.
\]
Furthermore,
\begin{align*}
    \frac{1 - p_i}{|S|} + \sum_{j \in S - i} \frac{1}{|S|} \cdot \frac{p_j}{|S| - 1} &= \frac{1}{|S|} \prn{\sum_{j \notin S} p_j + \sum_{j \in S - i} p_j} + \sum_{j \in S - i} \frac{1}{|S|} \cdot \frac{p_j}{|S| - 1} \\
    &= \frac{1}{|S|} \sum_{j \notin S} p_j + \sum_{j \in S - i} \frac{p_j}{|S|} \prn{1 + \frac{1}{|S| - 1}} = \frac{1}{|S|} \sum_{j \notin S} p_j + \sum_{j \in S - i} \frac{p_j}{|S|} \frac{|S|}{|S| - 1} \\
    &= \frac{1}{|S|} \sum_{j \notin S} p_j + \frac{1}{{|S| - 1}} \sum_{j \in S - i} p_j,
\end{align*}
where the first equality follows from the fact that $\sum_{j = 1}^n p_j = 1$. Therefore,
\begin{align}
    \Pr\brk{J = i \wedge Y^{(2)}_i = 0 \midd \cS = S} &+ \Pr\brk{J \neq i \wedge Y^{(2)}_J = 1 \wedge J' = i \midd \cS = S} \nonumber \\
    \label{eq:adv-order:two-samples-00} &\quad = \frac{1}{|S|} \sum_{j \notin S} p_j + \frac{1}{{|S| - 1}} \sum_{j \in S - i} p_j.
\end{align}
Combining \eqref{eq:adv-order:two-samples-0} and \eqref{eq:adv-order:two-samples-00}, we get
\[
    \Pr\brk{T = i \midd \cS - i \neq \emptyset} = \frac{1}{2}\sum_{\substack{S \subseteq [n] \\ i \in S, |S| \geq 2}} \Pr\brk{\cS = S \midd \cS - i \neq \emptyset} \prn{ \frac{1}{|S|}\sum_{j \notin S} p_j + \frac{1}{|S|-1}\sum_{j \in S-i} p_j}.
\]
To average this expression, we rewrite the inner summand using only the first samples of elements other than $i$ and $j$. Since $i \in S$, notice that $|S-i-j|+1$ equals $|S|$ if $j \notin S$ and $|S|-1$ if $j \in S-i$. Thus,
\begin{align*}
    \Pr\brk{T = i \midd \cS - i \neq \emptyset} &= \frac{1}{2} \sum_{\substack{S \subseteq [n] \\ i \in S, |S| \geq 2}} \Pr\brk{\cS = S \midd \cS - i \neq \emptyset} \sum_{j \neq i}\frac{p_j}{|S-i-j|+1} \\
    &= \frac{1}{2} \sum_{j \neq i} p_j \E\brk{\frac{1}{|\cS-i-j|+1} \midd \cS - i \neq \emptyset}.
\end{align*}
Combining both cases and averaging over their original distributions gives
\begin{align}
    \Pr\brk{T = i \midd Y_i^{(1)} = 1} &= \prn{1 - \frac{p_i}{2}} \Pr[\cS - i = \emptyset] \nonumber \\
    &\qquad + \frac{1}{2} \Pr[\cS - i \neq \emptyset] \sum_{j \neq i} p_j \E\brk{\frac{1}{|\cS-i-j|+1} \midd \cS - i \neq \emptyset} \nonumber \\
    &= \prn{1 - \frac{p_i}{2}} \Pr[\cS - i = \emptyset] + \frac{1}{2} \sum_{j \neq i} p_j \prn{\E\brk{\frac{1}{|\cS-i-j|+1}} - \Pr[\cS - i = \emptyset]} \nonumber \\
    &= \prn{1 - \frac{p_i}{2} - \frac{1}{2} \sum_{j \neq i} p_j} \Pr[\cS - i = \emptyset] + \frac{1}{2} \sum_{j \neq i} p_j \E\brk{\frac{1}{|\cS-i-j|+1}} \nonumber \\
    \label{eq:adv-order:two-samples-direct}&= \frac{1}{2} \Pr[\cS - i = \emptyset] + \frac{1}{2} \sum_{j \neq i} p_j \E\brk{\frac{1}{|\cS-i-j|+1}},
\end{align}
where the last equality follows from the fact that $\sum_{j = 1}^n p_j = 1$. It remains to show that the sum in \eqref{eq:adv-order:two-samples-direct} equals $\Pr[\cS-i \neq \emptyset]$. For every $j \neq i$, the random variable $|\cS-i-j|$ is independent of $Y_j^{(1)}$, and $\E[Y_j^{(1)}] = p_j$. Therefore,
\[
    \sum_{j \neq i}p_j\E\brk{\frac{1}{|\cS-i-j|+1}} = \sum_{j \neq i}\E\brk{\frac{Y_j^{(1)}}{|\cS-i-j|+1}}  = \E\brk{\sum_{j \neq i}\frac{Y_j^{(1)}}{|\cS-i-j|+1}}  = \Pr[\cS-i \neq \emptyset],
\]
where the last equality follows from the fact that, if $\cS-i = \emptyset$, the sum inside the expectation is zero, otherwise it has exactly $|\cS-i|$ non-zero summands, each equal to $1/|\cS-i|$, so the sum is one.

Substituting into \eqref{eq:adv-order:two-samples-direct}, we obtain
\[
    \Pr\brk{T = i \midd Y_i^{(1)} = 1}
    = \frac{1}{2}\prn{\Pr[\cS-i = \emptyset] + \Pr[\cS-i \neq \emptyset]}
    = \frac{1}{2}.
\]
Consequently, $\Pr[T = i] = p_i/2$ for every $i \in [n]$. Finally, since $T \in \set{0,1,\dots,n}$ and $\sum_{i = 1}^n p_i = 1$, we have
\[
    \Pr[T = 0] = 1 - \sum_{i = 1}^n \Pr[T = i]
    = 1 - \frac{1}{2}\sum_{i = 1}^n p_i
    = \frac{1}{2}. \qedhere
\]
\end{proof}

\subsection{Proof of Lemma~\ref{lem:sample-complexity:rigidity}}
\label{app:missing:rigidity}

\begin{replemma}{lem:sample-complexity:rigidity}
Consider any sequence of $\gamma$-balanced OCRSs and instances with positive activation probabilities summing to $k$, such that $\delta\to0$. Then, $H_2^{(\delta)}$ converges uniformly to $h_2$ on $[0,k]$. In particular, for any choice of prefix lengths $j_r$ along the sequence of instances $p^{(r)}$ such that
\[
    s_{j_r}^{(r)} = \sum_{i=1}^{j_r} p_i^{(r)} \longrightarrow t,
\]
we have
\[
    \E\brk{(N_{j_r}^{(r)}-1)^+}\longrightarrow h_2(t).
\]
\end{replemma}
\begin{proof}
We first analyze a fixed instance. Since $\Pr[A_i]=p_iC_i\geq\gamma p_i$,
\[
    H_{1,j}=\E[N_j]=\sum_{i=1}^j\Pr[A_i]\geq\gamma s_j.
\]
Also, by the definition of $H_{\ell,j}$,
\[
    H_{k,j}=\Pr[N_j=k],\qquad
    H_{\ell-1,j}-H_{\ell,j}=\Pr[N_j\geq\ell-1]
    \quad(2\leq\ell\leq k).
\]
For $2\leq\ell\leq k$, accepting element $j$ increases $(N_j-\ell+1)^+$ by one exactly when $N_{j-1}\geq\ell-1$. Therefore,
\begin{align*}
    H_{\ell,j}-H_{\ell,j-1}
    &=\Pr\brk{A_j\cap\set{N_{j-1}\geq\ell-1}}\\
    &=\Pr[A_j]-\Pr\brk{A_j\cap\set{N_{j-1}\leq\ell-2}}\\
    &\geq \gamma p_j-p_j\Pr[N_{j-1}\leq\ell-2]\\
    &=p_j\prn{\gamma-1+H_{\ell-1,j-1}-H_{\ell,j-1}},
\end{align*}
where the inequality follows from $A_j\subseteq\set{X_j=1}$ and the independence of $N_{j-1}$ and $X_j$. Since the increment is non-negative and requires $X_j=1$, we have
\begin{equation}\label{eq:sample-complexity:finite-increments}
    p_j\prn{\gamma-1+H_{\ell-1,j-1}-H_{\ell,j-1}}^+
    \leq H_{\ell,j}-H_{\ell,j-1}\leq p_j.
\end{equation}
The upper bound also holds for $\ell=1$.

Next, consider a sequence with $\delta\to0$. The functions $H_\ell^{(\delta)}$ are non-decreasing, $1$-Lipschitz, and bounded by $k$. By the Arzel\`a--Ascoli theorem, a subsequence converges uniformly to Lipschitz functions $H_1,\dots,H_k$. These limits satisfy $H_\ell(0)=0$ and $H_1(t)\geq\gamma t$. Moreover, almost everywhere on $[0,k]$,
\begin{equation}\label{eq:sample-complexity:limit-increments}
    H_\ell'(t)\geq\prn{\gamma-1+H_{\ell-1}(t)-H_\ell(t)}^+
    \quad(2\leq\ell\leq k).
\end{equation}
Indeed, on $(s_{j-1},s_j)$, the slope of $H_\ell^{(\delta)}$ is $(H_{\ell,j}-H_{\ell,j-1})/p_j$. Replacing the values at $s_{j-1}$ in \eqref{eq:sample-complexity:finite-increments} by the interpolated values at $t$ changes the lower bound on this slope by at most $2\delta$. Integrating, taking the uniform limit, and differentiating gives \eqref{eq:sample-complexity:limit-increments}.

Next, since $R_n$ is independent of $X_n$,
\[
    \gamma\leq C_n
       \leq\Pr\brk{R_n\midd X_n=1}
       =1-H_{k,n-1}.
\]
Consequently, $H_{k,n}\leq1-\gamma+p_n$, and hence, in the limit as $p_n \to 0$
\begin{equation}\label{eq:sample-complexity:limit-terminal}
    H_k(k)\leq1-\gamma.
\end{equation}
We now show that $H_2=h_2$. First, $H_\ell\geq h_\ell$ for every $\ell$, by induction. For $\ell=1$, this follows from $H_1(t)\geq\gamma t=h_1(t)$. If $H_{\ell-1}\geq h_{\ell-1}$, then wherever $H_\ell<h_\ell$, \eqref{eq:sample-complexity:benchmark} and \eqref{eq:sample-complexity:limit-increments} give $H_\ell'-h_\ell'\geq0$ almost everywhere. Since both functions start at zero, their difference cannot become negative.

Let $D_\ell(t)\coloneqq H_\ell(t)-h_\ell(t)\geq0$. Proposition~\ref{prop:sample-complexity:benchmark} and \eqref{eq:sample-complexity:limit-terminal} give
\[
    0\leq D_k(k)=H_k(k)-h_k(k)\leq(1-\gamma)-(1-\gamma)=0.
\]
To complete the proof of $H_2=h_2$, we show that $D_2(t_0)>0$ at any $t_0<k$ would imply $D_k(k)>0$, thereby arriving at a contradiction.

Since $D_{\ell-1}\geq0$, monotonicity and the $1$-Lipschitz property of the positive-part function in \eqref{eq:sample-complexity:benchmark} and \eqref{eq:sample-complexity:limit-increments} give $D_\ell'\geq-D_\ell$ almost everywhere for $2\leq\ell\leq k$. Multiplying by $e^t$ and integrating yields
\begin{equation}\label{eq:sample-complexity:slack-persists}
    D_\ell(t)\geq e^{-(t-u)}D_\ell(u)\qquad(0\leq u\leq t\leq k).
\end{equation}
Also, $\tilde y_k(k)>0$, so \eqref{eq:sample-complexity:tail-benchmark} implies that $\tilde y_{k-1}(t)>1-\gamma$ at some $t<k$. Since the functions $\tilde y_\ell$ are non-decreasing and ordered, $\tilde y_{\ell-1}(t)>1-\gamma$ for every $2\leq\ell\leq k$ and every $t$ sufficiently close to $k$. On this interval, the positive part in the equation for $h_\ell'$ can be removed, and hence
\begin{equation}\label{eq:sample-complexity:slack-propagates}
    D_\ell'(t)+D_\ell(t)\geq D_{\ell-1}(t)
    \quad(2\leq\ell\leq k)
\end{equation}
holds almost everywhere.

Suppose that $D_2(t_0)>0$ for some $t_0<k$. Choose $t_0<s<k$ so that \eqref{eq:sample-complexity:slack-propagates} holds on $[s,k]$. By \eqref{eq:sample-complexity:slack-persists}, $D_2(s)>0$. Multiplying \eqref{eq:sample-complexity:slack-propagates} by $e^t$ and integrating successively gives
\[
    D_\ell(t)\geq D_2(s)e^{-(t-s)}
       \frac{(t-s)^{\ell-2}}{(\ell-2)!}
    \qquad(s\leq t\leq k,\ 2\leq\ell\leq k),
\]
where the case $\ell=2$ follows from \eqref{eq:sample-complexity:slack-persists} and each subsequent case uses $D_\ell(s)\geq0$. For $\ell=t = k$, we obtain
\[
    D_k(k) \geq D_2(s) e^{-(k-s)} \frac{(k-s)^{k-2}}{(k-2)!} > 0,
\]
since $D_2(s) > 0$ and $s < k$. This contradicts $D_k(k)=0$, and thus $H_2=h_2$ on $[0,k]$, including $k$ by continuity.

Every uniformly convergent subsequence therefore has second component $h_2$. By compactness, we get uniform convergence of the full sequence $H_2^{(\delta)}$ to $h_2$. The fact that for any choice of prefix lengths $j_r$ along the sequence of instances $p^{(r)}$ such that $s_{j_r}^{(r)} = \sum_{i=1}^{j_r} p_i^{(r)} \longrightarrow t$ we have $\E\brk{(N_{j_r}^{(r)}-1)^+}\longrightarrow h_2(t)$ follows from uniform convergence and the continuity of $h_2$.
\end{proof}
\section{An Alternative Optimal ROCRS}
\label{app:alternative-proof}

In this section we give an alternative ROCRS with the same optimal balancedness as in Theorem~\ref{thm:cardinality:random-order}. Before that, we start with some definitions.

\begin{definition}[Poisson binomial distribution]
The \emph{Poisson binomial distribution} $\PB{p_1,\dots,p_n}$ is the distribution of a sum of independent Bernoulli random variables with success probabilities $p_1, \dots, p_n$. We also write such a sum as $\sum_{j = 1}^n \Ber{p_j}$, and use the same convention for sums of independent Poisson random variables.
\end{definition}

\begin{definition}[Discrete differences and convexity]
For a function $f : \N \to \R$, let
\[
    \Delta f(n) \coloneqq f(n+1)-f(n),
    \qquad \text{and} \qquad
    \Delta^2 f(n) \coloneqq \Delta f(n+1)-\Delta f(n).
\]
The function $f$ is \emph{convex} if $\Delta^2 f(n)\geq 0$ for every $n\in\N$, and \emph{concave} if $\Delta^2 f(n)\leq 0$ for every $n\in\N$.
\end{definition}

\begin{definition}[Convex order]
For random variables $X,Y$ supported on $\N$ with finite means, we say that $X$ \emph{dominates $Y$ in convex order}, and write $X\geq_{cx}Y$, if
\[
    \E[f(X)]\geq\E[f(Y)]
\]
for every convex function $f:\N\to\R$ for which both expectations are finite.
\end{definition}

For our alternative proof of Theorem~\ref{thm:cardinality:random-order}, we need the following characterization of stochastic convex order.

\subsection{Characterization of Convex Order}
\label{app:cardinality:hoeffding}

In this section, we generalize the following consequence of Hoeffding's theorem for Poisson binomial random variables \cite{hoeffding-poisson-binomial}.

\begin{theorem}[Hoeffding, \cite{hoeffding-poisson-binomial}]\label{thm:hoeffding-original-convex-order}
Let $X \sim \Poi{\lambda}$ and $Y \sim \PB{p_1, \dots, p_n}$, where $\lambda = \sum_{i = 1}^n p_i$. Then, for every convex $f : \N \to \R$ we have
\[
    \E[f(X)] \geq \E[f(Y)],
\]
and the inequality is reversed if $f$ is concave.
\end{theorem}

We generalize the above theorem in three ways:
\begin{enumerate}
    \item We allow $X$ and $Y$ to have different means.
    \item We allow $X$ and $Y$ to follow arbitrary distributions on $\N$.
    \item We allow for an arbitrary $f$ that is neither convex nor concave, provided it satisfies a mild second-order condition.
\end{enumerate}
Under the above assumptions, we characterize the exact conditions under which: (i) $\E[f(X)] \geq \E[f(Y)]$ and (ii) $X \geq_{cx} Y$, for all such $X,Y$ and $f$. In what follows, for two random variables $X,Y$ with finite mean, let
\[
    \delta \coloneqq \E[X] - \E[Y] \qquad \text{and} \qquad D_n \coloneqq \E[(X - n - 1)^+] - \E[(Y - n - 1)^+], \quad n \in \N.
\]

\begin{theorem}\label{thm:function-difference-characterization}
Let $X, Y$ be random variables with finite mean, supported on $\N$, $\delta \coloneqq \E[X] - \E[Y]$ and $D_n \coloneqq \E[(X - n - 1)^+] - \E[(Y - n - 1)^+]$ for all $n \in \N$. Then, for all functions $f : \N \to \R$ with
\[
    \sum_{n \geq 0} \prn{\E[(X - n - 1)^+] + \E[(Y - n - 1)^+]} \abs{\Delta^2 f(n)} < \infty,
\]
we have
\[
    \E[f(X)] - \E[f(Y)] = \delta \Delta f(0) + \sum_{n \geq 0} D_n \Delta^2 f(n).
\]
\end{theorem}
\begin{proof}
Note that, for all $m \geq 0$,
\begin{equation}\label{eq:hoeffding-proof-1}
    f(m) - f(0) = \sum_{i = 0}^{m-1} \prn{f(i+1) - f(i)} = \sum_{i = 0}^{m-1} \Delta f(i).
\end{equation}
Since $\Delta^2 f(i) = \Delta f(i+1) - \Delta f(i)$, we have
\[
    \Delta f(i) - \Delta f(0) = \sum_{n = 0}^{i-1} \prn{\Delta f(n+1) - \Delta f(n)} = \sum_{n = 0}^{i-1} \Delta^2 f(n),
\]
and thus
\[
    \sum_{i = 0}^{m-1} \Delta f(i) = m \Delta f(0) + \sum_{i = 0}^{m-1} \sum_{n = 0}^{i-1} \Delta^2 f(n).
\]
Substituting the above expression into \eqref{eq:hoeffding-proof-1} gives
\begin{align*}
    f(m) &= f(0) + m \Delta f(0) + \sum_{i = 0}^{m-1} \sum_{n = 0}^{i-1} \Delta^2 f(n).
\end{align*}
For a fixed $n \leq m-2$, the term $\Delta^2f(n)$ occurs in the double sum exactly $m-n-1$ times, once for each $i = n+1, n+2, \dots, m-1$. Thus, for all $m \in \N$,
\begin{equation}\label{eq:discrete-taylor}
    f(m) = f(0)+ m \Delta f(0) + \sum_{n \geq 0} (m-n-1)^+ \Delta^2 f(n).
\end{equation}
Therefore,
\begin{equation}\label{eq:hoeffding-proof-2}
    \E[f(X)] = f(0) + \E[X] \Delta f(0) + \E\brk{\sum_{n \geq 0} (X - n-1)^+ \Delta^2 f(n)}
\end{equation}
and
\begin{equation}\label{eq:hoeffding-proof-3}
    \E[f(Y)] = f(0) + \E[Y] \Delta f(0) + \E\brk{\sum_{n \geq 0} (Y - n-1)^+ \Delta^2 f(n)}.
\end{equation}
Subtracting \eqref{eq:hoeffding-proof-3} from \eqref{eq:hoeffding-proof-2} yields
\begin{align*}
    \E[f(X)] - \E[f(Y)] &= \delta \Delta f(0) + \E\brk{\sum_{n \geq 0} \prn{(X - n-1)^+ - (Y - n-1)^+} \Delta^2 f(n)} \\
    &= \delta \Delta f(0) + \sum_{n \geq 0} \E\brk{(X - n-1)^+ - (Y - n-1)^+} \Delta^2 f(n) \\
    &= \delta \Delta f(0) + \sum_{n \geq 0} D_n \Delta^2 f(n),
\end{align*}
where the second equality follows from the assumption that
\[
    \sum_{n \geq 0} \prn{\E[(X - n - 1)^+] + \E[(Y - n - 1)^+]} \abs{\Delta^2 f(n)} < \infty. \qedhere
\]
\end{proof}

Theorem~\ref{thm:function-difference-characterization} gives a simple characterization of the random variables and functions for which $\E[f(X)] \geq \E[f(Y)]$, under a mild assumption on $f$. It also yields the following characterization of convex order domination.

\begin{theorem}\label{thm:convex-order-characterization}
Let $X, Y$ be random variables with finite means, supported on $\N$. The following are equivalent:
\begin{enumerate}
    \item $X \geq_{cx} Y$.
    \item $\delta = 0$ and $D_n \geq 0$ for all $n \in \N$.
\end{enumerate}
\end{theorem}
\begin{proof}
First, assume that $X \geq_{cx} Y$. The functions $f(Z) = Z$ and $g(Z) = -Z$ are both convex, and together imply that
\[
    \begin{cases}
        \E[X] \geq \E[Y] \\
        \E[X] \leq \E[Y]
    \end{cases} \implies \E[X] = \E[Y] \iff \delta = 0.
\]
Furthermore, the function $h_n(Z) = (Z - n - 1)^+$ is also convex for all $n \in \N$, and thus $\E[h_n(X)] \geq \E[h_n(Y)]$, which implies $D_n \geq 0$.

Next, assume that $\delta = 0$ and $D_n \geq 0$ for all $n \in \N$. Let $f : \N \to \R$ be convex. As in the proof of Theorem~\ref{thm:function-difference-characterization}, we have
\[
    f(n) - f(0) - n \Delta f(0) = \sum_{m \geq 0} (n - m - 1)^+ \Delta^2 f(m),
\]
and thus
\begin{align*}
    \E[f(Y)] - f(0) - \E[Y] \Delta f(0) &= \sum_{m \geq 0} \E[(Y - m - 1)^+] \Delta^2 f(m) \\
    &\leq \sum_{m \geq 0} \E[(X - m - 1)^+] \Delta^2 f(m) \\
    &= \E[f(X)] - f(0) - \E[X] \Delta f(0),
\end{align*}
where the inequality follows from $D_n \geq 0$. Since $\delta = 0$, we have $\E[X] = \E[Y]$, and thus $\E[f(Y)] \leq \E[f(X)]$.
\end{proof}

Therefore, Theorem~\ref{thm:convex-order-characterization} recovers Theorem~\ref{thm:hoeffding-original-convex-order} as a corollary: since $\lambda = \sum_{i = 1}^n p_i$, we have $\E[X] = \E[Y] \implies \delta = 0$. Fix an integer $r \geq 0$, and let
\[
    V_i = \sum_{j = 1}^i \Poi{p_j} + \sum_{j = i+1}^n \Ber{p_j}.
\]
Thus, $V_n$ has the same distribution as $X$ by the additivity of Poisson random variables, and $V_0$ has the same distribution as $Y$. For $i = 1, \dots, n$, let $W_i = \sum_{j = 1}^{i-1} \Poi{p_j} + \sum_{j = i+1}^n \Ber{p_j}$, and notice that
\[
    \E[(V_i - r - 1)^+] - \E[(V_{i-1} - r - 1)^+] = \E\brk{\prn{\Poi{p_i} + W_i - r - 1}^+} - \E\brk{\prn{\Ber{p_i} + W_i - r - 1}^+}.
\]
Conditioning on $W_i = z$ for some $z \geq 0$, the expression above becomes
\[
    \sum_{z \geq 0} \Pr[W_i = z] \prn{\E\brk{\prn{\Poi{p_i} + z - r - 1}^+ - \prn{\Ber{p_i} + z - r - 1}^+}}.
\]
For $z \leq r$, since $\Ber{p_i} \leq 1$, we have
\[
    \E\brk{\prn{\Ber{p_i} + z - r - 1}^+} = 0 \leq \E\brk{\prn{\Poi{p_i} + z - r - 1}^+}.
\]
For $z > r$, both expressions are non-negative, and
\[
    \E\brk{\prn{\Poi{p_i} + z - r - 1} - \prn{\Ber{p_i} + z - r - 1}} = \E[\Poi{p_i}] - \E[\Ber{p_i}] = 0,
\]
since the two random variables have the same mean. Thus,
\[
    D_r = \E[(X - r - 1)^+] - \E[(Y - r - 1)^+] = \E[(V_n - r - 1)^+] - \E[(V_0 - r - 1)^+] \geq 0.
\]
Therefore, applying Theorem~\ref{thm:convex-order-characterization} to $X, Y$ yields Theorem~\ref{thm:hoeffding-original-convex-order}.

\subsection{The Main Proof}
\label{app:cardinality:random-order}

We use the notation of Section~\ref{sec:preliminaries} and assume $\sum_{i=1}^n p_i = k$. Recall the random-order acceptance-rate function from Section~\ref{sec:cardinality:random-order} (see also \cite{willma-random}): for $t \geq 0$,
\[
  \AR_k(t) = \min\set{1, \frac{k}{t+1}}.
\]
Also, let
\[
  \gamma_k = 1 - e^{-k}\frac{k^k}{k!},
  \qquad
  \delta_k = \E\brk{\AR_k(\Poi{k-1})},
  \qquad
  \alpha_k = \frac{\gamma_k}{\delta_k},
  \qquad
  \beta_k = 1 - \alpha_k.
\]

Consider the following generalization of Algorithm~\ref{alg:random-order:single-item} for general $k \in \N$. When we observe an element $i$ that is both online-active and sample-active and we still have capacity to accept it, we accept it with probability $\alpha_k$. Let $\sigma$ be the uniformly random arrival permutation, so that $\sigma(t)$ is the element arriving in position $t$. The set $\cA$ contains the accepted elements, and $S_t$ is the number of elements accepted after the first $t$ arrivals.

\begin{algorithm}[ht]
  \caption{The Conservative Scaling algorithm for general $k$.}
  \label{alg:cardinality:random-order-2}
  \begin{algorithmic}[1]
    \Require $\set{Y_1, \dots, Y_n}$
    \Ensure $\cA$
    \State $\cA \from \emptyset, S_0 \from 0$
    \For{arrival positions $t = 1, \dots, n$}
        \State $i \from \sigma(t)$
        \If{$S_{t-1} < k, X_i = 1$ and $Y_i = 0$}
            \State $q_i \from 1$
        \ElsIf{$S_{t-1} < k, X_i = 1$ and $Y_i = 1$}
            \State $q_i \from \alpha_k$
        \Else
            \State $q_i \from 0$
        \EndIf
        \State $Q_i \from \Ber{q_i}$
        \If{$Q_i = 1$}
            \State Accept $i$
            \State $\cA \from \cA + i$
            \State $S_t \from S_{t-1} + 1$
        \Else
            \State Reject $i$
            \State $S_t \from S_{t-1}$
        \EndIf
    \EndFor
  \end{algorithmic}
\end{algorithm}

For $k = 1$, we have $\delta_1 = 1$ and thus
\[
  \alpha_1 = \gamma_1 = 1 - \frac{1}{e},
\]
so the algorithm agrees with Algorithm~\ref{alg:random-order:single-item} for the $k = 1$ case.

For each element $i$, let $\rho_i \coloneqq p_i (1 - \beta_k p_i)$, $\xi_i \sim \Ber{\alpha_k}$, and
\[
    Z_i \coloneqq X_i \prn{(1 - Y_i) + Y_i \xi_i},
\]
where the $\xi_i$'s are independent across elements and independent of the samples, online activations, and arrival order. We call an element $i$ with $Z_i = 1$ a \emph{candidate}. Thus, an element is a candidate if it is online-active and either sample-inactive or sample-active with a successful coin $\xi_i$. All $Z_i$'s are independent, and
\[
    \Pr[Z_i = 1] = p_i \prn{(1 - p_i) + p_i \alpha_k} = p_i (1 - \beta_k p_i) = \rho_i.
\]
The parameter $q_i$ in Algorithm~\ref{alg:cardinality:random-order-2} is the coin bias chosen after observing $X_i$, $Y_i$, and the available capacity; $\rho_i$ is the unconditional probability that $i$ is a candidate. Recall that $R_i$ is the event that the algorithm reaches $i$. If $i = \sigma(t)$, then $R_i = \set{S_{t-1} < k}$. Coupling $\xi_i$ with the algorithm's coin whenever $i$ is online-active and sample-active gives $Q_i = \1_{R_i} Z_i$. Hence $A_i = R_i \cap \set{Z_i=1}$, or equivalently, the algorithm scans the random order and accepts the first $k$ candidates.

Fix an element $i$ with $p_i>0$. Its conditional probability of being a candidate is
\[
    \Pr\brk{Z_i = 1 \midd X_i = 1} = (1 - p_i) + p_i \alpha_k = 1 - \beta_k p_i.
\]
Also, let $G_i = \sum_{j \neq i} Z_j$ denote the (random) number of elements that $i$ has to compete with for a spot, and notice that $G_i$ is a sum of independent Bernoulli variables, each with parameter $\rho_j$, and all independent of $X_i$ and $Z_i$. In other words, $G_i \sim \PB{\rho_1, \dots, \rho_{i-1}, \rho_{i+1}, \dots, \rho_n}$. If $Z_i = 1$ and $G_i = t$, then there are exactly $t + 1$ candidate elements including $i$. Since the arrival order is uniformly random, $i$ is uniformly ranked among those $t + 1$ candidates. The event $A_i$ occurs exactly when $i$'s rank among candidates is at most $k$. Thus,
\[
    \Pr\brk{A_i \midd X_i = 1, Z_i = 1, G_i = t} = \min\set{1, \frac{k}{t+1}} = \AR_k(t).
\]
Therefore, its conditional acceptance probability is
\begin{equation}\label{eq:1}
    C_i = (1 - \beta_k p_i) \cdot \E[\AR_k(G_i)].
\end{equation}

To show that this is at least $\gamma_k$ for all $k \geq 1$, we will make use of two lemmas. The first is a corollary of Theorem~\ref{thm:function-difference-characterization} and its proof can be found in Section~\ref{sec:cardinality:poisson-binomial}.

\begin{lemma}\label{lem:poisson-binomial-difference}
For every $i \in [n]$, we have $\E[\AR_k(G_i)] \geq \E[\AR_k(\Poi{k - p_i})]$.
\end{lemma}

Notice that we cannot obtain the inequality in Lemma~\ref{lem:poisson-binomial-difference} from Theorem~\ref{thm:hoeffding-original-convex-order} in a black-box way, since $\AR_k(\cdot)$ is neither convex nor concave. Therefore, we develop a slightly more involved argument to prove the lemma.

The second lemma states that the reciprocal of $\E[\AR_k(\Poi{k - u})]$ is convex for $u \in [0,1]$ and its proof can be found in Section~\ref{sec:cardinality:convexity}.

\begin{lemma}\label{lem:convexity}
For $k \geq 2$, let
\[
	F_k(\lambda) = \frac{1}{\E\brk{\AR_k(\Poi{\lambda})}}.
\]
Then, $F_k(\lambda)$ is convex in $[k-1, k]$ or, in other words, $F_k(k - u)$ is convex for $u \in [0,1]$.
\end{lemma}

We are finally ready to show that the Conservative Scaling algorithm is $\gamma_k$-balanced for every $k \geq 1$.

\begin{theorem}\label{thm:balancedness}
The Conservative Scaling algorithm is $\gamma_k$-balanced for all $k \geq 1$.
\end{theorem}
\begin{proof}
The case $k = 1$ follows from Section~\ref{sec:single-item:random-order}. Let $k \geq 2$. Fix an element $i$ with $p_i>0$. Combining Lemma~\ref{lem:poisson-binomial-difference} with \eqref{eq:1}, we get
\begin{equation}\label{eq:2}
    C_i \geq (1 - \beta_k p_i) \E[\AR_k(\Poi{k - p_i})].
\end{equation}
Note that $1 - \beta_k p_i$ is the linear interpolation between $1$ at $p_i = 0$ and $\alpha_k$ at $p_i = 1$. Since, by Lemma~\ref{lem:convexity}, the function $F_k(\lambda) = 1/\E[\AR_k(\Poi{\lambda})]$ is convex in $[k-1, k]$, we have
\[
    F_k(k - p_i) \leq (1 - p_i) F_k(k) + p_i F_k(k-1).
\]
Using the fact that $F_k(k) = 1/\gamma_k$ and $F_k(k-1) = 1/\delta_k = \alpha_k/\gamma_k$, we have
\[
    F_k(k-p_i) \leq \frac{1 - p_i}{\gamma_k} + \frac{\alpha_k p_i}{\gamma_k} = \frac{1 - \beta_k p_i}{\gamma_k},
\]
which implies that
\[
    (1 - \beta_k p_i) \E[\AR_k(\Poi{k - p_i})] \geq \gamma_k. \qedhere
\]
\end{proof}

\subsubsection{The Poisson Binomial Lemma}
\label{sec:cardinality:poisson-binomial}

\begin{replemma}{lem:poisson-binomial-difference}
For every $i \in [n]$, we have $\E[\AR_k(G_i)] \geq \E[\AR_k(\Poi{k - p_i})]$.
\end{replemma}
\begin{proof}
The case $k = 1$ follows from the comparison in Section~\ref{sec:single-item:random-order}. Let $k \geq 2$. Recall that $G_i \sim \PB{\rho_1, \dots, \rho_{i-1}, \rho_{i+1}, \dots, \rho_n}$. Let $P_i \sim \Poi{k - p_i}$, $D_{s,i} \coloneqq \E[(G_i - s - 1)^+] - \E[(P_i - s - 1)^+]$, $\mu = \sum_{j \neq i} \rho_j$ and $\delta = \mu - (k - p_i)$. We claim that, for all $i \in [n]$,
\[
    \sum_{s \geq 0} \prn{\E[(G_i - s - 1)^+] + \E[(P_i - s - 1)^+]} \abs{\Delta^2 \AR_k(s)} < \infty.
\]
To see this, fix $i \in [n]$. For all $s \geq 0$, both $\E[(G_i - s - 1)^+]$ and $\E[(P_i - s - 1)^+]$ lie in $[0, k - p_i]$, and thus $\E[(G_i - s - 1)^+] + \E[(P_i - s - 1)^+] \leq 2(k - p_i)$. Since $\AR_k(s) \in [0,1]$ and is non-increasing in $s$, we have
\begin{align*}
    |\Delta^2 \AR_k(s)| &= \abs{(\AR_k(s+2) - \AR_k(s+1)) - (\AR_k(s+1) - \AR_k(s))} \\
    &= \abs{(\AR_k(s) - \AR_k(s+1)) - (\AR_k(s+1) - \AR_k(s+2))} \\
    &\leq (\AR_k(s) - \AR_k(s+1)) + (\AR_k(s+1) - \AR_k(s+2)) \\
    &= \AR_k(s) - \AR_k(s+2).
\end{align*}
Thus, for every $M \geq 0$,
\begin{align*}
    \sum_{s = 0}^M &\prn{\E[(G_i - s - 1)^+] + \E[(P_i - s - 1)^+]} \abs{\Delta^2 \AR_k(s)} \\
    &\qquad \qquad \leq 2 (k - p_i) \sum_{s = 0}^M (\AR_k(s) - \AR_k(s+2)) \\
    &\qquad \qquad = 2 (k - p_i) \prn{\AR_k(0) + \AR_k(1) - \AR_k(M+1) - \AR_k(M+2)} \\
    &\qquad \qquad \leq 4 (k - p_i).
\end{align*}
Letting $M \to \infty$, yields
\[
    \sum_{s \geq 0} \prn{\E[(G_i - s - 1)^+] + \E[(P_i - s - 1)^+]} \abs{\Delta^2 \AR_k(s)} \leq 4 (k - p_i) < \infty.
\]
Thus, by Theorem~\ref{thm:function-difference-characterization}, we have
\[
    \E[\AR_k(G_i)] - \E[\AR_k(P_i)] = \delta \Delta \AR_k(0) + \sum_{s \geq 0} D_{s,i} \Delta^2 \AR_k(s).
\]
Since $k \geq 2$, we have
\[
    \Delta \AR_k(0) = \AR_k(1) - \AR_k(0) = \min\set{1, \frac{k}{2}} - \min\set{1, k} = 1 - 1 = 0.
\]
Thus it suffices to show that
\begin{equation}\label{eq:curvature-non-neg}
    \sum_{s \geq 0} D_{s,i} \Delta^2 \AR_k(s) \geq 0.
\end{equation}
Let
\[
    J(X) \coloneqq \sum_{s \geq 0} \prn{\E[(X - s - 1)^+] - \E[(P_i - s - 1)^+]} \Delta^2 \AR_k(s).
\]
Since $\Delta\AR_k(0) = 0$, the preceding identity gives
\[
    J(X) = \E[\AR_k(X)] - \E[\AR_k(P_i)].
\]
Thus, it suffices to show that $J(G_i) \geq 0$. Start with the original underlying masses $p_j$, for $j \neq i$, and let $H_0 = G_i$. At each step choose a largest current underlying mass $x$,
replace it by two masses $x/2$, and assign to every current mass $u$ the Bernoulli success probability $u(1 - \beta_k u)$. Let $H_\ell$ be the resulting sum after $\ell$ steps.
The sum of the underlying masses remains $k - p_i$ even though the sum of the Bernoulli probabilities need not remain constant.

Observe that the largest underlying mass tends to zero. Moreover,
\[
    \E[H_\ell] = (k - p_i) - \beta_k \sum_j p_j^2 \longrightarrow k - p_i,
    \qquad \text{and} \qquad
    \max_j p_j(1 - \beta_k p_j) \longrightarrow 0,
\]
where the sums and maximum are over the current masses. By the law of small numbers we get $H_\ell \Rightarrow P_i$. Since $\AR_k$ is bounded,
\[
    J(H_\ell) = \E[\AR_k(H_\ell)] - \E[\AR_k(P_i)] \longrightarrow 0.
\]
It therefore suffices to prove $J(H_\ell) \geq J(H_{\ell + 1})$.

Let $E^{(\ell)}_{-i}$ index the current masses, where $E^{(0)}_{-i} = [n] - i$. Choose $j_x \in \argmax_{j \in E^{(\ell)}_{-i}} p_j$, set $x = p_{j_x}$ and let $T$ be the sum of the independent Bernoulli variables with parameters $\rho_j = p_j (1 - \beta_k p_j)$ for $j \in E^{(\ell)}_{-i} - j_x$.

Let
\[
    q = x (1 - \beta_k x), \qquad h = \frac{x}{2} \prn{1 - \beta_k \frac{x}{2}},
\]
and note that
\[
    H_\ell = T + \Ber{q}, \qquad H_{\ell+1} = T + \Ber{h} + \Ber{h}.
\]
For $s \in \N$, when $T \leq s-1$, we have $H_{\ell}, H_{\ell+1} \leq s+1$, and thus
\[
    (H_\ell - s - 1)^+ = (H_{\ell+1} - s - 1)^+ = 0.
\]
Furthermore, when $T = s$, we have $H_\ell \leq s+1$, and thus
\[
    (H_\ell - s - 1)^+ - (H_{\ell+1} - s - 1)^+ = - (\Ber{h} + \Ber{h} - 1)^+.
\]
Finally, when $T \geq s+1$, we have
\[
    (H_\ell - s - 1)^+ - (H_{\ell+1} - s - 1)^+ = (H_\ell - s - 1) - (H_{\ell+1} - s - 1) = \Ber{q} - \Ber{h} - \Ber{h}.
\]
Putting everything together, we obtain
\begin{equation}\label{eq:pb-3}
    \E[(H_\ell - s - 1)^+] - \E[(H_{\ell+1} - s - 1)^+] = (q - 2h) \Pr[T \geq s+1] - h^2 \Pr[T = s].
\end{equation}
Note that
\[
    q - 2h = x(1 - \beta_k x) - 2 \frac{x}{2} \prn{1 - \beta_k \frac{x}{2}} = - \beta_k x^2 + \beta_k \frac{x^2}{2} = - \beta_k \frac{x^2}{2},
\]
and thus \eqref{eq:pb-3} becomes
\[
    \E[(H_\ell - s - 1)^+] - \E[(H_{\ell+1} - s - 1)^+] = -\prn{\beta_k \frac{x^2}{2} \Pr[T \geq s+1] + h^2 \Pr[T = s]}. 
\]
Using the above equality, we get that
\begin{align}
    J(H_\ell) - J(H_{\ell+1}) &= \sum_{s \geq 0} \prn{\E[(H_\ell - s - 1)^+] - \E[(H_{\ell+1} - s - 1)^+]} \Delta^2 \AR_k(s) \nonumber \\
    &= - \sum_{s \geq 0} \prn{\beta_k \frac{x^2}{2} \Pr[T \geq s+1] + h^2 \Pr[T = s]} \Delta^2 \AR_k(s) \nonumber \\
    \label{eq:pb-4} &= \beta_k \frac{x^2}{2} \E\brk{- \Delta \AR_k(T)} + h^2 \E\brk{- \Delta^2 \AR_k(T)},
\end{align}
where the last equality follows from the fact that
\[
    \sum_{s \geq 0} \Pr[T \geq s+1] \Delta^2 \AR_k(s) = \E\brk{\sum_{s = 0}^{T-1} \Delta^2 \AR_k(s)} = \E\brk{\Delta \AR_k(T) - \Delta \AR_k(0)} = \E[\Delta \AR_k(T)].
\]
Next, we show that
\begin{equation}\label{eq:pb-5}
    \E[- \Delta^2 \AR_k(T)] \geq 2 \beta_k \E[\Delta \AR_k(T)].
\end{equation}
Substituting this into \eqref{eq:pb-4} yields
\[
    J(H_\ell) - J(H_{\ell+1}) \geq \beta_k \prn{\frac{x^2}{2} - 2 h^2} \E[- \Delta \AR_k(T)] \geq 0,
\]
where the inequality follows from the facts that $2h^2 \leq x^2 / 2$ and that $\E[- \Delta \AR_k(T)] \geq 0$, since $\AR_k$ is non-increasing. Thus, to show $J(H_\ell) \geq J(H_{\ell+1})$, it suffices to show \eqref{eq:pb-5}, which is equivalent to
\begin{equation}\label{eq:pb-6}
    \E[\Delta^2 \AR_k(T)] + 2 \beta_k \E[\Delta \AR_k(T)] \leq 0.
\end{equation}
To show \eqref{eq:pb-6}, we will define a new random variable $Z$ that point-wise dominates $\Delta^2 \AR_k(T) + 2 \beta_k \Delta \AR_k(T)$ and that satisfies $\E[Z] = 0$. Note that, for all $s \in \N$
\[
    \Delta^2 \AR_k(s) + 2 \beta_k \Delta \AR_k(s) =
    \begin{cases}
        0 & \text{for } s \leq k - 3 \\
        -\frac{1}{k+1} & \text{for } s = k - 2 \\
        \frac{2k}{(s+1)(s+2)} \prn{\frac{1}{s+3} - \beta_k} & \text{for } s \geq k - 1 \\
    \end{cases}.
\]
Let
\[
    g(s, t) \coloneqq
    \begin{cases}
        0 & \text{for } s < k - 1 \\
        \frac{k}{(s+1)(s+2)(2s - k + 2 - t)} & \text{for } s \geq k - 1
    \end{cases},
\]
where $s$ corresponds to the number of successes for a realization of $T$ and $t$ to the sum of the $p_j$'s for those successes. Also let $F = E^{(\ell)}_{-i} - j_x$. For $j \in F$, write $B_j \sim \Ber{\rho_j}$ for the independent draws defining $T$, so that $T=\sum_{j \in F} B_j$. In the following computation of $Z$, all element indices range over $F$, and zero-mass terms are omitted. We define $Z$ as
\[
    Z \coloneqq \sum_{j \in F} \frac{p_j}{\rho_j} \prn{B_j - \rho_j} \cdot g\prn{1 + \sum_{u \in F-j} B_u, \; p_j + \sum_{u \in F-j} p_u B_u}.
\]
Notice that the arguments of $g$ describe the outcome obtained by forcing the $j$-th element to succeed, which then depends only on the other elements. Therefore, if we condition on $(B_u)_{u \in F-j}$, $g\prn{1 + \sum_{u \in F-j} B_u, \; p_j + \sum_{u \in F-j} p_u B_u}$ becomes a constant. In addition,
\[
    \E\brk{B_j - \rho_j \midd (B_u)_{u \in F-j}} = \rho_j - \rho_j = 0,
\]
and thus every term in the sum defining $Z$ has expectation equal to zero, which implies that $\E[Z] = 0$.

Fix a realization $s$ of $T$ and let $t = \sum_{j : B_j = 1} p_j$. Notice that, if $B_j = 1$, the arguments of $g$ for that $j$ are $(s,t)$. However, if $B_j = 0$, the arguments of $g$ for that $j$ are $(s+1, t+p_j)$, since the arguments of $g$ for the $j$-th term in the sum defining $Z$ are forcing the $j$-th Bernoulli to be a success. Thus, for the fixed realization, $Z$ becomes
\[
    Z = \sum_{j : B_j = 1} \frac{p_j (1 - \rho_j)}{\rho_j} \cdot g\prn{s, t} - \sum_{j : B_j = 0} p_j \cdot g\prn{s + 1, t + p_j}.
\]
If $s < k - 2$, then $Z = 0$ by the definition of $g$. For $s = k - 2$, we have
\begin{align*}
    Z &= - \sum_{j : B_j = 0} p_j \cdot g\prn{k - 1, t + p_j} = - \sum_{j : B_j = 0} \frac{p_j}{(k+1)(2(k - 1) - k + 2 - t - p_j)} \\
    &= - \frac{1}{k+1} \sum_{j : B_j = 0} \frac{p_j}{k - t - p_j} \geq - \frac{1}{k+1} \sum_{j : B_j = 0} \frac{p_j}{k - t - x} \\
    &\geq - \frac{1}{k+1},
\end{align*}
where the last inequality follows from the fact that $t + \sum_{j : B_j = 0} p_j \leq k - x$. Finally, for $s \geq k - 1$, we have
\begin{align*}
    Z &= \frac{k}{(s+1)(s+2)} \prn{\sum_{j : B_j = 1} \frac{\frac{p_j (1 - \rho_j)}{\rho_j}}{2s - k + 2 - t} - \sum_{j : B_j = 0} \frac{p_j (s+1)}{(s+3)(2s - k + 4 - t - p_j)}} \\
    &\geq \frac{k}{(s+1)(s+2)} \prn{\frac{1}{2s - k + 2 - t} \sum_{j : B_j = 1} \frac{p_j (1 - \rho_j)}{\rho_j} - \frac{(k - x - t)(s+1)}{(s+3)(2s - k + 4 - t - x)}} \\
    &\geq \frac{k}{(s+1)(s+2)} \prn{\frac{s - (1 - \beta_k) t}{2s - k + 2 - t} - \frac{(k - x - t)(s+1)}{(s+3)(2s - k + 4 - t - x)}},
\end{align*}
where the first inequality follows from the facts that $p_j \leq x$ for all $j$ and $\sum_{j : B_j = 0} p_j \leq k - x - t$, and the second inequality follows from the fact that
\[
    \frac{p_j (1 - \rho_j)}{\rho_j} = \frac{1 - p_j(1 - \beta_k p_j)}{1 - \beta_k p_j} = \frac{1}{1 - \beta_k p_j} - p_j \geq 1 - (1 - \beta_k) p_j,
\]
as well as the definitions of $s = \sum_{j : B_j = 1} 1$ and $t = \sum_{j : B_j = 1} p_j$. For $Z$ to dominate $\Delta^2 \AR_k(s) + 2 \beta_k \Delta \AR_k(s)$ for $s \geq k-1$, it suffices to show that
\[
    \frac{k}{(s+1)(s+2)} \prn{\frac{s - (1 - \beta_k) t}{2s - k + 2 - t} - \frac{(k - x - t)(s+1)}{(s+3)(2s - k + 4 - t - x)}} \geq \frac{2k}{(s+1)(s+2)} \prn{\frac{1}{s+3} - \beta_k},
\]
or, equivalently, that
\begin{equation}\label{eq:pb-7}
    \frac{s - (1 - \beta_k) t}{2s - k + 2 - t} - \frac{(k - x - t)(s+1)}{(s+3)(2s - k + 4 - t - x)} \geq \frac{2}{s+3} - 2 \beta_k.
\end{equation}
Let
\begin{align*}
    r(s, t, k, x) &\coloneqq (s-k+2)\prn{(s-1)(2s-k+2-t) - (s+3)(2-x)} \\
    &\qquad + \beta_k (s+3)(4s-2k+4-t) (2s-k+4-x-t).
\end{align*}
Re-arranging \eqref{eq:pb-7}, it suffices to show that $r(s, t, k, x) \geq 0$. Observe that $r$ is increasing in $\beta_k$ and, keeping $x+t$ fixed, in $x$.

Next, observe that $t \leq s x$ and that, if we show
\begin{equation}\label{eq:beta-lower-bound}
    \beta_k \geq \frac{2}{(k+1)(k+2)},
\end{equation}
since we are in the case $s \geq k - 1$, this implies $\beta_k \geq 2/((s+2)(s+3))$. Furthermore, $x + t \leq k$, which implies $x + t \leq s+1$. Thus, we can also replace $(x, t)$ by
\[
    (x', t') = \prn{\frac{x+t}{s+1}, \frac{s(x+t)}{s+1}},
\]
which decreases $x$ and increases $t$, keeping $x+t$ unchanged and satisfying $t' = s x'$. Substituting these bounds in the definition of $r$ yields
\begin{align*}
    r(s, t, k, x) &\geq (k-2)(k-x-t) + \frac{s (2s-k+4-x-t)}{(s+1)(s+2)} \prn{2(k-x-t)+(s-k+1)(s^2+s+2)} \\
    &\geq 0,
\end{align*}
where the last expression is non-negative because $k \geq 2$, $s \geq k-1$ and $x+t \leq k$.

Therefore, to show $r(s,t,k,x) \geq 0$, it suffices to show \eqref{eq:beta-lower-bound}. Observe that
\[
    \gamma_k = 1 - e^{-k} \frac{k^k}{k!} = e^{-k} \prn{e^k - \frac{k^k}{k!}} = e^{-k} \prn{\sum_{s = 0}^{k-1} \frac{k^s}{s!} + \sum_{s = k}^\infty \frac{k^{s+1}}{(s+1)!}} = \E[\AR_k(\Poi{k})],
\]
and thus
\[
    \beta_k = 1 - \frac{\gamma_k}{\delta_k} = 1 - \frac{\E[\AR_k(\Poi{k})]}{\E[\AR_k(\Poi{k-1})]}.
\]
Thus, \eqref{eq:beta-lower-bound} is equivalent to
\begin{align*}
    (k+1)(k+2)\prn{\E[\AR_k(\Poi{k-1})] - \E[\AR_k(\Poi{k})]} &\geq 2 \E[\AR_k(\Poi{k-1})] &\iff \\
    k(k+3) \E[\AR_k(\Poi{k-1})] &\geq (k+1)(k+2) \E[\AR_k(\Poi{k})] &\iff \\
    \frac{k(k+3)}{(k+1)(k+2)} \E[\AR_k(\Poi{k-1})] &\geq \E[\AR_k(\Poi{k})]. &
\end{align*}
Observe that
\begin{align*}
    \E[\AR_k(\Poi{k})] &= e^{-k} \sum_{j \geq 0} \frac{k^j}{j!} \min\set{1, \frac{k}{j+1}} = e^{-k} \sum_{j = 0}^{k-1} \frac{k^j}{j!} + e^{-k} \sum_{j \geq k+1} \frac{k^j}{j!} \\
    &= 1 - e^{-k} \frac{k^k}{k!},
\end{align*}
and
\begin{align*}
    \E[\AR_k(\Poi{k-1})] &= e^{-(k-1)} \sum_{j \geq 0} \frac{(k-1)^j}{j!} \min\set{1, \frac{k}{j+1}} \\
    &= e^{-(k-1)} \sum_{j = 0}^{k-1} \frac{(k-1)^j}{j!} + e^{-(k-1)} \sum_{j \geq k} \frac{k (k-1)^j}{(j+1)!} \\
    &= e^{-(k-1)} \sum_{j = 0}^{k-1} \frac{(k-1)^j}{j!} + \frac{k}{k-1} e^{-(k-1)} \sum_{j \geq k+1} \frac{(k-1)^j}{j!}.
\end{align*}
Next, let $a_k = e^{-(k-1)} (k-1)^{k-1}/(k-1)!$. We have that
\begin{align*}
     \frac{e^{-(k-1)} \sum_{j = 0}^{k-1} \frac{(k-1)^j}{j!}}{a_k} &= 1 + \sum_{j = 0}^{k-2} \prod_{\ell = j+1}^{k-1} \frac{\ell}{k-1} \leq 1 + \sum_{j = 0}^{k-2} \frac{j+1}{k-1} = \frac{k+2}{2} &\iff \\
     2 e^{-(k-1)} \sum_{j = 0}^{k-1} \frac{(k-1)^j}{j!} &\leq (k+2) a_k,
\end{align*}
which implies that
\begin{align}
    \frac{k(k+3)}{(k+1)(k+2)} e^{-(k-1)} \sum_{j = 0}^{k-1} \frac{(k-1)^j}{j!} &= \prn{1 - \frac{2}{(k+1)(k+2)}} e^{-(k-1)} \sum_{j = 0}^{k-1} \frac{(k-1)^j}{j!} \nonumber \\
    \label{eq:pb-8} &\geq e^{-(k-1)} \sum_{j = 0}^{k-1} \frac{(k-1)^j}{j!} - \frac{a_k}{k+1}.
\end{align}
Also, notice that
\[
    \frac{k}{k-1} e^{-(k-1)} \sum_{j \geq k+1} \frac{(k-1)^j}{j!} \geq \frac{k}{k-1} e^{-(k-1)} \frac{(k-1)^{k+1}}{(k+1)!} = \frac{k}{k(k+1)} e^{-(k-1)} \frac{(k-1)^k}{(k-1)!} = \frac{k-1}{k+1} a_k,
\]
which implies that
\begin{align}
    \frac{k^2(k+3)}{(k-1)(k+1)(k+2)} e^{-(k-1)} \sum_{j \geq k+1} \frac{(k-1)^j}{j!} &= \prn{1 + \frac{k^2 + k + 2}{(k-1)(k+1)(k+2)}} e^{-(k-1)} \sum_{j \geq k+1} \frac{(k-1)^j}{j!} \nonumber \\
    \label{eq:pb-9} &\geq e^{-(k-1)} \sum_{j \geq k+1} \frac{(k-1)^j}{j!} + \frac{(k-1)(k^2 + k + 2)}{k(k+1)^2(k+2)} a_k.
\end{align}
However, note that
\begin{align}
    e^{-(k-1)} \sum_{j = 0}^{k-1} \frac{(k-1)^j}{j!} + e^{-(k-1)} \frac{(k-1)^k}{k!} + e^{-(k-1)} \sum_{j \geq k+1} \frac{(k-1)^j}{j!} &= 1 &\iff \nonumber \\
    e^{-(k-1)} \sum_{j = 0}^{k-1} \frac{(k-1)^j}{j!} + e^{-(k-1)} \sum_{j \geq k+1} \frac{(k-1)^j}{j!} - 1 &= - e^{-(k-1)} \frac{(k-1)^k}{k!} &\iff \nonumber \\
    \label{eq:pb-10} e^{-(k-1)} \sum_{j = 0}^{k-1} \frac{(k-1)^j}{j!} + e^{-(k-1)} \sum_{j \geq k+1} \frac{(k-1)^j}{j!} - 1 &= - \frac{k-1}{k} a_k. &
\end{align}

Together, \eqref{eq:pb-8}, \eqref{eq:pb-9}, and \eqref{eq:pb-10} imply
\[
    \frac{k(k+3)}{(k+1)(k+2)} \delta_k - \gamma_k \geq \prn{\frac{k^2+k+4}{(k+1)^2(k+2)} - 1} a_k + e^{-k} \frac{k^k}{k!}.
\]
For $k \geq 3$, we use the fact that $\log(1+x) \geq2x/(2+x)$ for $x \geq 0$ to obtain
\[
    \frac{e^{-k} k^k/k!}{a_k} = \frac{1}{e} \prn{\frac{k}{k-1}}^{k-1} \geq e^{-1/(2k-1)} \geq 1 - \frac{1}{2k-1}.
\]
Thus,
\[
    \frac{k(k+3)}{(k+1)(k+2)} \delta_k - \gamma_k \geq \frac{(k-3)(k^2+2)}{(2k-1)(k+1)^2(k+2)} a_k \geq 0.
\]
For $k = 2$, we have $\delta_2 = 2-3/e$ and $\gamma_2 = 1 - 2/e^2$, and thus
\[
    5 \delta_2 - 6\gamma_2 = 4 - 15/e + 12/e^2 > 0.
\]
Thus \eqref{eq:beta-lower-bound} holds for every $k \geq 2$, which completes the proof.
\end{proof}

\subsubsection{The Convexity Lemma}
\label{sec:cardinality:convexity}

In this section we show the following lemma.
\begin{replemma}{lem:convexity}
For $k \geq 2$, let
\[
    F_k(\lambda) = \frac{1}{\E\brk{\AR_k(\Poi{\lambda})}}.
\]
Then, $F_k(\lambda)$ is convex in $[k-1, k]$ or, in other words, $F_k(k - u)$ is convex for $u \in [0,1]$.
\end{replemma}
\begin{proof}
Let $N \sim \Poi{\lambda}$, $L = \Pr[N < k]$, $x = \Pr[N = k]$ and $H = \Pr[N>k]$. Let $\Phi_k(\lambda) = \E\brk{\AR_k(\Poi{\lambda})}$. We have that
\[
    \Phi_k(\lambda) = \sum_{r = 0}^{k-1} \Pr[N = r] + \sum_{r = k}^{\infty} \frac{k}{r + 1} \Pr[N = r],
\]
and
\[
    \sum_{r = k}^{\infty} \frac{k}{r+1} \Pr[N = r] = \sum_{r = k}^{\infty} \frac{k}{r+1} e^{-\lambda} \frac{\lambda^r}{r!} = \frac{k}{\lambda} \sum_{r = k}^{\infty} e^{-\lambda} \frac{\lambda^{r+1}}{(r+1)!}
=
\frac{k}{\lambda}\Pr[N > k].
\]
Thus,
\[
    \Phi_k(\lambda) = L + \frac{k}{\lambda} H.
\]

We next compute the derivative of $\Phi_k$ with respect to $\lambda$. Note that
\[
    \der{L}{\lambda} = \sum_{r = 0}^{k-1} \frac{1}{r!} \der{\prn{e^{-\lambda} \lambda^r}}{\lambda} = \sum_{r = 0}^{k-2} e^{-\lambda} \frac{\lambda^r}{r!} - \sum_{r = 0}^{k-1} e^{-\lambda} \frac{\lambda^r}{r!} = - \Pr[N = k-1],
\]
and
\[
    \der{H}{\lambda} = \sum_{r = k+1}^\infty \frac{1}{r!} \der{\prn{e^{-\lambda} \lambda^r}}{\lambda} = \sum_{r = k}^\infty e^{-\lambda} \frac{\lambda^r}{r!} - \sum_{r = k+1}^\infty e^{-\lambda} \frac{\lambda^r}{r!} = \Pr[N = k] = x.
\]
Also,
\[
    x = e^{-\lambda} \frac{\lambda^k}{k!} = \frac{\lambda}{k} e^{-\lambda} \frac{\lambda^{k-1}}{(k-1)!} = \frac{\lambda}{k} \Pr[N = k-1].
\]
Thus,
\[
    \der{\Phi_k}{\lambda} = \der{L}{\lambda} + \frac{k}{\lambda} \der{H}{\lambda} - \frac{k H}{\lambda^2} = - \Pr[N = k - 1] + \frac{k}{\lambda} x - \frac{k H}{\lambda^2} = - \frac{k H}{\lambda^2}.
\]
Therefore, the second derivative of $\Phi_k$ is
\[
    \der{^2\Phi_k}{\lambda^2} = \frac{2 k H}{\lambda^3} - \frac{k}{\lambda^2} \der{H}{\lambda} = \frac{2 k H}{\lambda^3} - \frac{k x}{\lambda^2}.
\]

Since $F_k(\lambda) = 1/\Phi_k(\lambda)$, we have that
\[
    \der{^2F_k}{\lambda^2} = \frac{2 \prn{\der{\Phi_k}{\lambda}}^2 - \Phi_k \der{^2\Phi_k}{\lambda^2}}{\Phi_k^3}.
\]
Since $\Phi_k(\lambda) > 0$, it suffices to prove that
\[
    2 \prn{\der{\Phi_k}{\lambda}}^2 - \Phi_k \der{^2\Phi_k}{\lambda^2} \geq 0.
\]
Recall that
\[
    \Phi_k = L + \frac{k H}{\lambda}, \qquad \der{\Phi_k}{\lambda} = - \frac{k H}{\lambda^2}, \quad \text{ and } \quad \der{^2\Phi_k}{\lambda^2} = \frac{2 k H}{\lambda^3} - \frac{k x}{\lambda^2}.
\]
Thus,
\begin{align*}
    2 \prn{\der{\Phi_k}{\lambda}}^2 - \Phi_k \der{^2\Phi_k}{\lambda^2} &=  2 \frac{k^2 H^2}{\lambda^4} - \prn{L + \frac{k H}{\lambda}} \prn{\frac{2 k H}{\lambda^3} - \frac{k x}{\lambda^2}} \\
    &= 2 \frac{k^2H ^2}{\lambda^4} - \frac{2 k L H}{\lambda^3} + \frac{k x L}{\lambda^2} - \frac{2 k^2 H^2}{\lambda^4} + \frac{k^2 x H}{\lambda^3} \\
    &= \frac{k}{\lambda^3} \prn{\lambda x L + k x H - 2 L H}.
\end{align*}
Therefore, for $F_k$ to be convex in $[k-1, k]$, it suffices to show that, for $k-1 \leq \lambda \leq k$, we have $\lambda  x L + k x H \geq 2 L H$. If we divide by $e^{-2\lambda}$, this is equivalent to
\[
    \frac{\lambda^k}{k!} \prn{\lambda \sum_{r = 0}^{k-1} \frac{\lambda^r}{r!} + k \sum_{r = k+1}^\infty \frac{\lambda^r}{r!}} \geq 2 \prn{\sum_{r = 0}^{k-1} \frac{\lambda^r}{r!}} \prn{\sum_{r = k+1}^\infty \frac{\lambda^r}{r!}}.
\]
Let
\[
    \Psi_k(\lambda) = \lambda \frac{\frac{\lambda^k}{k!}}{\sum_{r = k+1}^\infty \frac{\lambda^r}{r!}} + k \frac{\frac{\lambda^k}{k!}}{\sum_{r = 0}^{k-1} \frac{\lambda^r}{r!}}.
\]
It suffices to show that $\Psi_k(\lambda) \geq 2$. Notice that
\[
    \sum_{r = k+1}^\infty \frac{\lambda^r}{r!} = \frac{\lambda^k}{k!} \prn{\frac{\lambda}{k+1} + \frac{\lambda^2}{(k+1)(k+2)} + \dots},
\]
thus
\[
    \Psi_k(\lambda) = \lambda \frac{1}{\sum_{r = k+1}^\infty \frac{\lambda^{r-k}}{\prod_{j = k+1}^r j}} + k \frac{\frac{\lambda^k}{k!}}{\sum_{r = 0}^{k-1} \frac{\lambda^r}{r!}}.
\]
But
\[
    \sum_{r = k+1}^\infty \frac{\lambda^{r-k}}{\prod_{j = k+1}^r j} \leq \sum_{r = k+1}^\infty \prn{\frac{\lambda}{k+1}}^{r-k} = \sum_{r = 1}^\infty \prn{\frac{\lambda}{k+1}}^r = \frac{\lambda}{k + 1 - \lambda}.
\]
Also, we claim that
\[
    k \frac{\frac{\lambda^k}{k!}}{\sum_{r = 0}^{k-1} \frac{\lambda^r}{r!}} \geq \lambda - k + 1.
\]
Indeed,
\begin{align*}
    \frac{\lambda^k}{(k - 1)!} - \prn{\lambda - k + 1} \sum_{r = 0}^{k - 1} \frac{\lambda^r}{r!} &= \frac{\lambda^k}{(k - 1)!} - \prn{\lambda - k + 1} \prn{\frac{\lambda^{k-1}}{(k-1)!} + \sum_{r = 0}^{k - 2} \frac{\lambda^r}{r!}} \\
    &= \frac{\lambda^k}{(k - 1)!} - \frac{\lambda^k}{(k-1)!} + \frac{\lambda^{k-1}}{(k-2)!} - \prn{\lambda - k + 1} \sum_{r = 0}^{k - 2} \frac{\lambda^r}{r!} \\
    &= \frac{\lambda^{k-1}}{(k-2)!} + \sum_{r = 0}^{k - 2} \prn{k - 1 - \lambda} \frac{\lambda^r}{r!} \\
    &= \sum_{r = 0}^{k - 2} \prn{k - 1} \frac{\lambda^r}{r!} + \frac{\lambda^{k-1}}{(k-2)!} - \sum_{r = 1}^{k - 1} r \frac{\lambda^r}{r!} \\
    &= \sum_{r = 0}^{k - 2} \prn{k - 1} \frac{\lambda^r}{r!} - \sum_{r = 1}^{k - 2} r \frac{\lambda^r}{r!} \\
    &= \sum_{r = 0}^{k - 2} \prn{k - 1 - r} \frac{\lambda^r}{r!} \geq 0.
\end{align*}
Putting everything together, we get that
\[
    \Psi_k(\lambda) \geq \lambda \frac{1}{\frac{\lambda}{k + 1 - \lambda}} + (\lambda - k + 1) = (k + 1 - \lambda) + (\lambda - k + 1) = 2,
\]
and the claim follows.
\end{proof}
\section{Monotonicity}
\label{app:monotonicity}

In this section, we prove that all our schemes are monotone. At the end of this appendix we show how the balance guarantees can be extended to the case $\sum_i p_i \leq k$. First, we show that, for every fixed $p \in \cP_k$ and every $i \in A_1 \subseteq A_2 \subseteq [n]$,
\begin{equation}\label{eq:monotonicity}
    \Pr\brk{i \in \pi_p(A_1)} \geq \Pr\brk{i \in \pi_p(A_2)},
\end{equation}
where the probabilities are over the samples, the internal randomness of the scheme, and, for an ROCRS, the uniformly random arrival order. The same arguments establish monotonicity for the offline CRSs obtained by running our ROCRSs in an independently generated uniformly random order.

We show \eqref{eq:monotonicity} for each scheme in its own subsection; the extension of the balance guarantees is given at the end of this appendix.

\subsection{The Conservative Scaling ROCRS}
\label{sec:monotonicity:single-item-random-order}

We start with the ROCRS of Section~\ref{sec:single-item:random-order}.

\begin{proposition}\label{prop:monotonicity:single-item-random-order}
Algorithm~\ref{alg:random-order:single-item} is a monotone ROCRS.
\end{proposition}
\begin{proof}
For each element $j$, draw an independent coin $\xi_j \sim \Ber{1-1/e}$, independently of the samples, online activations, and arrival order, which plays the role of $q_j$ in the algorithm. Let
\[
    H \coloneqq \set{j \in [n] \midd Y_j = 0 \text{ or } \xi_j = 1}.
\]
On online-active set $W$, the algorithm accepts the first element of $W \cap H$, if this set is non-empty. Condition on the samples, the coins, and the arrival order, and consider $i \in A_1 \subseteq A_2$. If the algorithm accepts $i$ on $W = A_2$, then $i \in H$ and no element of $A_2 \cap H$ precedes $i$. Since $A_1 \cap H \subseteq A_2 \cap H$, the same is true on $W = A_1$. By unconditioning, we obtain \eqref{eq:monotonicity}.
\end{proof}

Next, consider the Conservative Scaling ROCRS for general $k$, found in Appendix~\ref{app:alternative-proof}, which directly generalizes the $k = 1$ variant. Its parameter $\alpha_k$ depends only on the capacity $k$ and thus the same argument as above with $\xi_j \sim \Ber{\alpha_k}$ yields the following.

\begin{proposition}\label{prop:monotonicity:general-random-order}
Algorithm~\ref{alg:cardinality:random-order-2} is a monotone ROCRS.
\end{proposition}

\subsection{The Single-Item OCRSs}
\label{sec:monotonicity:two-samples}

We continue with the two schemes in Section~\ref{sec:single-item:adversarial:two-samples}, namely the single-sample offline CRS applied on the first samples in order to select $T$, as well as the two-sample OCRS that accepts the first online-active element strictly after $T$.

\begin{proposition}\label{prop:monotonicity:two-samples-offline}
Algorithm~\ref{alg:adv-order:two-samples} is a monotone CRS.
\end{proposition}
\begin{proof}
We interpret $T=0$ as returning the empty set and $T=i$ as returning $\set{i}$. Consider $i \in A_1 \subseteq A_2$ and observe that if $A_1 = A_2$ then \eqref{eq:monotonicity} trivially holds with equality, so assume that $A_1 \subsetneq A_2$. Since $i \in A_1$, we have $|A_2|\geq 2$. Also, the algorithm distinguishes between $\cS - i = \emptyset$ and $\cS - i \neq \emptyset$, for it to accept $i$ on $\cS = A_2$, it must be that we are in the $\cS - i \neq \emptyset$ branch of the execution and, in addition, that $Q = 0$. In this case, there are two ways we can have $T = i$: either $J = i$ and $Y^{(2)}_i = 0$ or $Y^{(2)}_J = 1$ and $J' = i$, and these are mutually exclusive events. Let $\mu \coloneqq \sum_{j=1}^n p_j \leq 1$. By the algorithm's description,
\[
    \Pr\brk{T = i \midd \cS = A_2} = \frac{1}{2} \prn{\frac{1-\mu}{|A_2|} + \frac{1}{|A_2|}\sum_{j \notin A_2} p_j + \frac{1}{|A_2|-1}\sum_{j \in A_2-i}p_j},
\]
where the probability is over the $Y^{(2)}_j$'s and the internal random coins. We distinguish two cases:
\begin{itemize}
    \item If $A_1 - i \neq \emptyset$, condition on $Q = 0$. By the same analysis as above, we have
    \begin{align*}
        \Pr\brk{T = i \midd \cS = A_1, Q = 0} &= \frac{1-\mu}{|A_1|} + \frac{1}{|A_1|} \sum_{j \notin A_1} p_j + \frac{1}{|A_1| - 1} \sum_{j \in A_1 - i} p_j \\
        &= \frac{1-\mu}{|A_1|} + \frac{1}{|A_1|} \sum_{j \notin A_2} p_j + \frac{1}{|A_1|} \sum_{j \in A_2 \setminus A_1} p_j + \frac{1}{|A_1| - 1} \sum_{j \in A_1 - i} p_j \\
        &\geq \frac{1-\mu}{|A_2|} + \frac{1}{|A_2|} \sum_{j \notin A_2} p_j + \frac{1}{|A_2| - 1} \sum_{j \in A_2 \setminus A_1} p_j + \frac{1}{|A_2| - 1} \sum_{j \in A_1 - i} p_j \\
        &= \frac{1-\mu}{|A_2|} + \frac{1}{|A_2|} \sum_{j \notin A_2} p_j + \frac{1}{|A_2| - 1} \sum_{j \in A_2 - i} p_j \\
        &= \Pr\brk{T = i \midd \cS = A_2, Q = 0},
    \end{align*}
    where the inequality follows from the facts $|A_1| \leq |A_2| - 1$ and $\mu \leq 1$. Unconditioning on $Q$ gives the desired inequality.

    \item If $A_1 - i = \emptyset$, by the description of the algorithm, we have
    \[
        \Pr\brk{T = i \midd \cS = A_1} = 1 - \frac{p_i}{2} \geq \frac{1}{2} \geq \Pr\brk{T = i \midd \cS = A_2},
    \]
    where the first inequality follows from the fact that $p_i \leq 1$ and the second one from the fact that, when $\cS = A_2$, the event $\set{T = i}$ requires $\set{Q = 0}$ which happens with probability $1/2$.
\end{itemize}
Therefore, for all $i \in A_1 \subseteq A_2$, \eqref{eq:monotonicity} holds.
\end{proof}

\begin{proposition}\label{prop:monotonicity:two-samples-online}
The two-sample OCRS that runs Algorithm~\ref{alg:adv-order:two-samples} to select a random $T$ and accepts the first online-active element arriving strictly after $T$ is a monotone OCRS.
\end{proposition}
\begin{proof}
Condition on the samples and all random choices used to construct $T$, which is independent of the online activations. For an online-active set $W$ containing $i$, the algorithm accepts $i$ exactly when
\[
    T < i \qquad \text{and} \qquad
    W \cap \set{T+1,\dots,i-1} = \emptyset.
\]
If these conditions hold when $W = A_2$, they also hold when $W = A_1$ for every $A_1 \subseteq A_2$ containing $i$. By unconditioning, we obtain \eqref{eq:monotonicity}.
\end{proof}

Since the $0.48$-balanced OCRS of Section~\ref{sec:single-item:adversarial:single-sample} also accepts an element $i \in W$ exactly when $T < i$ and $W$ contains no element strictly between $T$ and $i$, and $T$ is chosen independently from all online activations, the same argument as above yields the following.

\begin{proposition}\label{prop:monotonicity:single-sample}
Algorithm~\ref{alg:adv-order:single-sample} is a monotone OCRS.
\end{proposition}

\subsection{The Almost-Greedy ROCRS for General \texorpdfstring{$k$}{k}}
\label{sec:monotonicity:almost-greedy}

Clearly, the simple ROCRS from Section~\ref{sec:sample-complexity:random-order} that greedily accepts the first $k$ online-active elements without using any samples is monotone. For the Almost-Greedy ROCRS, we compare acceptance probabilities by averaging over the samples, the coins, and the arrival order, using the rank calculation from Section~\ref{sec:cardinality:random-order}.

\begin{proposition}\label{prop:monotonicity:almost-greedy}
Algorithm~\ref{alg:cardinality:random-order} is a monotone ROCRS.
\end{proposition}
\begin{proof}
Fix $i \in A_1 \subseteq A_2$, and let $\mu = \sum_{j=1}^n p_j \leq k$. If $A_1=A_2$, then \eqref{eq:monotonicity} trivially holds with equality, so assume that $A_1 \subsetneq A_2$. If $|A_1| \leq k-1$, the algorithm accepts $i$ with probability one when $W=A_1$, which proves \eqref{eq:monotonicity}. Henceforth, assume that $|A_1| \geq k$, so $|A_2| \geq k+1$.

For either $W=A_1$ or $W=A_2$ with $|W| \geq k+1$, the relative arrival order of the elements of $W$ is uniformly random. Thus, each possible rank of $i$ has probability $1/|W|$. The first $k-1$ ranks always lead to acceptance, contributing $(k-1)/|W|$ to the acceptance probability. If $i$ has rank $k$, it is rejected exactly when its sample is active and its acceptance coin fails. These events are independent of each other and of the arrival order, so their joint probability is $p_i \cdot k/(2k+1)$.

If $i$ has rank $k+1$, it is accepted exactly when the element of rank $k$ is rejected. Conditional on $i$ having rank $k+1$, each $j \in W-i$ is equally likely to have rank $k$, and that element is rejected with probability $kp_j/(2k+1)$. Hence, the conditional acceptance probability at rank $k+1$ is the average of these rejection probabilities over $W-i$. At any later rank, the algorithm has already accepted $k$ elements and rejects $i$. Adding the contributions, we obtain
\begin{align}
    \Pr\brk{i \in \pi_p(W)}
    &= \frac{k-1}{|W|}
       +\frac{1}{|W|}\prn{1-\frac{kp_i}{2k+1}}
       +\frac{1}{|W|(|W|-1)}\sum_{j \in W-i}\frac{kp_j}{2k+1}
       \nonumber \\
    &= \frac{k}{|W|}\prn{1-\frac{p_i}{2k+1}
        +\frac{\sum_{j \in W-i}p_j}{(2k+1)(|W|-1)}}.
    \label{eq:monotonicity:almost-greedy-probability}
\end{align}

If $|A_1|=k$, then $i$ is rejected on $W=A_1$ only when it has rank $k$ and its sample and coin cause rejection. The probability of this event is $(1/k) \cdot p_i \cdot k/(2k+1)=p_i/(2k+1)$, so its acceptance probability is $1-p_i/(2k+1)$. To bound the acceptance probability on $W=A_2$, we use $p_i \geq 0$ and $\sum_{j \in A_2-i}p_j \leq |A_2|-1$ in \eqref{eq:monotonicity:almost-greedy-probability}. This yields
\begin{align*}
    \Pr\brk{i \in \pi_p(A_2)} &\leq \frac{k}{|A_2|}\prn{1+\frac{1}{2k+1}} \leq \frac{k}{k+1}\prn{1+\frac{1}{2k+1}} = 1-\frac{1}{2k+1} \leq 1-\frac{p_i}{2k+1} \\
    &= \Pr\brk{i \in \pi_p(A_1)},
\end{align*}
where the second inequality follows from $|A_2| \geq k+1$ and the third inequality follows from $p_i \leq 1$. This proves \eqref{eq:monotonicity} in this case.

Finally, suppose that $|A_1| \geq k+1$. For both $W=A_1$ and $W=A_2$, partitioning the ground set into $\set{i}$, $W-i$, and the elements outside $W$ gives
\[
    \mu = p_i+\sum_{j \in W-i}p_j+\sum_{j \notin W}p_j.
\]
Factoring out $k/(2k+1)$ from \eqref{eq:monotonicity:almost-greedy-probability}, substituting the above identity for $p_i$ and combining the two coefficients of $\sum_{j \in W-i}p_j$, using
\[
    \frac{1}{|W|} + \frac{1}{|W| (|W| - 1)} = \frac{1}{|W|-1},
\]
yields
\begin{align*}
    \Pr\brk{i \in \pi_p(W)}
    &= \frac{k}{2k+1}\prn{
        \frac{2k+1-p_i}{|W|}
        +\frac{\sum_{j \in W-i}p_j}{|W|(|W|-1)}} \\
    &= \frac{k}{2k+1}\prn{
        \frac{2k+1-\mu+\sum_{j \notin W}p_j}{|W|}
        +\frac{\sum_{j \in W-i}p_j}{|W|-1}}.
\end{align*}
Therefore,
\begin{align*}
    \Pr\brk{i \in \pi_p(A_1)}
    &= \frac{k}{2k+1}\prn{
        \frac{2k+1-\mu+\sum_{j \notin A_2}p_j}{|A_1|}
        +\frac{\sum_{j \in A_2 \setminus A_1}p_j}{|A_1|}
        +\frac{\sum_{j \in A_1-i}p_j}{|A_1|-1}} \\
    &\geq \frac{k}{2k+1}\prn{
        \frac{2k+1-\mu+\sum_{j \notin A_2}p_j}{|A_2|}
        +\frac{\sum_{j \in A_2 \setminus A_1}p_j}{|A_2|-1}
        +\frac{\sum_{j \in A_1-i}p_j}{|A_2|-1}} \\
    &= \Pr\brk{i \in \pi_p(A_2)},
\end{align*}
where the inequality follows from the facts that $|A_1| \leq |A_2|-1$ and $2k+1-\mu \geq k+1$. Thus, \eqref{eq:monotonicity} holds in every case.
\end{proof}

\subsection{The Count Test OCRS for General \texorpdfstring{$k$}{k}}
\label{sec:monotonicity:count-test}

Finally, consider the Count Test OCRS from Section~\ref{sec:cardinality:adv-order}. We use the equivalent description of the algorithm, found in the proof of Theorem~\ref{thm:cardinality:adv-order}, in which the algorithm chooses candidate positions uniformly at random.

\begin{proposition}\label{prop:monotonicity:count-test}
Algorithm~\ref{alg:cardinality:adv-order} is a monotone OCRS for every fixed parameter $r$.
\end{proposition}
\begin{proof}
Fix $R$ and all samples, and let $N = R+\sum_{j=1}^n Y_j$. For an online-active set $W$, let $B_t(W)$ be the difference between $R+\sum_{j=1}^t Y_j$ and the number of candidates among the first $t$ elements. The same recursion as in the proof of Theorem~\ref{thm:cardinality:adv-order} gives
\[
    B_0(W) = R, \qquad \text{and} \qquad
    B_t(W) = \prn{B_{t-1}(W)+Y_t-\1_{\set{t \in W}}}^+.
\]
For $A_1 \subseteq A_2$, by induction on $t$, we have that
\[
    B_t(A_1) \geq B_t(A_2) \qquad \text{for every } t \in \set{0,\dots,n}.
\]
For the base case, we have $B_0(A_1) = B_0(A_2) = R$. Assume the inequality holds for all $t' < t$. For $t$, we have
\[
    B_t(A_1) = \prn{B_{t-1}(A_1) + Y_t - \1_{\set{t \in A_1}}}^+ \geq \prn{B_{t-1}(A_2) + Y_t - \1_{\set{t \in A_2}}}^+ = B_t(A_2),
\]
where the inequality follows by the induction hypothesis and the fact that $\1_{\set{t \in A_1}} \leq \1_{\set{t \in A_2}}$. Therefore, if $i \in A_1$ passes the count test under $A_2$, then
\[
    B_{i-1}(A_1)+Y_i \geq B_{i-1}(A_2)+Y_i \geq 1,
\]
so it also passes under $A_1$.

Conditional on $R$, the samples, and $W$, the list of candidates is fixed. The algorithm chooses $\min\set{k,N}$ of the $N$ possible candidate positions uniformly at random, so every candidate has acceptance probability $\min\set{k,N}/N$ when $N>0$. Let
\[
    a_N \coloneqq
    \begin{cases}
        1, & 0 \leq N \leq k, \\
        k/N, & N > k.
    \end{cases}
\]
Therefore, for every $i \in W$, we have
\[
    \Pr\brk{i \in \pi_p(W) \midd R,Y_1,\dots,Y_n} = a_N \1_{\set{B_{i-1}(W)+Y_i \geq 1}},
\]
which implies
\begin{align*}
    \Pr\brk{i \in \pi_p(A_1) \midd R,Y_1,\dots,Y_n} &= a_N \1_{\set{B_{i-1}(A_1)+Y_i \geq 1}} \\
    &\geq a_N \1_{\set{B_{i-1}(A_2)+Y_i \geq 1}} = \Pr\brk{i \in \pi_p(A_2) \midd R,Y_1,\dots,Y_n},
\end{align*}
where the inequality follows from the fact that the factor $a_N$ is the same for $A_1$ and $A_2$, since $N$ does not depend on $A_1, A_2$. By unconditioning, we obtain \eqref{eq:monotonicity}.
\end{proof}

\subsection{Extension}

\paragraph{Random order.}
For each of the two Conservative Scaling schemes, Almost-Greedy, and the zero-sample greedy scheme, append independent dummy elements whose probabilities sum to $k - \sum_i p_i$, splitting this mass among several elements if necessary. Keep the probabilities and samples of the original elements unchanged. In the resulting instance, forcing all dummy elements to be online-inactive leaves the distribution of every original acceptance decision unchanged, since the samples of inactive elements are never used, and the relative order of the original elements is still uniform. For the fixed resulting probability vector, the monotonicity of the active-set shows that allowing the dummy elements to be online-active can only decrease an original element's acceptance probability. The guarantee for total mass $k$ therefore implies the same guarantee for the original instance.

\paragraph{The two-sample single-item OCRS.}
Let $\mu=\sum_i p_i\leq1$ and recall that
$\cS=\{j:Y_j^{(1)}=1\}$.
Repeating the calculation in
Appendix~\ref{app:missing:self-contained}, with
$1-p_i=(1-\mu)+\sum_{j\ne i}p_j$, gives
\[
 \Pr[T=i]=\frac{p_i}{2}
     \left(1+(1-\mu)\E\left[\frac1{|\cS-i|+1}\right]\right).
\]
Indeed, conditional on $Y_i^{(1)}=1$, the singleton branch has
probability $1-p_i/2$ of choosing $i$, and the other branch
contributes, on each set $S$ containing $i$,
\[
 \frac12\left(\frac{1-\mu}{|S|}
    +\sum_{j\ne i}\frac{p_j}{|S-i-j|+1}\right).
\]
Averaging and using
$\sum_{j\ne i}p_j\E[1/(|\cS-i-j|+1)]
=\Pr[\cS-i\ne\emptyset]$ proves the formula.
Also,
\[
 \sum_i p_i\E\left[\frac1{|\cS-i|+1}\right]
     =\Pr[\cS\ne\emptyset],\qquad
 \Pr[T=0]=\frac12+\frac{1-\mu}{2}\Pr[\cS=\emptyset].
\]
Consequently $\Pr[T=0]\geq1/2$ and
$\Pr[T=i]\geq p_i/2$ for every $i$. Substitution in the
non-negative sum in the proof of
Lemma~\ref{lem:adv-order:ideal-threshold} yields $C_i\geq1/2$.

\paragraph{The one-sample single-item OCRS.}
Again write $\mu=\sum_i p_i\leq1$. The proof of
Lemma~\ref{lem:adv-order:threshold-deficit} works with
$q_i=\sum_{j\ne i}p_j\leq1-p_i$ and gives
\[
 \eps_i\leq\frac{p_iq_i(1-e^{-q_i})}{12}
       \leq\frac{1-1/e}{12}p_i(1-p_i).
\]
Use the ideal threshold with
$\Pr[T^*=0]=1-\mu/2$ and $\Pr[T^*=i]=p_i/2$.
The same telescoping identity gives
\[
 \E[g_\ell(T^*)]
     =\frac12+\frac{1-\mu}{2}g_\ell(0)\geq\frac12.
\]
The actual threshold moves $\eps_i$ mass from each $i$ to $0$.
The same loss calculation as in
Theorem~\ref{thm:adv-order:single-sample}, followed by
Lemma~\ref{lem:adv-order:g-bound}, therefore gives
\[
 C_\ell\geq\frac12-
   \sum_{i<\ell}\eps_i\bigl(g_\ell(i)-g_\ell(0)\bigr)
 \geq\frac12-\frac{1-1/e}{32}>0.48.
\]
Finally, Theorem~\ref{thm:cardinality:adv-order} already applies directly when
$\sum_i p_i \leq k$.

\end{document}